\documentclass[12pt,a4paper,english]{article}
\usepackage{longtable}
\usepackage{lscape}
\pdfoutput=1
\usepackage{graphicx}
\usepackage{epsfig}
\usepackage{dsfont}
\usepackage{amsfonts}
\usepackage{amsmath}
\usepackage{amsthm}
\usepackage{amssymb}
\usepackage{bbold}
\usepackage[english]{babel}
\usepackage[utf8]{inputenc}
\usepackage[left=0.8in,right=0.8in,top=1.0in,bottom=1.2in]{geometry}
\usepackage{calc}
\usepackage{cancel}
\usepackage{float}
\usepackage{slashed}
\usepackage{mathrsfs}
\usepackage{pdfpages}
\usepackage{tabu}
\usepackage{simplewick}
\usepackage{hyperref}
\usepackage{bm}
\usepackage{multirow}
\usepackage{hhline}
\usepackage{arydshln}
\usepackage{fancyhdr}
\usepackage{datetime}
\usepackage{booktabs}
\usepackage{cite}

\numberwithin{equation}{section}
\numberwithin{figure}{section}
\numberwithin{table}{section}
\newcommand{\be} {\begin{equation}}
\newcommand{\ee} {\end{equation}}
\newcommand{\bea} {\begin{eqnarray}}
\newcommand{\eea} {\end{eqnarray}}
\newcommand{\ba} {\begin{align}}
\newcommand{\ea} {\end{align}}
\newcommand{\no} {\nonumber}

\newcommand{\cL}{{\cal L}}

\newcommand{\cO}{{\cal O}}
\newcommand{\cC}{{\cal C}}
\newcommand{\cM}{{\cal M}}
\newcommand{\cA}{{\cal A}}
\newcommand{\ord}[1]{\mathcal{O}({#1})}
\newcommand{\cUq}[1][]{\beta_{q}^{#1}}
\newcommand{\cUu}[1][]{\beta_{u}^{#1}}
\newcommand{\cUd}[1][]{\beta_{d}^{#1}}
\newcommand{\cUqs}[1][]{(\beta_{q}^{\,#1})^\ast}
\newcommand{\cUus}[1][]{(\beta_{u}^{\,#1})^\ast}
\newcommand{\cUds}[1][]{(\beta_{d}^{\,#1})^\ast}
\newcommand{\cGq}[1][]{\kappa_q^{#1}}
\newcommand{\cGu}[1][]{\kappa_u^{#1}}
\newcommand{\cGd}[1][]{\kappa_d^{#1}}
\newcommand{\cZq}[1][]{\xi_q^{#1}}
\newcommand{\cZl}[1][]{\xi_\ell^{#1}}
\newcommand{\cZnu}[1][]{\xi_\nu^{#1}}
\newcommand{\cZe}[1][]{\xi_e^{#1}}
\newcommand{\cZu}[1][]{\xi_u^{#1}}
\newcommand{\cZd}[1][]{\xi_d^{#1}}
\newcommand{\eftU}[2][]{B_{#2}^{#1}}
\newcommand{\eftUF}[3][]{\big[B_{#2}^{#1}\big]_{#3}}
\newcommand{\eftG}[2][]{K_{#2}^{#1}}
\newcommand{\eftGF}[3][]{\big[K_{#2}^{#1}\big]_{#3}}
\newcommand{\eftZp}[2][]{\Xi_{#2}^{#1}}
\newcommand{\eftZpF}[3][]{\big[\Xi_{#2}^{#1}\big]_{#3}}
\newcommand{\eftWC}[2]{\mathcal{C}_{#1}^{\rm #2}}
\newcommand{\eftWCF}[3]{\big[\mathcal{C}_{#1}^{\rm #2}\big]_{#3}}
\newcommand{\eftWCFS}[4]{\big[\mathcal{C}_{#1}^{\rm #2}\left(#3\right)\big]_{#4}}
\newcommand{\eftOp}[2]{\mathcal{O}_{#1}^{#2}}

\newcommand{\GeV}{~\text{GeV}}

\renewcommand{\Re}{{\rm Re}}
\renewcommand{\Im}{{\rm Im}}
\newcommand{\PSC}{{\rm PS}^3}
\newcommand{\Ufive}{\mathrm{U(2)}^5}

\newcommand{\lsim}{\stackrel{<}{_\sim}}

\newcommand{\ttaumu}{s_\tau}
\newcommand{\stau}{s_\tau}
\newcommand{\ttaumuR}{\theta^R_{\tau\mu}}

\newcommand{\tbsR}{\theta^R_{b s}} 
\newcommand{\etaF}{\eta_S} 

\makeatletter
\g@addto@macro\bfseries{\boldmath}
\makeatother

\begin{document}
%%%%%%%%%%%%%%%%%%%%%%%%%%%%%%%%%%%%%%%%%%%%%%%%%%%%%%%%%%%%%%%%%%

\begin{flushright}
 ZU-TH-18/18  \\
\end{flushright}

\thispagestyle{empty}

\bigskip

\begin{center}
\vspace{1.5cm}
     {\Large\bf 
 Low-energy signatures of the $\boldsymbol{\PSC}$ model: \\[0.2 cm]
     from $B$-physics anomalies to LFV}
       \\ [1cm] 

   {\bf Marzia Bordone, Claudia Cornella,  Javier Fuentes-Mart\'{\i}n, Gino Isidori}    \\[0.5cm]
  {\em Physik-Institut, Universit\"at Z\"urich, CH-8057 Z\"urich, Switzerland}  \\
\end{center}
\vspace{1cm}

\centerline{\large\bf Abstract}
\begin{quote}
The three-site Pati-Salam gauge model provides a consistent description for  
the hints of lepton-flavor non-universality observed in $B$ decays, connecting the present pattern 
of ``anomalies" to the origin of the Standard Model Yukawa couplings. 
We present here a detailed analysis of the model predictions for a series of low-energy observables, 
mainly in $B$ and $\tau$ physics. The model is in good agreement with present data and 
predicts a well-defined pattern of non-standard effects in low-energy observables 
that could allow us to test it in the near future.
Particularly interesting are the predictions of large $\tau\to\mu$ Lepton Flavor Violating processes,
such as $\tau \to \mu\gamma$, $\tau \to 3\mu$, $B\to K \tau \mu$, and  $B_s\to\tau\mu$.  
Also $\mu \to 3 e$,  $\mu \to e\gamma$,  and $K_L \to \mu e$ decays could be not far from the 
present exclusion bounds, although this conclusion is more model dependent. 

\end{quote}

\newpage 
\tableofcontents
\newpage

%%%%%%%%%%%%%%%%%%%%%%%%%%%%%%%%%%%%%%%%%%%%%%%%%%%%%%%%%%%%%%%%%%
\section{Introduction}
%%%%%%%%%%%%%%%%%%%%%%%%%%%%%%%%%%%%%%%%%%%%%%%%%%%%%%%%%%%%%%%%%%

In a recent paper~\cite{Bordone:2017bld} we have proposed a model based on the flavor non-universal gauge group 
$\PSC=[\mathrm{SU(4)}\times \mathrm{SU(2)_L }\times \mathrm{SU(2)_R}]^3$ as an interesting 
framework to describe the hints of lepton-flavor non-universality observed in $B$ meson decays,
both in neutral currents~\cite{Aaij:2014ora,Aaij:2017vbb} and 
in charged currents~\cite{Lees:2013uzd,Aaij:2015yra,Hirose:2016wfn,Aaij:2017deq}. Besides the phenomenological success, 
the virtue of this model is the natural link between the pattern of ``anomalies'' observed so far
and the hierarchical structure of quark and  lepton mass matrices: both structures follow from the same 
dynamical breaking of the flavor symmetry present in the model.  
This, together with the unification of quarks and lepton quantum numbers 
\`a la Pati-Salam~\cite{Pati:1974yy}, makes the model quite interesting and worth being further investigated.
The purpose of this paper is to analyze in more detail the rich low-energy phenomenology of the model,
which presents several distinctive features with respect to other models 
proposed so far for a combined explanation of the two sets of anomalies.

The link between the anomalies and Yukawa couplings in the $\PSC$ model follows from an approximate
$\Ufive$ flavor symmetry~\cite{Barbieri:2011ci,Barbieri:2012uh,Blankenburg:2012nx} that, as shown in a series of recent papers, 
provides a natural starting point to address this problem~\cite{Greljo:2015mma,Barbieri:2015yvd,Bordone:2017anc,Buttazzo:2017ixm}.
Interestingly enough, in the $\PSC$ model the $\Ufive$  flavor symmetry is an accidental symmetry of the gauge sector of the theory
(below about $100$~TeV) and its breaking is controlled by the spontaneous symmetry breaking $\PSC\to{\rm SM}$.
The main TeV-scale mediator responsible for the $B$ anomalies is a vector leptoquark field, $U\sim(\mathbf{3},\mathbf{1})_{2/3}$, which 
has already been identified as an excellent single mediator for the anomalies
(assuming pure left-handed couplings) in Ref.~\cite{Barbieri:2015yvd,Calibbi:2015kma,Buttazzo:2017ixm},
and has indeed been at the center of a series of explicit 
model-building attempts~\cite{Barbieri:2016las,DiLuzio:2017vat,Calibbi:2017qbu,Barbieri:2017tuq,Blanke:2018sro,Greljo:2018tuh}.\footnote{Interesting recent attempts to explain the anomalies not based 
on vector leptoquark mediators have been presented in 
Ref.~\cite{Marzocca:2018wcf,Greljo:2018ogz,Asadi:2018wea,Dorsner:2017ufx,
Megias:2017vdg,Aloni:2017ixa,Descotes-Genon:2017ptp,Fraser:2018aqj,Camargo-Molina:2018cwu, Azatov:2018knx}.}
The difference of the $\PSC$ model with respect to these previous attempts is twofold: on the one hand,
two other TeV-scale fields can mediate flavor-changing processes: a color octet and a $Z'$ 
(as also in~\cite{DiLuzio:2017vat}); on the other hand, all these TeV fields are 
not only coupled to left-handed currents, but also to right-handed currents. 

In this paper we present a systematic analysis of the low-energy phenomenology of the model. We focus mainly on the effects of the 
TeV-scale gauge mediators in processes involving the transition of the $b$ quark and $\tau$ lepton into lighter fermions, since they 
are the most directly connected to the anomalies. In particular, we show that if the anomalies were to be confirmed, the model would 
predict a rather characteristic pattern of correlations among these observables. Processes involving only the light families, such as 
those in $K$ and $D$ physics and $\mu\to e$ transitions, are controlled by subleading free parameters (more precisely subleading 
breaking terms of the $\Ufive$ symmetry) which are constrained neither by the anomalies nor by the Yukawa couplings and are 
therefore more model dependent. As far as these transitions are concerned, we investigate the consistency of the model and the 
constraints on these subleading effects arising from neutral meson mixing and $\mu\to e$ Lepton Flavor Violating (LFV) observables.

The paper is organized as follows: in Section~\ref{sect:model} we summarize the key features of the model, focusing in particular 
on the flavor structure of the massive gauge bosons at the TeV scale. In Section~\ref{sect:LEFT} we briefly illustrate the procedure 
adopted to integrate out the heavy fields and build a corresponding low-energy effective theory.  In Section~\ref{sect:pheno} we 
present a detailed analytical discussion of the most interesting observables, 
namely $\Delta F=2$ amplitudes,  $b\to c \ell \nu$ decays, $b\to s \ell\ell$ decays, and LFV processes.
The results of a global fit and a general discussion of the low-energy phenomenology 
is presented in Section~\ref{sect:Fit}.
The results are summarized in the conclusions. 
A series of technical details about  the model, the construction of the low-energy effective theory, 
and expressions for the observables are reported in the various appendices.

%%%%%%%%%%%%%%%%%%%%%%%%%%%%%%%%%%%%%%%%%%%%%%%%%%%%%%%%%%%%%%%%%%
\section{The $\PSC$ model}\label{sect:model}
%%%%%%%%%%%%%%%%%%%%%%%%%%%%%%%%%%%%%%%%%%%%%%%%%%%%%%%%%%%%%%%%%%

In this section we briefly summarize the main features of the model, with particular 
attention to its flavor structure, that plays a key role in low-energy flavor-changing observables, and to the 
spectrum of exotic gauge bosons at the TeV scale.

\subsection{High-scale dynamics}

The gauge symmetry holding at high energies 
is $\mathrm{PS}^3\equiv\mathrm{PS}_1\times\mathrm{PS}_2\times\mathrm{PS}_3$, where
$\mathrm{PS}_i = \mathrm{SU(4)}_i\times \mathrm{[SU(2)_L]}_i\times \mathrm{[SU(2)_R]}_i$.
The fermion content is the same as in the SM plus three right-handed neutrinos, such that each fermion family
is embedded in left- and right-handed multiplets of a given $\mathrm{PS}_i$ subgroup:
$(\mathbf{4},\mathbf{2},\mathbf{1})_i$ and  $(\mathbf{4},\mathbf{1},\mathbf{2})_i$.
At this level the index $i=1,2,3$ can be identified with 
the generation index. The SM gauge group is a subgroup of the diagonal group,  $\mathrm{PS}_{\rm diag} = \mathrm{PS}_{1+2+3}$. 
The spontaneous symmetry breaking (SSB)  $\mathrm{PS}^3 \to {\rm SM}$ occurs in a series of steps at different energy scales,
with appropriate scalar fields acquiring non-vanishing vacuum expectation values (VEVs), as described in Ref.~\cite{Bordone:2017bld}.

As far as low-energy physics is concerned, we can ignore what happens above the scale where the initial gauge group
is spontaneously broken to $\mathrm{SM}_{1+2} \times \mathrm{PS}_3$. This SSB scale ($\Lambda_{12}$) is chosen sufficiently high 
to neglect the effect of the $d\ge 6$ effective operators generated at this scale,
even for rare processes such as
$K_L \to \mu e$ or $K$--$\bar K$ mixing. The key aspect of the $\mathrm{SM}_{1+2} \times \mathrm{PS}_3$ {\em local}
symmetry is the corresponding accidental $\Ufive$ {\em global} flavor symmetry~\cite{Barbieri:2011ci,Blankenburg:2012nx}
\be
\Ufive = \mathrm{U(2)}_{q}\times  \mathrm{U(2)}_{\ell}\times \mathrm{U(2)}_{u}\times \mathrm{U(2)}_{d}\times \mathrm{U(2)}_{e}\,,
\ee
acting on the 
first two generations of SM fermions, in the limit where we ignore the scalar sector of the theory.

The SSB $\mathrm{SM}_{1+2} \times \mathrm{PS}_3 \to \mathrm{SM}$
occurs below the scale $\Lambda_{23} = {\rm few} \times 10~{\rm TeV}$ 
via an appropriate set of scalar (link) fields acquiring a non-trivial VEV:\footnote{For simplicity, we classify the link fields according 
to their transformation properties under $[\mathrm{SU(2)_R}]_{1+2}$, rather than $\mathrm{[U(1)_Y]}_{1+2}$. We also 
changed notation for the link fields with respect to Ref.~\cite{Bordone:2017bld}, given we focus only in the last step of the breaking chain.}
\begin{align}
\label{eq:linkfields}
\begin{aligned}
\Phi_L&\sim(\mathbf{1},\mathbf{2},\mathbf{1})_{1+2} \times(\mathbf{1},\mathbf{\bar 2},\mathbf{1})_3~, \qquad 
\Phi_R\sim(\mathbf{1},\mathbf{1},\mathbf{2})_{1+2}\times(\mathbf{1},\mathbf{1},\mathbf{\bar 2})_3~,\\
\Omega_1  & \sim(\mathbf{1},\mathbf{2},\mathbf{1})_{1+2} \times(\mathbf{\bar 4} ,\mathbf{\bar 2},\mathbf{1})_3~, \qquad 
\Omega_3     \sim(\mathbf{3},\mathbf{2},\mathbf{1})_{1+2} \times(\mathbf{\bar 4} ,\mathbf{\bar 2},\mathbf{1})_3~.
\end{aligned}
\end{align}
The VEV of such fields obey a hierarchical pattern, $\langle \Phi_{L,R} \rangle  >  \langle \Omega_{1,3} \rangle$,
such that the heavy fields with masses proportional to  $\langle \Phi_{L,R} \rangle = \cO(10~{\rm TeV})$ 
can safely be decoupled due to their heavy mass and the $\Ufive$
flavor symmetry. 

The gauge bosons responsible for the flavor anomalies, and potentially relevant in many flavor observables, 
are those acquiring mass in the last step of the breaking chain, 
\be
\mathrm{SU(4)}_3 \times \mathrm{SU(3)}_{1+2}  \times \mathrm{SU(2)_L} \times \mathrm{ U(1)^\prime} \to~{\rm SM}~,
\label{eq:4321}
\ee
triggered by $\langle \Omega_{1,3} \rangle \not=0$ around the TeV scale.
The 15 broken generators give rise to the following massive spin-1 fields: a 
leptoquark, $U\sim(\mathbf{3},\mathbf{1})_{2/3}$, a coloron, $G^\prime\sim(\mathbf{8},\mathbf{1})_0$, and a $Z^\prime\sim(\mathbf{1},\mathbf{1})_0$. 
As we discuss below, these are not the only TeV-scale fields: the spectrum contains additional scalars and fermions with masses of
the order of a few TeV. However, these play no direct role in low-energy observables. 

Finally, the breaking of the electroweak symmetry takes place through the VEV of four SM-like Higgs fields (or two 
fields transforming as bi-doublets under  $\mathrm{SU(2)_L}\times \mathrm{SU(2)_R}$) that, before the breaking of 
$\mathrm{PS}_3$, are embedded in the following two scalars:
\begin{align}
H_1\sim(\mathbf{1},\mathbf{2},\mathbf{\bar 2})_3~, \qquad H_{15}\sim(\mathbf{15},\mathbf{2},\mathbf{\bar 2})_3\,,
\end{align}
with $\langle  H_{15} \rangle$ aligned along the $T^{15}$ generator of $\mathrm{SU(4)_3}$. 
Being singlets of $\mathrm{SM}_{1+2}$, these fields allow us to extend the $\mathrm{U(2)^5}$ symmetry also to the Yukawa sector, 
which remains exact at the level of renormalizable operators. 

\subsection{Yukawa couplings and breaking of the $\Ufive$ flavor symmetry}
\label{sect:MainYuk}
The Yukawa couplings for the light generations and, more generally,  the breaking of the $\Ufive$ symmetry,
arise from higher-dimensional operators involving the link fields  $\Omega_{1,3}$ and $\Phi_{L,R}$,
generated at the scale $\Lambda_{23}$~\cite{Bordone:2017bld}. 
Taking into account the effect of operators up to $d=7$, quark and charged-lepton Yukawa couplings assume the following general 
parametric structure
\be
Y_f \sim
\begin{pmatrix}
\frac{\langle\Phi_L\rangle \langle\Phi_{R}^\dagger \rangle}{\Lambda_{23}^2}&   \frac{\langle\Omega_{a} \rangle}{\Lambda_{23}}\\
  \frac{\langle\Phi_L\rangle \langle\Phi_{R}^\dagger \rangle \langle\Omega_{a}\rangle}{\Lambda_{23}^3} 
\phantom{\Big]^{1}}
& y^f_{3} 
\end{pmatrix}~,
\label{eq:Y1}
\ee
with $a=3\,(1)$ for quarks (leptons). Here, the 11 (12)  entry of this matrix should be understood as a $2\times2$ matrix (2-component vector)
in flavor space (see appendix~\ref{app:Yukawa}).

The only entries in Eq.~(\ref{eq:Y1}) induced by renormalizable interactions below the scale $\Lambda_{23}$ 
are the Yukawa couplings for the third generation,
which arise from 
\bea
\begin{aligned}
\mathcal{L}_{\rm Yuk}   &= 
 y_{1}     \bar \Psi^3_L  H_1  \Psi^3_R   
+ y_{15}  \bar \Psi^3_L  H_{15}  \Psi^3_R    + 
 y^\prime_{1}   \bar \Psi^3_L  H^c_1  \Psi^3_R  +
  y^\prime_{15} \bar \Psi^3_L  H^c_{15}  \Psi^3_R   
  + {\rm h.c.}~,
\label{eq:d4Yuk}
\end{aligned}
\eea
where $(\Psi^3_{L(R)})^\intercal =  [(q_{L(R)}^3)^\intercal, (\ell_{L(R)}^3)^\intercal]$ denote the PS multiplets of third-generation fermions.
Here $(q_R^3)^\intercal=(t_R, b_R)$,  $(\ell_R^3)^\intercal=(\tau_R, \nu^\tau_R)$, 
and $q^3_L$ and $\ell^3_L$ indicate the SM left-handed doublets.\footnote{In the absence of tuning, this Lagrangian predicts $y_t$ and 
$y_{\nu_\tau}$ to be of similar size. As pointed out in~\cite{Greljo:2018tuh}, this prediction can be made compatible with realistic light-neutrino 
masses by means of an appropriate inverse seesaw mechanism.}  
The $y^f_{3}$ couplings in 
Eq.~(\ref{eq:Y1}) are combinations of the $y^{(\prime)}_{1(15)}$ weighted by the VEVs of 
$H_1$ and $H_{15}$ normalised to $v=246\GeV$.
The leading terms controlling the left-handed mixing between third and second generations
are generated by the following dimension-five operators 
\be
\mathcal{L}_{\rm \Omega}^{d=5}  = 
 \frac{ y_{q3}  }{\Lambda_{23}}\,  \bar q^2_L  H_1  \Omega_3 \Psi^3_R    +
 \frac{ y_{\ell3}  }{\Lambda_{23}}\, \bar \ell^{\,2}_L  H_1   \Omega_1  \Psi^3_R    
+ 
 \frac{y_{q3}^\prime}{\Lambda_{23}}    \bar q^2_L  H_1^c  \Omega_3  \Psi^3_R     +
 \frac{ y_{\ell3}^\prime   }{\Lambda_{23}}\, \bar \ell^{\,2}_L H_1^c  \Omega_1 \Psi^3_R  
  + {\rm h.c.}
\label{eq:d5spurions}
\ee
The upper index on the left-handed doublets denotes the second family (in the interaction basis) that, by construction, 
is defined as the fermion combination appearing in these operators (see appendix~\ref{app:Yukawa}).
Similarly, operators of  $d=6$ and $7$ involving also the link fields $\Phi_{L,R}$ 
are responsible for the subleading terms in (\ref{eq:Y1}).

The dynamical origin of these higher-dimensional operators
is not relevant to analyze low-energy phenomenology. 
The only important point is the $\Ufive$  symmetry breaking structure they induce.
This is highlighted by re-writing each Yukawa  matrix in terms of three normalized $\Ufive$ breaking spurions
$\{V_L$, $V_R$, $X_{LR}\}$, with hierarchical ordered coefficients 
($|\epsilon^f_R| \ll |\epsilon^f_{LR}| \ll |\epsilon^f_L| \ll 1$):
\be
Y_f = y_3^f 
\begin{pmatrix}
 \epsilon^f_{LR}\, X_{LR} & \epsilon^{f}_L\, V_L  \\[2pt]  
\epsilon^f_{R}\, V^\intercal_R  & 1
\end{pmatrix}~.
\label{eq:Y2}
\ee
Here $V_L$ and $V_R$ are unit vectors in the $\mathrm{U(2)}_{q+\ell}$ and $\mathrm{U(2)}_{u+d+e}$ space, 
while $X_{LR}$ is a bi-fundamental spurion of  $\Ufive$. \\

We define the interaction basis for the left-handed doublets as the basis where the second generation is identified by the 
direction of leading spurion $V_{L}$  in flavor space  (i.e.~in this basis $V_{L}$ is aligned to the second generation). 
We move from the interaction to the mass basis by means of the rotations 
\be
 L_u^\dagger Y_u R_u  = {\rm diag}(y_u,y_c,y_t)~ ,\quad 
 L_d^\dagger Y_d R_d  = {\rm diag}(y_d,y_s,y_b)~,  \quad 
 L_e^\dagger Y_e R_e  = {\rm diag}(y_e,y_\mu,y_\tau)~,
\label{eq:rotations}
\ee
where the $y_i$ are  real and positive and $V_{\rm CKM}=L_u^\dagger L_d$. The left-handed rotation matrices, generated by the leading spurions, play a prominent role in the phenomenological analysis. As discussed in detail in appendix~\ref{app:Yukawa}, the known structure of the SM Yukawa couplings determines only some of the (complex) coefficients $\epsilon^f_{L,R,LR}$.  In particular three real parameters and two phases in the quark sector can be expressed in terms of CKM matrix elements, leaving us with the mixing angles and phases listed in Table \ref{tab:mixingList}. In the left-handed sector we end up with three mixing angles ($s_b$, $\stau$, $s_{e}$) and four CP-violating phases, out of which only two play a relevant role ($\phi_b$ and $\alpha_d$). The other two phases ($\phi_\tau$ and $\alpha_e$) are set to zero for simplicity. The left-handed mixing angles, which are nothing but the magnitudes of the $\epsilon^f_L$ parameters in the down and charged-lepton sector, are expected to be small, the natural size being set by $|V_{ts}|$. The subleading right-handed rotations in the lepton sector, controlled by the parameter $\epsilon^e_R$, play an important role in the rare $B_s\to\bar \mu\mu$ decay and in LFV transitions. Right-handed rotations in the quark sector, controlled by $\epsilon_R^d$ and $\epsilon_R^u$, do not significantly affect the phenomenology and thus are neglected in the following.
\begin{table}[t]
\centering
\renewcommand{\arraystretch}{1.2}
\begin{tabular}{ccccc}
\toprule
 & \multicolumn{3}{c}{Parameters} & Natural size\\
\midrule
\multirow{2}{*}{\centering Left-handed mixing} & $\epsilon_{L}^{u,d},\,\epsilon_{LR}^{u,d}$ & $\stackrel{\rm CKM}{\longrightarrow}$  & $s_b$, $\phi_b$, $\alpha_d$ &  $s_b = \mathcal{O}(|V_{ts}|)$\\
 &$\epsilon_{L}^{e}$, $\epsilon_{LR}^e$ & $\longrightarrow$ & $s_\tau$, $s_e$, $\phi_\tau$, $\alpha_e$ & $s_\tau = \mathcal{O}(|V_{ts}|)$, $s_e \ll s_{\tau}$\\
\midrule
\multirow{2}{*}{\centering Right-handed mixing} & \multicolumn{3}{c}{$\epsilon_R^d$, $\epsilon_R^u$}  & $|\epsilon_R^d|=\mathcal{O}(\frac{m_{s}}{m_{b}} s_{b})$, $|\epsilon_R^u|=\mathcal{O}(\frac{m_{c}}{m_{t}} |V_{cb}| )$ \\
& \multicolumn{3}{c}{$\epsilon_R^e$} & $|\epsilon_R^e|=\mathcal{O}(\frac{m_{\mu}}{m_{\tau}} s_{\tau})$ \\
\bottomrule
\end{tabular}
\caption{Flavor mixing parameters arising from the $\mathrm{U(2)}^5$-breaking spurions in the Yukawa sector. The mixing parameters in the left-handed sector ($\epsilon_{L,LR}^f$) are parameterized in terms of mixing angles and phases after removing terms fixed by known 
CKM elements.
The parameters $\phi_\tau$, $\alpha_e$, and $\epsilon_R^{u,d}$ are listed for completeness but are set to zero in the phenomenological analysis
since they play a marginal role (see main text).
}\label{tab:mixingList}
\end{table}

\subsubsection{Additional $\Ufive$ breaking from non-Yukawa operators}\label{sec:noYukBreak}
An additional  important aspect to analyze low-energy physics is the fact that the $\Ufive$ breaking 
 is not confined only to the Yukawa sector, 
but it appears also in other effective operators. 
Among them, those with phenomenological implications at low energies are
the $d=6$ operators bilinear in the light fermion fields and in the $\Omega_{1,3}$ link fields:
\bea
\begin{aligned}
\mathcal{L}_{\rm \Omega}^{d=6} &=  
\frac{ c_{q \ell}  }{\Lambda^2_{23}}  (X_{q\ell})_{ij}  \,   {\rm Tr}[ i \Omega_1^\dagger  D^\mu \Omega_3 ]
( \bar q^i_L  \gamma_\mu \ell^j_L)
+ \frac{ c_{qq}  }{\Lambda^2_{23}}  (X_{qq})_{ij}  \,
 {\rm Tr}[ i \Omega_3^\dagger   D^\mu \Omega_3]
( \bar q^i_L  \gamma_\mu q^j_L) \\
&+  \frac{ c_{\ell\ell}  }{\Lambda^2_{23}}  (X_{\ell\ell})_{ij} 
 {\rm Tr}[ i \Omega_1^\dagger   D^\mu \Omega_1]
( \bar \ell^i_L  \gamma_\mu \ell^j_L)~+~{\rm h.c.}\,,
\end{aligned}
\label{eq:d6eps}
\eea
(with $i,j=1,2$). These operators introduce three new bi-fundamental spurions of $\Ufive$, 
$X_{q\ell} \sim 2_{q}\times\bar 2_\ell$, $X_{\ell\ell} \sim 2_{\ell} \times \bar 2_\ell$, and $X_{qq}\sim 2_{q}\times\bar 2_{q}$
that, as shown below, modify the couplings of the TeV-scale vectors to the  SM fermions.
In order to simplify the phenomenological discussion, it is convenient 
to define a {\em minimal breaking structure} for these additional spurions
\be
\left. X_{qq} \right|_{\rm min} =  \mathbb{1}~, \qquad  \left.  X_{\ell\ell} \right|_{\rm min}  = \left. X_{q\ell} \right|_{\rm min} =  {\rm diag}(0,1)~,
\label{eq:minU2}
\ee
corresponding to $\Ufive$ symmetric couplings for quark currents, and breaking terms 
aligned to those appearing in the Yukawa couplings for lepton currents.
As we show in Section~\ref{sect:pheno}, such minimal breaking structure helps evading the tight bounds from neutral meson mixing
while maximizing the impact on the $b\to s\ell\ell$ anomalies.  In the limit where we neglect deviations from this structure, 
the relevant parameters controlling the breaking of $\Ufive$ in the coupling of the TeV-scale leptoquark and $Z^{\prime}$ are
\be
\epsilon_U ~=~ c_{q \ell}\,  \frac{ \omega_1 \omega_3  }{\Lambda^2_{23} }\,, \qquad \epsilon_\ell ~=~ c_{\ell \ell}\,  \frac{ \omega_1^{2}}{\Lambda^2_{23}}\,,
\ee
with $\omega_{1,3}$ defined in (\ref{eq:omegadef}).   
For completeness we also mention the $\Ufive$-preserving parameter
\be
\epsilon_q ~=~  c_{qq}\,  \frac{\omega_3^{2}}{\Lambda^2_{23} }\,, 
\ee
which however does not play any role in the phenomenological analysis. Deviations from the minimal $\Ufive$ breaking stucture of Eq.\ref{eq:minU2} are possible, and are unavoidably generated when considering the product of two or more spurions, hence they are expected to be small. 
Leading and sub-leading $\Ufive$-breaking parameters are summarized in Table \ref{tab:epsilonList}, together with their expected relative size
(see Eq.(\ref{eq:DeltaU}) for the definition of the subleading terms).
 Analogous sub-leading $\mathrm{U(2)}_\ell$ breaking parameters could also be present; however, their effect is irrelevant and thus we do not consider them here.

In appendix~\ref{sect:Omegaeffops} we present an  explicit dynamical realization of $\mathcal{L}_{\rm \Omega}^{d=5}$ and $\mathcal{L}_{\rm \Omega}^{d=6}$ in terms of 
heavy fields to be integrated out. In particular, we show how these operators and the minimal breaking structure
can be generated by integrating out an appropriate set of TeV-scale vector-like fermions
with renormalizable interactions at the scale of unbroken $\mathrm{SM}_{1+2} \times \mathrm{PS}_3$.
A discussion about the possible deviations from the minimal breaking structure in Eq.~(\ref{eq:minU2}), is also presented.
\begin{table}[t]
\centering
\renewcommand{\arraystretch}{1.2}
\begin{tabular}{ccccc}
\toprule
Breaking & Leading & Sub-leading & Sub-sub-leading & Natural size \\
\midrule
$\mathrm{U(2)}_q\times\mathrm{U(2)}_\ell$ & $\epsilon_{U}$ &  $\tilde\epsilon^d_U$, $\tilde\epsilon^e_U$ & $\Delta \epsilon_{U}$ & $\tilde\epsilon^{d,e}_U   = \cO(\epsilon_U s_{d,e}),\,\Delta\epsilon_U =  \cO(\epsilon_U s_e s_d)$  \\
$\mathrm{U(2)}_{q}$ &- & $\tilde \epsilon_{q}$  & $\Delta \epsilon_{q}$ & $ \tilde\epsilon_q   = \cO(\epsilon_U \ttaumu ),  \,\Delta\epsilon_q  =  \cO(\epsilon^2_U)$ \\
$\mathrm{U(2)}_{\ell}$ & $\epsilon_{\ell}$  & - & - & $\epsilon_{\ell} = \cO(\epsilon_U)$  \\
\bottomrule
\end{tabular}
\caption{$\Ufive$ breaking parameters arising from non-Yukawa operators. Only $\epsilon_U$ is used as free parameter in the fit. All the 
subleading terms are set to zero after checking that bounds set by present data are less stringent than the expected natural size.
} \label{tab:epsilonList}
\end{table}
In principle, also $d=6$ operators involving right-handed light fermion fields could be relevant at low-energies.
 However, it is easy to conceive ultraviolet completions where such operators are not generated (or are
extremely suppressed), as  in the example presented in the appendix~\ref{sect:Omegaeffops}.
As argued in Ref.~\cite{Bordone:2017bld} (see discussion in Sec.~II.B of this reference),
all other \textit{$\mathrm{U(2)}$-violating operators at} $d=6$ operators either contribute to the Yukawa couplings or have 
negligible impact at low energies. In particular, given the connection of $\mathrm{U(2)}$-violating terms with the link fields, 
$\mathrm{U(2)}^5$-violating four-fermion operators are forbidden in our model.

\subsection{The model at the TeV scale}\label{sec:modelTeV}
Here we focus on the last step of the breaking chain before reaching the SM, namely Eq.~(\ref{eq:4321}).
  With an obvious notation, we denote the gauge couplings 
before the symmetry breaking 
by $g_{i}$, with $i=1\ldots 4$, and the gauge fields of $\mathrm{SU(4)_3}$, 
$\mathrm{SU(3)_{1+2}}$,  and $\mathrm{U(1)^\prime}$ by $H^a_\mu$,  $A^a_\mu$, and $B^\prime_\mu$, respectively. The symmetry breaking in Eq.~(\ref{eq:4321}) occurs via the 
VEVs of $\Omega_{1,3}$ along the SM direction, that we normalize as
\begin{align}
\langle\Omega^\intercal_{3}\rangle&=\frac{1}{\sqrt{2}}
\begin{pmatrix}
\omega_3 & 0 & 0\\
0 & \omega_3 & 0\\
0 & 0 & \omega_3\\
0 & 0 & 0\\
\end{pmatrix}
\,,\quad
\langle\Omega^\intercal_{1}\rangle=\frac{1}{\sqrt{2}}
\begin{pmatrix}
0\\
0\\
0\\
\omega_1\\
\end{pmatrix}
\,,
\label{eq:omegadef}
\end{align}
with $\omega_{1,3} = \mathcal{O}(\mathrm{TeV})$. 
These scalar fields can be decomposed under the unbroken SM subgroup as $\Omega_3\sim(\textbf{8},\textbf{1})_0\oplus(\textbf{1},\textbf{1})_0\oplus(\textbf{3},\textbf{1})_{2/3}$ and $\Omega_1\sim(\mathbf{\bar 3},\textbf{1})_{-2/3}\oplus(\textbf{1},\textbf{1})_0$. As a result, after removing the Goldstones, we end up with a real color octect, one real and one complex singlet, and a complex leptoquark.

The gauge spectrum, which coincides with the one originally proposed in Ref.~\cite{DiLuzio:2017vat}, contains the following massive fields 
\bea
\begin{aligned}
 & U_\mu^{1,2,3}=\frac{1}{\sqrt{2}}\left(H_{\mu}^{9,11,13}-iH_{\mu}^{10,12,14}\right)~, \quad 
 G_\mu^{\prime\, a} =\frac{1}{ \sqrt{  g_4^2+g_3^2  } } \left( g_3 A_{\mu}^a- g_4 H_{\mu}^a \right)~, \\
&  Z^\prime_\mu = \frac{1}{ \sqrt{g_4^2 + \frac{2}{3}\, g_1^2 }  } \left( g_4  H_{\mu}^{15}- \sqrt{\frac{2}{3}} \,g_1  B^\prime_\mu \right)~, 
\end{aligned}
\eea
with masses 
\be
M_{U}=\frac{g_4}{2}\sqrt{\omega_1^2+\omega_3^2}\,, \quad 
M_{G^\prime}= \sqrt{ \frac{  g_4^2+g_3^2 }{2} } \,   \omega_3  \,,\quad 
M_{Z^\prime}= \frac{1}{2} \sqrt{ \frac{3}{2} g_4^2+ g_1^2} \sqrt{\omega_1^2+\frac{\omega_3^2}{3}}\,.
\ee
The combinations orthogonal to $G_\mu^{\prime\, a}$ and $ Z^\prime_\mu$ are the 
SM gauge fields $G_\mu^a$ and $ B_\mu$, whose couplings are 
$g_c=g_3 g_4/\sqrt{  g_4^2+g_3^2 }  $ and $g_Y=  g_1 g_4 / \sqrt{g_4^2 + \frac{2}{3}\, g_1^2 }$. For later convenience, we introduce the effective couplings 
\be
 g_U \equiv g_4~, \qquad 
 g_{G^\prime} \equiv \sqrt{g_U^2-g_c^2}~,   \qquad
 g_{Z^\prime}\equiv\frac{1}{2\sqrt{6}}\,\sqrt{g_U^2-\frac{2}{3}g_Y^2}\,,
\ee
that control the strength of the interactions with third-generation fermions.
Note that in the limit $g_4\gg g_3$ (hence $g_U \gg g_c$), one has $g_U\approx g_{G^\prime}\approx 2\sqrt{6}\,g_{Z^\prime}$. 

The interactions of the heavy gauge bosons with SM fermions (and right-handed neutrinos) are described by the following Lagrangian 
\be
 {\cal L}_{\rm int} ~\supset~ \frac{g_U}{\sqrt{2}}  \left( U_\mu J_U^\mu+{\rm h.c.}\right) -g_{G^\prime}\, G^{\prime \,a}_\mu J_{G^\prime}^{\mu\,a}
 -g_{Z^\prime}\, Z^\prime_\mu J_{Z^\prime}^\mu \,,
 \ee
where
\bea
\begin{aligned}
  J_U^\mu & \supset \overline{q}_L  N^L_U \gamma^\mu \ell_L +  \overline{u}_R N^R_U \gamma_\mu \nu_R +\overline{d}_R N^R_U \gamma_\mu e_R~, \\
J_{G^\prime}^{\mu\,a} & \supset  \overline{q}_L  N^L_{G^\prime} \gamma^\mu T^a q_L  + 
\overline{u}_R  N^R_{G^\prime} \gamma^\mu T^a u_R +\overline{d}_R   N^R_{G^\prime} \gamma^\mu T^a d_R~,  \\
J_{Z^\prime}^\mu &\supset  3\,\overline{\ell}_L N^\ell_{Z^\prime} \gamma^\mu \ell_L  + 3\,  \overline{\nu}_R  N^\nu_{Z^\prime}  \gamma^\mu \nu_R
-\overline{q}_L N^q_{Z^\prime} \gamma^\mu q_L  \\
&\quad + 3\, \overline{e}_R N^e_{Z^\prime} \gamma^\mu e_R
-\overline{u}_R N^u_{Z^\prime} \gamma^\mu u_R -\overline{d}_R  N^d_{Z^\prime} \gamma^\mu d_R~,
\end{aligned}
\label{eq:UGZcurr0}
\eea
and the $N$'s are $3\times3$ matrices in flavor space.
In the absence of $\Ufive$ breaking, these matrices assume the following form in the interaction basis
\begin{align}
\begin{aligned}
N^{L,R}_{U}&= N_U \equiv \mathrm{diag}\left(0,0,1\right)~, \\
N_{Z^\prime}^\ell&= N_{Z^\prime}^q = N_{Z^\prime} \equiv \mathrm{diag}\left(-\frac{2}{3} \left(\frac{g_1}{g_4}\right)^2,-\frac{2}{3} \left(\frac{g_1}{g_4}\right)^2,1\right)~,\\
N_{G^\prime}^{L,R} &=N_{G^\prime} \equiv \mathrm{diag}\left(-\left(\frac{g_3}{g_4}\right)^2,-\left(\frac{g_3}{g_4}\right)^2,1\right)~, \\
N_{Z^\prime}^{\nu(e)}&=N_{Z^\prime} \pm \frac{2}{3} \left(\frac{g_1}{g_4}\right)^2\,\mathbb{1}~, \qquad\qquad 
N_{Z^\prime}^{u(d)}=N_{Z^\prime}\mp 2 \left(\frac{g_1}{g_4}\right)^2 \mathbb{1}~. 
\end{aligned}
\end{align}
The inclusion of the effective operators of $\mathcal{L}_{\rm \Omega}^{d=6}$ in Eq.~(\ref{eq:d6eps}) modifies these flavor couplings into
\begin{align}\label{eq:CoupMod}
\begin{aligned}
& N^L_{U}\to    \begin{pmatrix}  \epsilon_U X_{ql}  & 0 \\ 0 & 1 \end{pmatrix}~, \qquad 
N_{Z^\prime}^\ell \to N_{Z^\prime} +\begin{pmatrix}   \epsilon_{\ell} X_{\ell\ell}  & 0 \\ 0 & 0 \end{pmatrix}~, \\
&  N_{Z^\prime}^q (N^L_{G^\prime}) \to N_{Z^\prime} (N_{G^\prime}) +\begin{pmatrix}  \epsilon_q X_{qq}  & 0 \\ 0 & 0 \end{pmatrix}~. 
\end{aligned}
\end{align}
As discussed in appendix~\ref{sect:Omegaeffops}, the natural size for the $\epsilon_{\ell,q,U}$ parameters is  $10^{-3} \lsim  | \epsilon_{\ell,q,U} | \lsim 10^{-2}$.
In the limit where we adopt the  minimal breaking structure in Eq.~(\ref{eq:minU2})
the $Z^\prime$ and $G^\prime$  couplings to quarks remain $\mathrm{ U(2)_q}$ symmetric. 
Additional modifications to the couplings in Eq.~\eqref{eq:CoupMod} arise when considering
 deviations from the minimal breaking structure (see Table~\ref{tab:epsilonList}). In this case one finds
\bea
N^L_{G^\prime} (N_{Z^\prime}^q)  
  &\to& \left. N^L_{G^\prime} (N_{Z^\prime}^q) \right|_{\mathrm{U(2)_q-symm}} 
  + \begin{pmatrix}  0 & 0 & 0 \\ 0 & \Delta\epsilon_q  &  \tilde\epsilon_q  \\ 0 & \tilde\epsilon^{\,*}_q  & 0 \end{pmatrix}, 
    \label{eq:DeltaU} \\ 
 N^L_{U} &\to&   \begin{pmatrix}  \Delta\epsilon_U &   \tilde\epsilon^d_U & 0 \\ \tilde\epsilon^{e}_U &  \epsilon_U &  0  \\ 0 & 0  & 1 \end{pmatrix}. \no
\eea 
These subleading effects are specially relevant in two cases:
i)~$\mathrm{U(2)_{q}}$ violating terms in the $Z^\prime$ and $G^\prime$ couplings to quarks, 
which are severely constrained by $\Delta F=2$ amplitudes;
ii)~non-vanishing entries of the $U$ couplings involving the first family, which 
receive important constraints from $K_L \to \mu e$.

When discussing low-energy observables, the heavy vectors are integrated out and the overall strength of their interactions 
is controlled by three effective Fermi-like couplings 
\be
 C_U \equiv \frac{g_U^2v^2}{4M_U^2} =  \frac{v^2}{ \omega_1^2 + \omega_3^2}~, 
 \qquad   C_{G^\prime} \equiv \frac{g_{G^\prime}^2 v^2}{4M_{G^\prime}^2}~, \qquad 
 C_{Z^\prime} \equiv \frac{g_{Z^\prime}^2v^2}{4M_{Z^\prime}^2}~,
\label{eq:effFermi}
\ee
which span a limited range depending on the values of $\omega_1$ and $\omega_3$ and, to a smaller extent, $g_U$.
These effective couplings (or better $\omega_1$ and $\omega_3$), 
together with the flavor parameters listed in Tables~\ref{tab:mixingList} and~\ref{tab:epsilonList}, are the free parameters
used in the phenomenological analysis of the low-energy observables.

%%%%%%%%%%%%%%%%%%%%%%%%%%%%%%%%%%%%%%%%%%%%%%%%%%%%%%%%%%%%%%%%%%
\section{Construction of the low-energy EFT}\label{sect:LEFT}
%%%%%%%%%%%%%%%%%%%%%%%%%%%%%%%%%%%%%%%%%%%%%%%%%%%%%%%%%%%%%%%%%%

The construction of the EFT relevant for low-energy phenomenology  occurs in three steps: i)~we integrate out the 
TeV fields at the tree-level, matching the theory into the so-called SM effective field theory (SMEFT), for which we adopt the Warsaw operator 
basis~\cite{Grzadkowski:2010es};  ii)~the SMEFT operators are evolved down to the electroweak scale using the one-loop
Renormalization Group (RG) equations  
in Ref.~\cite{Jenkins:2013wua,Jenkins:2013zja,Alonso:2013hga}. At this point, all the ingredients necessary to check 
possible modifications of the on-shell $W$ and $Z$ couplings are available. For all the other observables 
a third step is needed: iii)~the heavy SM fields are integrated out and the theory is matched into a low-energy effective field theory (LEFT)
containing only light SM fields~\cite{Jenkins:2017jig}. The key points of these three steps are briefly illustrated below.

\subsection{Matching heavy gauge boson contributions to the SMEFT}\label{sec:Model2SMEFT}
Moving from the interaction basis to the  quark down-basis, defined in~\eqref{eq:down_basis}, and the mass-eigenstate basis of charged 
leptons, the currents in Eq.~(\ref{eq:UGZcurr0}) assume the form 
\begin{align}
\begin{aligned}
J_U^\mu &\supset \overline{q}_L  \,\cUq \gamma^\mu \ell_L + \overline{u}_R\,\cUu \gamma^\mu \nu_R + \overline{d}_R\, \cUd \gamma^\mu e_R\,,\\
J_{G^\prime}^{\mu\,a}  &\supset \overline{q}_L \,\cGq \gamma^\mu T^a q_L + \overline{u}_R \,\cGu \gamma^\mu T^a u_R  + \overline{d}_R \,\cGd \gamma^\mu T^a d_R \,, \\
J_{Z^\prime}^\mu &\supset 3 \, \overline{\ell}_L \,\cZl \gamma^\mu \ell_L - \overline{q}_L \,\cZq \gamma^\mu q_L + 3 \, \overline{\nu}_R \,\cZnu \gamma^\mu \nu_R + 3 \,\overline{e}_R \,\cZe \gamma^\mu e_R \\
&\quad-\overline{u}_R \,\cZu \gamma^\mu u_R -\overline{d}_R \,\cZd \gamma^\mu d_R+2\left(\frac{g_1}{g_4}\right)^2\phi^\dagger \,i\overleftrightarrow{D}_{\!\!\!\mu}\,\phi\,,
\end{aligned}
\label{eq:currents}
\end{align}
where the new flavor structures
are expressed in terms of the $N$'s and the unitary rotation matrices that diagonalize the Yukawa couplings:
\begin{align}
\begin{aligned}
\cUq &= L_d^\dagger N_U^L L_\ell\,, & \cGq &= L_d^\dagger N_{G^\prime}^L L_d\,, & \cZq &= L_d^\dagger N_{Z^\prime}^q L_d\,, & \cZl &= L_e^\dagger N_{Z^\prime}^\ell L_e\,,\\
\cUu &= R_u^\dagger N_U^R R_\nu\,, & \cGu &= R_u^\dagger N_{G^\prime}^R R_u\,, & \cZu &= R_u^\dagger N_{Z^\prime}^u R_u \,, & \cZe &= R_e^\dagger N_{Z^\prime}^e R_e \,,\\
\cUd &= -R_d^\dagger N_U^R R_e\,, & \cGd &= R_d^\dagger N_{G^\prime}^R R_d\,,  &  \cZd &= R_d^\dagger N_{Z^\prime}^d R_d\,, &  \cZnu &= R_\nu^\dagger N_{Z^\prime}^\nu R_\nu\,.
\end{aligned}
\end{align}
The relative sign in $\beta_d$ follows from the phase choice discussed  in appendix~\ref{app:Yukawa}.  This phase choice fixes 
the sign of the scalar contribution to $\Delta R_{D^{(*)}}$, see Eqs.~\eqref{eq:bctanuLag} and~\eqref{eq:DRDCU}, and therefore it plays a key 
role in the explanation of the $R_{D^{(*)}}$ anomalies. Also note that, in the case of the $Z'$ current, we have included also the contribution 
of the SM Higgs ($\phi$), which is obtained combining the four SM-like Higgses of the model.

By integrating out $U$, $Z^{\prime}$ and $G^{\prime}$ at the tree level we obtain the effective Lagrangians 
\begin{align}
\begin{aligned}
\mathcal{L}_{\rm EFT}^U &= -\frac{4\,G_F}{\sqrt{2}}\,C_U\,J_{U}^\mu J_{U\,\mu}^\dagger= -\frac{2}{v^2}\,C_U \sum_k \eftU{k}\, Q_k\,,   \\
\mathcal{L}_{\rm EFT}^{G^\prime}  &= -\frac{4\,G_F}{\sqrt{2}}\,C_{G^\prime}\,(J_{G^\prime}^\mu)^2=-\frac{2}{v^2}\,C_{G^\prime} \sum_k \eftG{k}\, Q_k\,,\\
\mathcal{L}_{\rm EFT}^{Z^\prime}&= -\frac{4\,G_F}{\sqrt{2}}\,C_{Z^\prime}\,(J_{Z^\prime}^\mu)^2=-\frac{2}{v^2}\,C_{Z^\prime} \sum_k \eftZp{k}\, Q_k\,,  \\
\end{aligned}
\end{align}
where $Q_{k}$ denote the SMEFT operators in the Warsaw basis~\cite{Grzadkowski:2010es}, 
plus additional dimension six operators involving right-handed neutrinos, reported in Table~\ref{eq:operators_nuR}. 
More compactly,
\begin{align}
\begin{aligned}
\mathcal{L}_{\rm SMEFT} = - \frac{4\,G_F}{\sqrt{2}} \sum_k \mathcal{C}_{k} Q_{k} \qquad\;\; \mathcal{C}_{k}= C_U \eftU{k} + C_{G^\prime} \eftG{k} +  C_{Z^\prime} \eftZp{k} \,.
\end{aligned}
\label{eq:lag_smeft}
\end{align}
Tables \ref{tab:no4ferm}, \ref{tab:4ferm}, and \ref{tab:4ferm_nuR} contain the tree level matching results for the SMEFT Wilson coefficients $\mathcal{C}_k$. 

\subsection{From the SMEFT to the LEFT}

After matching, we perform the RG evolution of the resulting Wilson coefficients using DsixTools~\cite{Celis:2017hod}. RG effects are particularly important for the scalar operators and for dimension-six operators in the $\psi^2\phi^2\,D$ category. The latter introduce modifications to the $W$ and $Z$ after SSB (see e.g.~\cite{Jenkins:2017jig})\footnote{Contributions to other dimension-six operators that could potentially induce $W$ and $Z$ coupling modifications, such as those of the class $X^2 H^2$ or $Q_{HD}$,  are negligible in our model.} which are tightly constrained by electroweak precision data at LEP as well as by universality tests in lepton decays~\cite{Feruglio:2016gvd,Feruglio:2017rjo,Cornella:2018tfd}. 
NP effects below the electroweak scale are conveniently described in terms of a low-energy effective field theory (LEFT) in which the $W$, the $Z$, the $t$ and the Higgs have been integrated out:
\begin{align}
\mathcal{L}^{\rm LEFT}=-\frac{4 G_F}{\sqrt{2}}\sum_k\,\eftWC{k}{}\mathcal{O}_k\,.
\end{align}
We then proceed by matching the SMEFT to the LEFT and provide the expressions for the relevant observables in terms of its Wilson coefficients. 
We adopt the same operator basis for the LEFT as in Table 7 of Ref.~\cite{Jenkins:2017jig}, where the matching conditions between the SMEFT and the LEFT can also be found.

%%%%%%%%%%%%%%%%%%%%%%%%%%%%%%%%%%%%%%%%%%%%%%%%%%%%%%%%%%%%%%%%%%
\section{The key low-energy observables}\label{sect:pheno}
%%%%%%%%%%%%%%%%%%%%%%%%%%%%%%%%%%%%%%%%%%%%%%%%%%%%%%%%%%%%%%%%%%

In what follows we provide simplified expressions for the most relevant low-energy observables, and discuss their role in constraining 
the model and in offering future test of this framework. This simplified expressions are mainly for illustration purposes; 
for all figures and numerical estimates throughout the paper we use the full expressions quoted in appendix~\ref{app:obs}.

\subsection{$\Delta F=2$ transitions}\label{subsec:DF2}

\begin{figure}[p]
\centering
\includegraphics[width=0.45\textwidth]{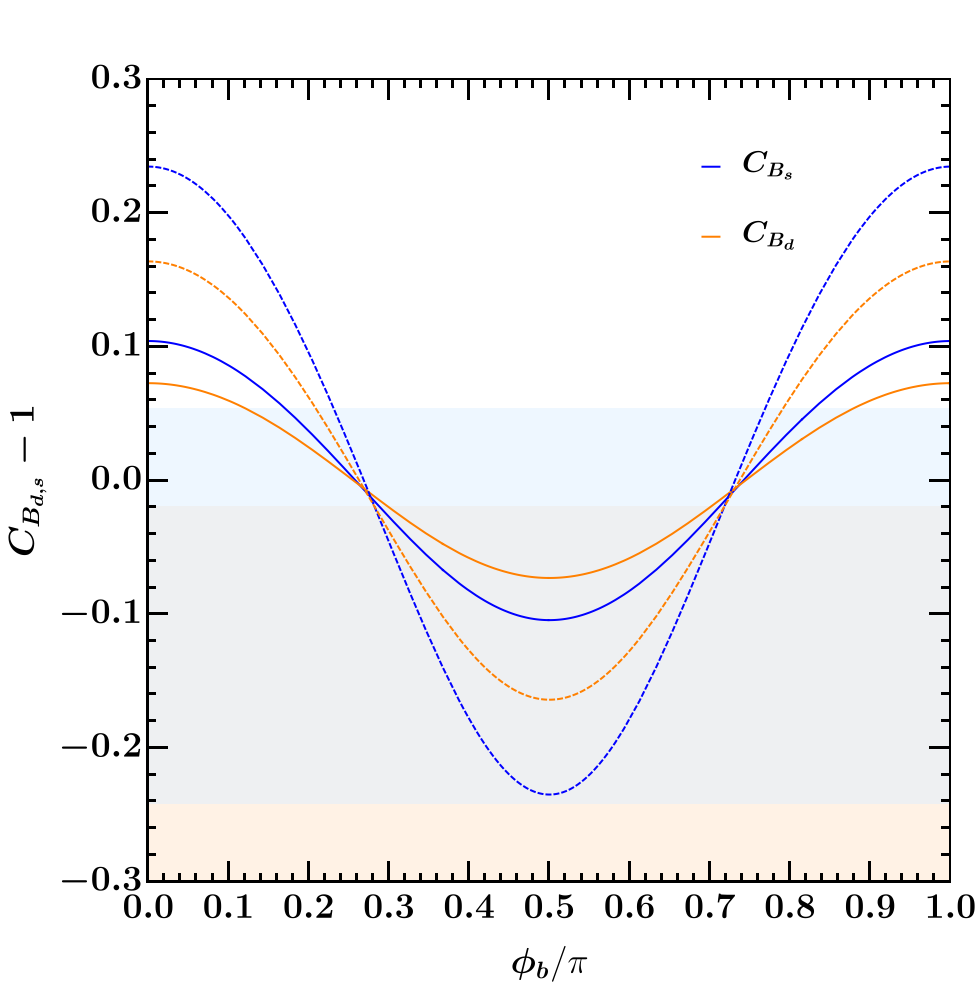}\qquad\quad\includegraphics[width=0.45\textwidth]{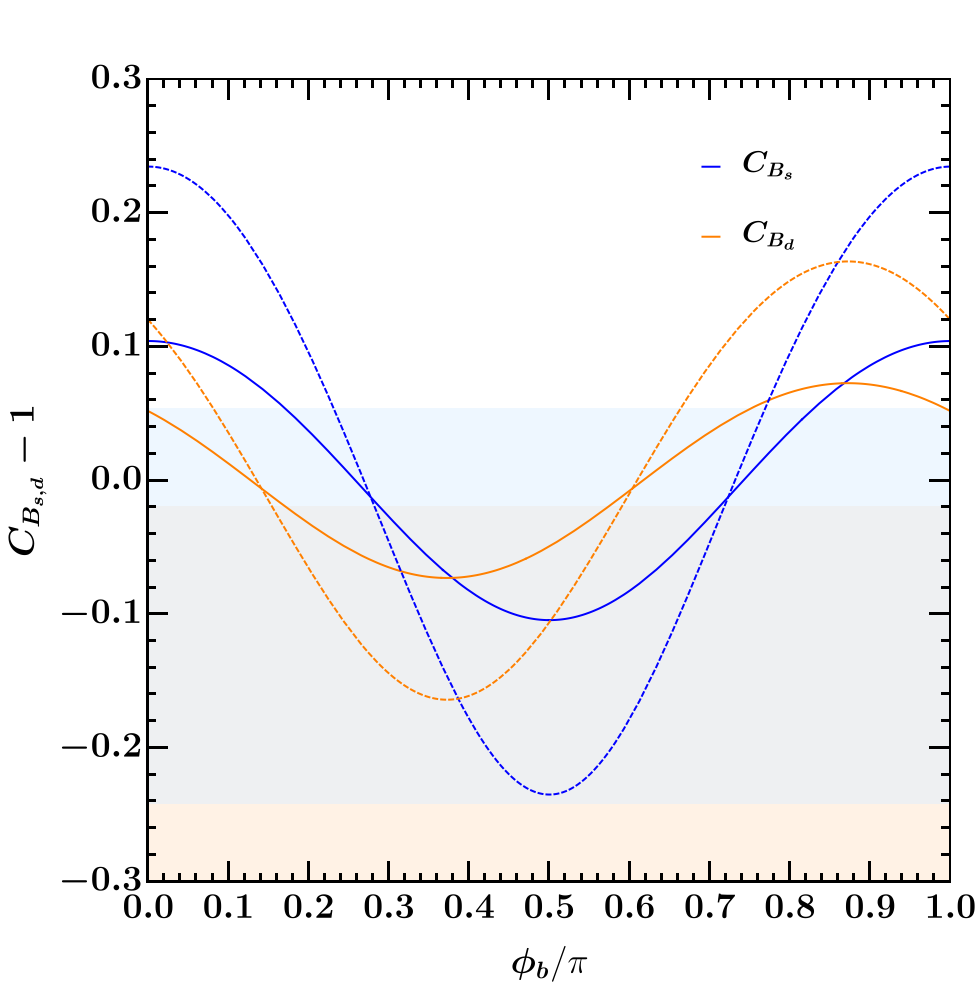}\\
\includegraphics[width=0.45\textwidth]{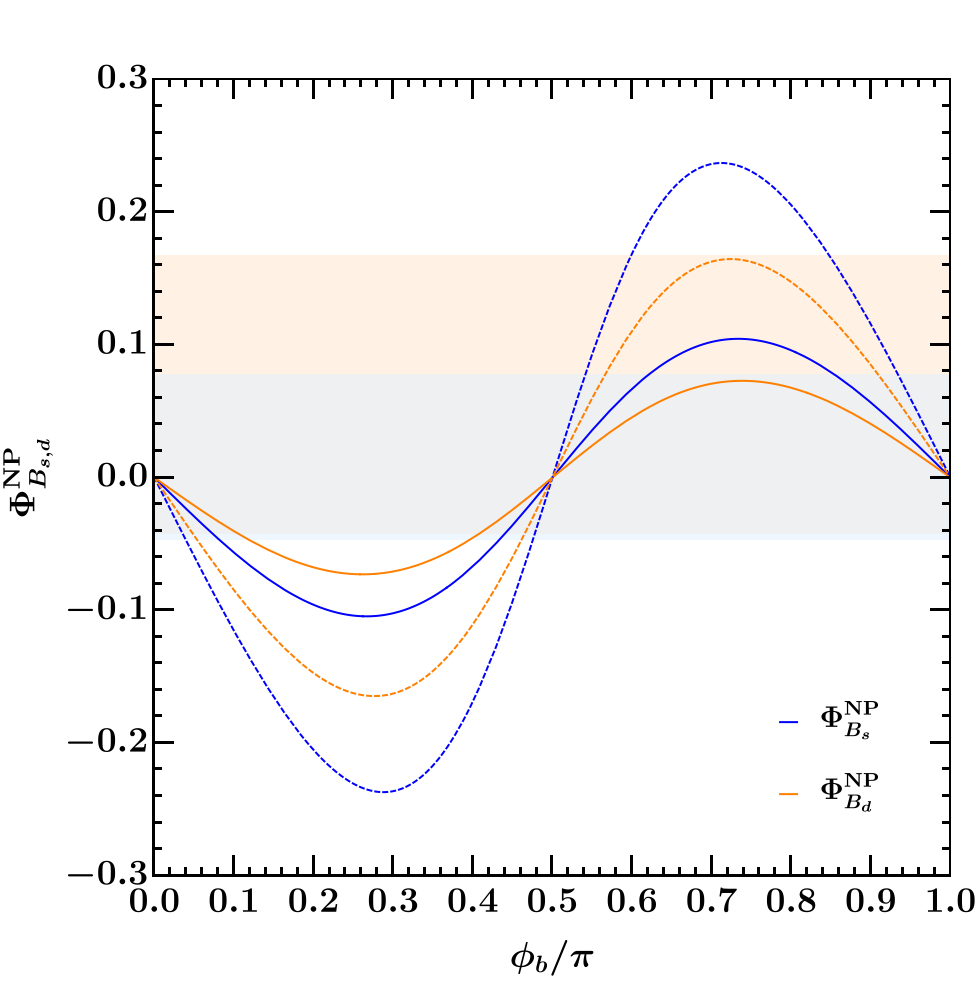}\qquad\quad\includegraphics[width=0.45\textwidth]{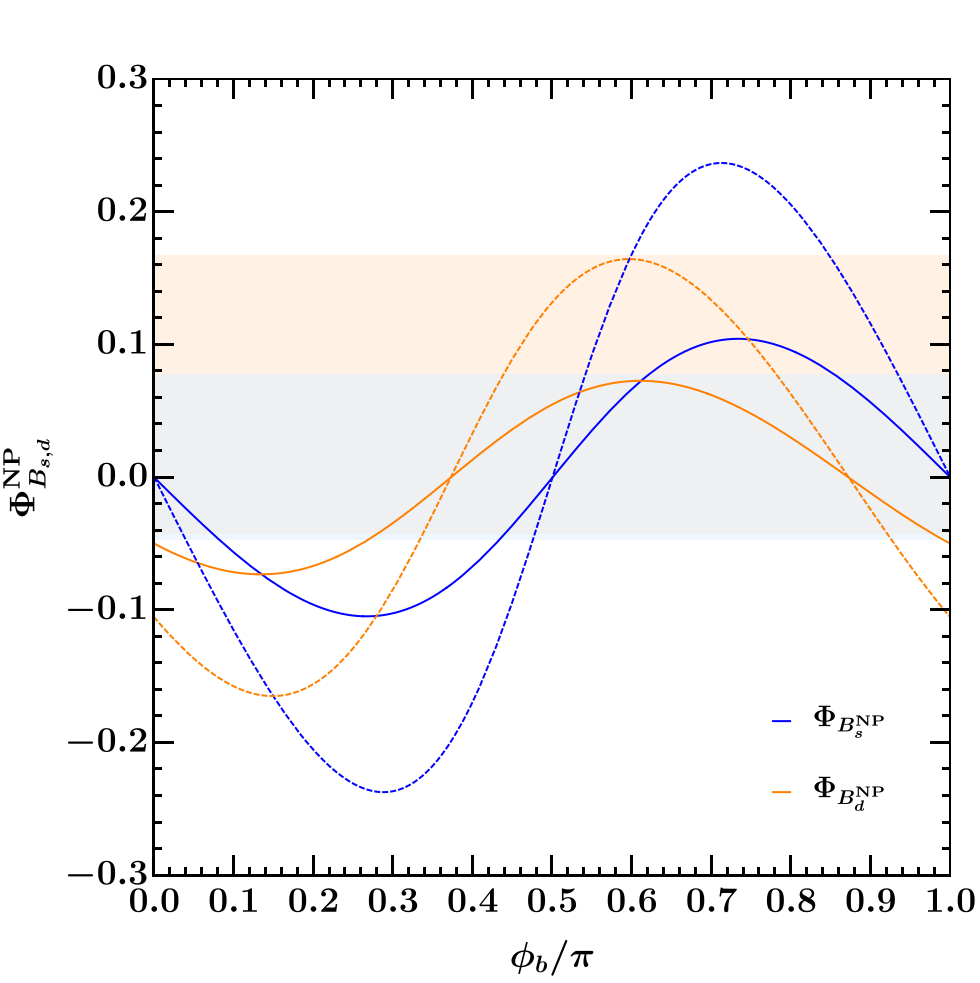}
\caption{NP effects in $B_{s,d}-\bar B_{s,d}$ mixing as function of the phase $\phi_b$ for $\Delta\alpha_d=0,\pi$ (left) and $\alpha_d=0,\pi$ (right). The blue and orange bands correspond to the $95\%$ CL experimental bounds for $B_s$ and $B_d$ mixing, respectively. We use the following inputs: $s_b=0.10\,|V_{ts}|$ (solid), $s_b=0.15\,|V_{ts}|$ (dashed), $\epsilon_R^d=0$, $g_4=3.0$, $M_{Z^\prime}=1.75$~TeV, and $M_{G^\prime}=2.5$~TeV.}\label{fig:DF2}
\end{figure}
As in any extension of the SM with non-trivial flavor strucutre, also in the $\PSC$ framework
$\Delta F=2$ amplitudes provide one of the most significant constraints on model parameters, particularly 
on the new sources of flavor violation in the quark sector. 
These amplitudes receive  tree-level contributions mediated by 
the $Z^\prime$ and $G^\prime$, whose strength is controlled by the $\Ufive$ breaking spurions. 
To a good approximation, the three down-type $\Delta F=2$ amplitudes can be written as 
\begin{align}\label{eq:M12simp}
\begin{aligned}
\mathcal{M}(K^0\to\bar K^0)&\approx\left|\mathcal{M}_{\rm SM}^{(tt)}\right|\left[\frac{(V_{td}V_{ts}^*)^2}{\left|V_{td}V_{ts}^*\right|^2}+e^{-2i\alpha_d}\frac{c_d^4\,[s_b^2+2\,s_b\,\Re(\tilde\epsilon_q\,e^{-i\phi_b})+\Delta\epsilon_q]^2}{|V_{ts}|^4}F_0\right]+\mathcal{M}_{\rm SM}^{(tc+cc)}\,,\\[5pt]
\mathcal{M}(B_d\to\bar B_d)&\approx\left|\mathcal{M}_{\rm SM}\right|\frac{(V_{td}V_{tb}^*)^2}{\left|V_{td}V_{tb}^*\right|^2}\left[1+\frac{c_d^2\,(s_b\,e^{-i\phi_b}+\tilde\epsilon_q^{*})^2}{|V_{ts}|^2}\,F_0\,e^{-2i \Delta \alpha_d} \right]\,,\\[5pt]
\mathcal{M}(B_s\to\bar B_s)&\approx\left|\mathcal{M}_{\rm SM}\right|\frac{(V_{ts}V_{tb}^*)^2}{\left|V_{ts}V_{tb}^*\right|^2}\left[1+\frac{c_d^2\,(s_b\,e^{-i\phi_b}+\tilde\epsilon_q^{*})^2}{|V_{ts}|^{2}}\,F_0\,(1+f(\tbsR))\right]\,,
\end{aligned}
\end{align}
where
\begin{align}
F_0=\frac{16\pi^2}{\sqrt{2}\,G_FM_W^2\, S_0(x_t)}\,\left(C_{Z^\prime}+\frac{C_{G^\prime}}{3}\right)\,,
\end{align}
and $S_0(x_t=m_t^2/M_W^2)\approx2.4$ denotes the SM one-loop function (in the $\Delta S=2$ case we 
normalize the NP amplitude to the short-distance top-quark SM contribution).

As far as left-handed flavor-mixing parameters are concerned, $s_b$ and $\phi_b$ arise from the leading $\mathrm{U(2)_q}$ breaking term
in the quark sector; $\Delta \alpha_d = \alpha_d- (\pi-\mathrm{Arg}\left\{V_{td}/V_{ts}\right\})$ denotes the phase difference between the
leading quark spurion and subleading terms describing light-quark masses
(see appendix~\ref{app:Yukawa}); $c_d=1 +\cO(|V_{us}|^2)$;  $\Delta \epsilon_q$ and $\tilde\epsilon_q$, defined in Eq.~(\ref{eq:DeltaU}),
encode the effect  of the subleading breaking terms in the $Z^\prime$ and $G^\prime$ couplings.

Finally, $f(\tbsR)$ describes the contributions from the right-handed flavor rotations in~\eqref{eq:RHrot1}. Using the inputs in~\cite{Bazavov:2016nty} for the bag 
parameters of non-SM operators, we find
\begin{align}
f(\tbsR)&\approx\frac{16\,C_{Z^\prime}+22\,C_{G^\prime}}{3\,C_{Z^\prime}+C_{G^\prime}}\, \frac{(\tbsR)^*}{c_d\,s_b\,e^{-i\phi_b}}
+ \cO[(\tbsR)^2]~.
\end{align}
As shown in appendix~\ref{app:Yukawa}, in the limit where we neglect contributions to the Yukawa couplings from $d=7$  effective operators, i.e.~when we set $\epsilon_R^d=0$, the  
right-handed rotation angle is unambiguously fixed to $\tbsR=m_s/m_b\,s_b\,e^{i\phi_b}$, that in turn implies $f(\tbsR)\approx0.4$ for typical values of $C_{Z^\prime}/C_{G^\prime}$. 

\paragraph*{CP violation in Kaon mixing.}
The most significant constraints on the subleading parameters $\Delta \epsilon_q$ and $\tilde\epsilon_q$, which describe the 
deviations from the exact $\mathrm{U(2)_q}$ limit in the $Z^\prime$ and $G^\prime$ left-handed couplings, arise from the 
CP-violating observable $\epsilon_K \propto \Im[\mathcal{M}(K^0\to\bar K^0)]$, that can be decomposed as 
\begin{align}
\epsilon_K&\approx\epsilon_K^{\rm SM}-\sqrt{2}\,\epsilon_K^{\rm SM,\,(tt)}\,\sin (2\alpha_d)\,\left[s_b^2+2\,s_b\,\,\Re(\tilde\epsilon_q\,e^{-i\phi_b})+\Delta\epsilon_q\right]^2\,\frac{c_d^4\,F_0}{|V_{ts}|^4}\,,
\end{align}
where $\epsilon_K^{\rm SM,\,(tt)}$ corresponds to the top-mediated SM contribution. 
The NP contribution to $\epsilon_K$ vanishes for $\alpha_d\to 0$.
Setting $\Delta\epsilon_q=\tilde\epsilon_q=0$, and choosing the other parameters in their natural range, we find that $\epsilon_K$ is well within 
its current bound, irrespective of the value of $\alpha_d$. 
Allowing for $\Delta\epsilon_q, \tilde\epsilon_q \not =0$, imposing modifications in $|\epsilon_K|$ of up to $\mathcal{O}(15\%)$, and barring 
accidental cancellations with generic values of $\alpha_d$, we find 
\be
| \Delta\epsilon_q | \lesssim0.1\,|V_{ts}|^2~, \qquad | \tilde\epsilon_q | \lesssim0.3\,|V_{ts}|~.
\ee
Similar limits, although slightly less stringent, are obtained from $B_{s,d}-\bar B_{s,d}$ and $D-\bar D$ mixing. 
Despite being stringent, these limits are below the natural size of these subleading breaking terms 
inferred in Table~\ref{tab:epsilonList} (setting $|\epsilon_U| \leq 10^{-2}$). 
This result implies that: i)~it is perfectly consistent to focus on the scenario $\Delta\epsilon_q=\tilde\epsilon_q=0$; 
ii)~once the symmetry breaking terms assume their natural size, no fine-tuning on the CP-violating phases is necessary  
in order to satisfy the $\epsilon_K$ constraint.

\paragraph{$\Delta B=2$ observables.}
Setting $\Delta\epsilon_q=\tilde\epsilon_q=0$, the physical observables sensitive to $\Delta B=2$ amplitudes, 
namely the mass differences ($\Delta M_q)$ and the CP violating asymmetries $S_{\psi K_S}$ and $S_{\psi \phi}$
can be expressed as 
\begin{align}
\begin{aligned}
C_{B_d}&\equiv\frac{\Delta M_d}{\Delta M_d^{\rm SM}}\approx\left|1+\frac{c_d^2\, s^2_b\,e^{-2i(\phi_b+\Delta\alpha_d)}}{|V_{ts}|^2}\,F_0\right|\,,\\
C_{B_s}&\equiv\frac{\Delta M_s}{\Delta M_s^{\rm SM}}\approx\left|1+\frac{c_d^2\, s^2_b\,e^{-2 i\phi_b}}{|V_{ts}|^2}\,F_0\,\left( 1+f(\tbsR) \right)\right|\,,\\
\end{aligned}
\end{align}
and 
\begin{align}
\begin{aligned}
S_{\psi K_s}&=\sin\left(2\beta+\Phi_{B_d}\right)\,, \qquad  
 \Phi_{B_d}\approx\mathrm{Arg}\left(1+\frac{c_d^2 \,s^2_b\,e^{-2i(\phi_b+\Delta\alpha_d)} }{|V_{ts}|^2}\,F_0\right)\,,\\ 
S_{\psi \phi}&=\sin\left(2|\beta_s|-\Phi_{B_s}\right)\,,~  \quad
 \Phi_{B_s}\approx\mathrm{Arg}\left(1+\frac{c_d^2\, s^2_b\,e^{-2i\phi_b} }{|V_{ts}|^2}\,F_0\,\left(1+f(\tbsR)\right)\right)\,.
\end{aligned}
\end{align}
Current lattice data~\cite{Bazavov:2016nty} point to a deficit in the experimental values of $\Delta M_{d,s}$ with respect to the SM prediction (or equivalently to values of $C_{B_{s,d}}$ smaller than one). As show in Figure~\ref{fig:DF2}, the presence of the free phase $\phi_b$ allows the model to accommodate this deficit, even for small departures from $\phi_b=\pi/2$, while satisfying the bounds from CP violation 
(see Ref.~\cite{DiLuzio:2017fdq} for a similar discussion). 
The mixing angle $s_b$ is constrained to be  up to 0.2 $|V_{ts}|$ (depending on $\phi_b$), indicating a mild alignment of the leading $\mathrm{U(2)_q}$ breaking spurion in the down sector.
As we discuss in Section~\ref{subsec:b2sll}, in our framework the vector leptoquark provides a good fit of the semileptonic anomalies 
irrespective of the value of $\phi_b$ (contrary to the case discussed in Ref.~\cite{DiLuzio:2017fdq}).
We thus conclude that the model leads to a good description of $\Delta B=2$ observables, possibly 
improved compared to the SM case. We also note that using previous lattice determinations of the SM prediction for $\Delta M_{d,s}$, consistent with the experimental value but with larger errors (see e.g.~\cite{Artuso:2015swg,Lenz:2011ti,DiLuzio:2017fdq}), does not affect the results of our phenomenological analysis.

\paragraph*{CP violation in $D$ mixing.}
Last but not least, we analyze the bounds from $\Delta C=2$ amplitudes. 
Following the analysis from UTfit~\cite{Bona:2017cxr,UTfit2018,Carrasco:2014uya}, the constraint obtained from the non-observation of CP-violation in the $D-\bar D$ transition can be expressed as 
\be
\Im (C_1^D)  =  \frac{4 G_F}{ \sqrt{2} } \Im\left(\eftWCF{uu}{V, LL}{2121}(\mu_t) \right)    = (-0.03\pm0.46) \times 10^{-14}\ {\rm GeV}^{-2}~.
\ee
Taking into account also the subleading breaking terms, we find the following simplified expression for this Wilson coefficient:
\begin{align}
 \mathrm{Im}\left(C_1^D\right) &\approx  \frac{4 G_F}{ \sqrt{2} }\, \Im\left\{(V_{ub}^*V_{cb})^2\left[\left(1+c_d\,(s_b\,e^{-i\phi_b}+\tilde\epsilon_q^{\,*})\,\frac{V_{tb}}{|V_{ts}|}\,\Lambda_u^*\right)\left(1+c_d\,(s_b\,e^{i\phi_b}+\tilde\epsilon_q))\,\frac{V_{tb}^*}{|V_{ts}|}\,\Lambda_c\right)\right.\right.  \no\\
&\quad  \left.\left.+\,\Delta\epsilon_q\,c_d^2\,\frac{|V_{tb}|^2}{|V_{ts}|^2}\,\Lambda_u^*\,\Lambda_c\right]^2\right\}\left(C_{Z^\prime}+\frac{C_{G^\prime}}{3}\right)  \no\\
& = 
\frac{4 G_F}{ \sqrt{2} }\, \left(C_{Z^\prime}+\frac{C_{G^\prime}}{3}\right)\, \Im\left\{(V_{ub}^*V_{cb})^2 
\left[ 1+ \cO(s_b, \tilde\epsilon_q, \Delta\epsilon_q) \right] \right\}\,,
\end{align}
where we have defined
\begin{align}
\begin{aligned}
\Lambda_{i}&=\frac{V_{is}|V_{ts}|-V_{id}\left|V_{td}\right|e^{i\alpha_d}}{V_{ib}V_{tb}^*}
%\\
~= ~ \left\{  \begin{array}{ll}    1  +\cO(\lambda^2) &  (i=c)    \\ 
 1-\frac{V_{ud}V_{td}^*}{V_{ub} }\left[1-e^{i\Delta \alpha_d}\right]   +\cO(\lambda^2)   & (i=u) 
\end{array}  \right.\,,
\end{aligned}
\end{align}
which in the limit $\Delta \alpha_d \to 0$ reduces to the $\mathrm{U(2)}$ symmetric result $\Lambda_c = \Lambda_u=1$. 
Contrary to down-type observables, in this case non-vanishing NP contributions are generated also in the $s_b\to 0$  limit. 

\begin{figure}[t]
\centering
\includegraphics[width=0.45\textwidth]{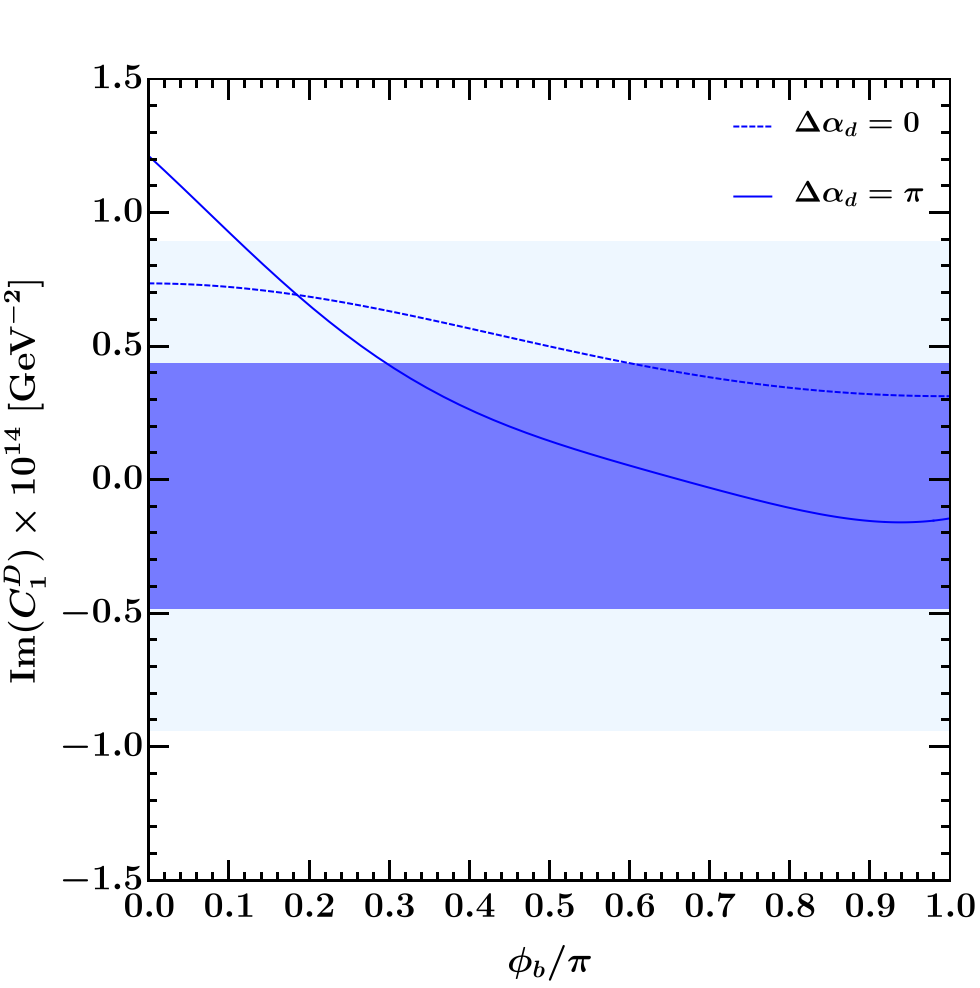}
\caption{ Model contributions to $\mathrm{Im}(C_1^D)$ as function of $\phi_b$. We use the following inputs: $s_b=0.10\,|V_{ts}|$, $g_4=3.0$, $M_{Z^\prime}=1.75$~TeV, and $M_{G^\prime}=2.5$~TeV.  The dark- and light-blue bands correspond to the $68\%$ and $95\%$ CL bound from UTfit~\cite{UTfit2018},  respectively.}\label{fig:DDbar}
\end{figure}

Setting to zero the subleading breaking terms ($\Delta\epsilon_q= \tilde\epsilon_q =0$), we find that the experimental bound 
is satisfied over a wide range of $\{s_b,\phi_b\}$ values compatible with the $\Delta B=2$ constraints.
Note in particular that in the limit where $\Delta\alpha_d=\pi$, we have $\Lambda_u=1.1-4.6\,i$. In this case the large imaginary piece of $\Lambda_u$, together with the values of $s_b$ and $\phi_b$ introduced to explain the deficit in $\Delta B=2$ transitions, yields a partial cancellation in $C_1^D$, both in the real and in the imaginary part. This is shown in Figure~\ref{fig:DDbar} where we plot the $Z^\prime$ and $G^\prime$ mediated tree-level contributions to the imaginary part of $C_1^D$ together with the current bound from UTfit. A similar behaviour is also obtained when $\alpha_d=\pi$, in which case $\Lambda_u=0.2-4.4\,i$.

\subsection{LFU tests in charged lepton decays}
Beside $\Delta F=2$ observables, another very relevant set of constraints on the model is posed by LFU tests in charged-lepton decays. 
These provide an important bound on the overall strength of leptoquark interactions, yielding an upper limit on the possible NP contribution to $R_{D^{(*)}}$. Such tests are constructed by performing ratios of the partial widths of a lepton decaying to lighter leptons or hadrons
(see appendix~\ref{app:LFU}). In our model, both the $\mu$ vs $e$ and the $\tau$ vs $\mu$ ratios are modified: the former is dominated by the tree-level exchange of a $Z^\prime$, the latter by a leptoquark loop. Setting $M_U=2$~TeV to evaluate the leptoquark loop we find\footnote{In the $\tau$ vs $\mu$ ratio we include the full RG running from $M_U$ to $m_t$ using DsixTools~\cite{Celis:2017hod}. Because of the large running effects in the top Yukawa coupling, we find differences of $\mathcal{O}(20\%)$ in the NP contribution when comparing the full RG result to the non-RG improved one-loop expression. 
We also include the non-logarithmic terms computed in~\cite{Bordone:2017bld}.}
\begin{align}
\left(\frac{g_\mu}{g_e}\right)_\ell &\approx 1 + 9 \, C_{Z^{\prime}}\, \stau^{2}\,,  \\
\left(\frac{g_\tau}{g_\mu}\right)_{\ell,\pi,K}&\approx1 - 0.063\,C_U\,.
\end{align}
The high-precision measurements of these effective couplings only allow for per mille modifications of the ratios. This in turn implies a strong bound on the possible value of $C_U$. Taking the HFLAV average in the $\tau$ vs $\mu$ ratio~\cite{Amhis:2016xyh}
\begin{align}
\left(\frac{g_\tau}{g_\mu}\right)_{\ell+\pi+K}&=1.0000\pm0.0014\,.
\end{align}
we find the following limit on $C_U$ at $95\%$~CL:
\begin{align}
C_U\lesssim0.04\stackrel{M_U=2~\mathrm{TeV}}{\Longrightarrow}g_4\lesssim3.2\,.
\label{eq:LFUbound}
\end{align}
This bound is shown in Figure~\ref{fig:CC} together with the NP enhancement in $b\to c(u)\ell\nu$ transitions. On the other hand, we find that possible modifications in the $\mu$ vs $e$ ratio are of $\mathcal{O}(10^{-4})$ and thus do not yield any relevant constraint. We also find that tests of LFU from precision $Z$- and $W$-pole measurements at LEP do not lead to stringent bounds. In particular we note that the $Z^\prime$ tree-level contribution to $Z$ anomalous couplings, given in terms of the $\psi^2\phi^2 D$ SMEFT operators in Table~\ref{tab:no4ferm},  is found to be well below the present limits.

\subsection{$b\to c(u)\tau\nu$}
\label{subsec:b2ctaunu}

\begin{figure}[t]
\centering
\includegraphics[width=0.45\textwidth]{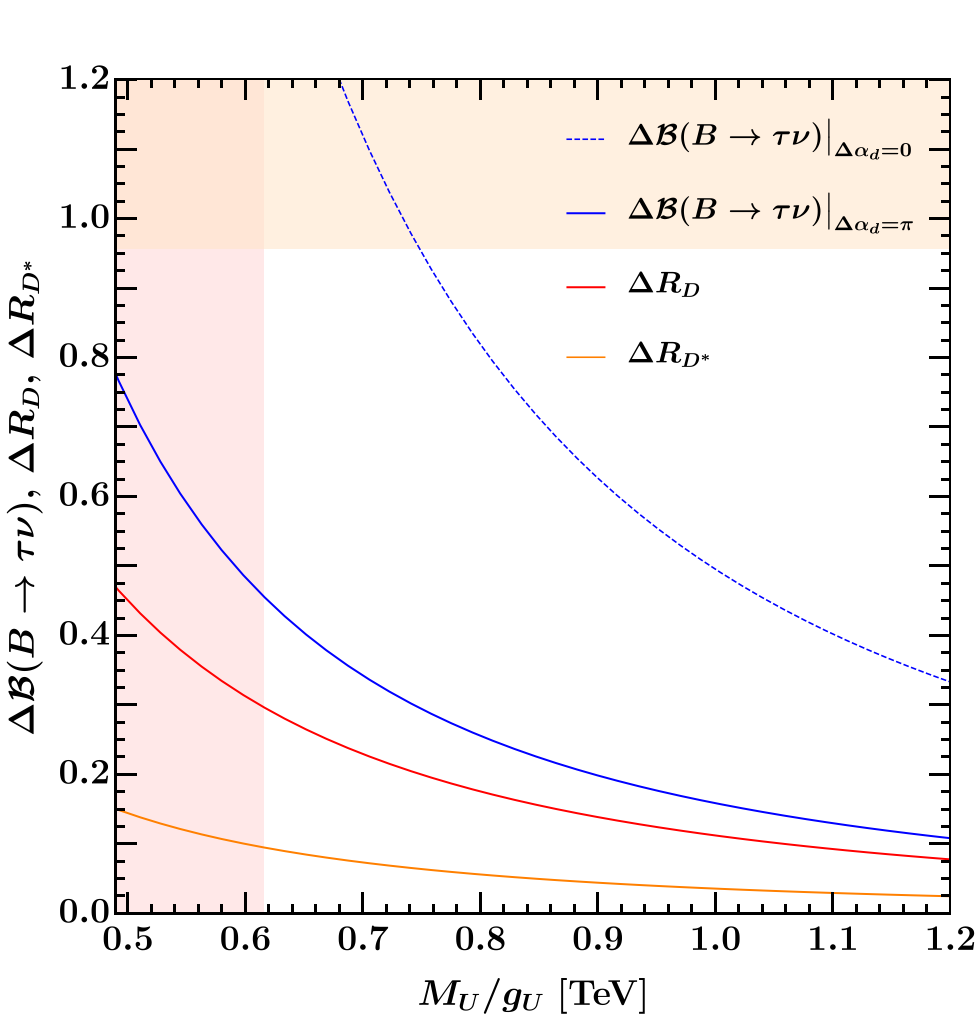}\qquad\quad\includegraphics[width=0.45\textwidth]{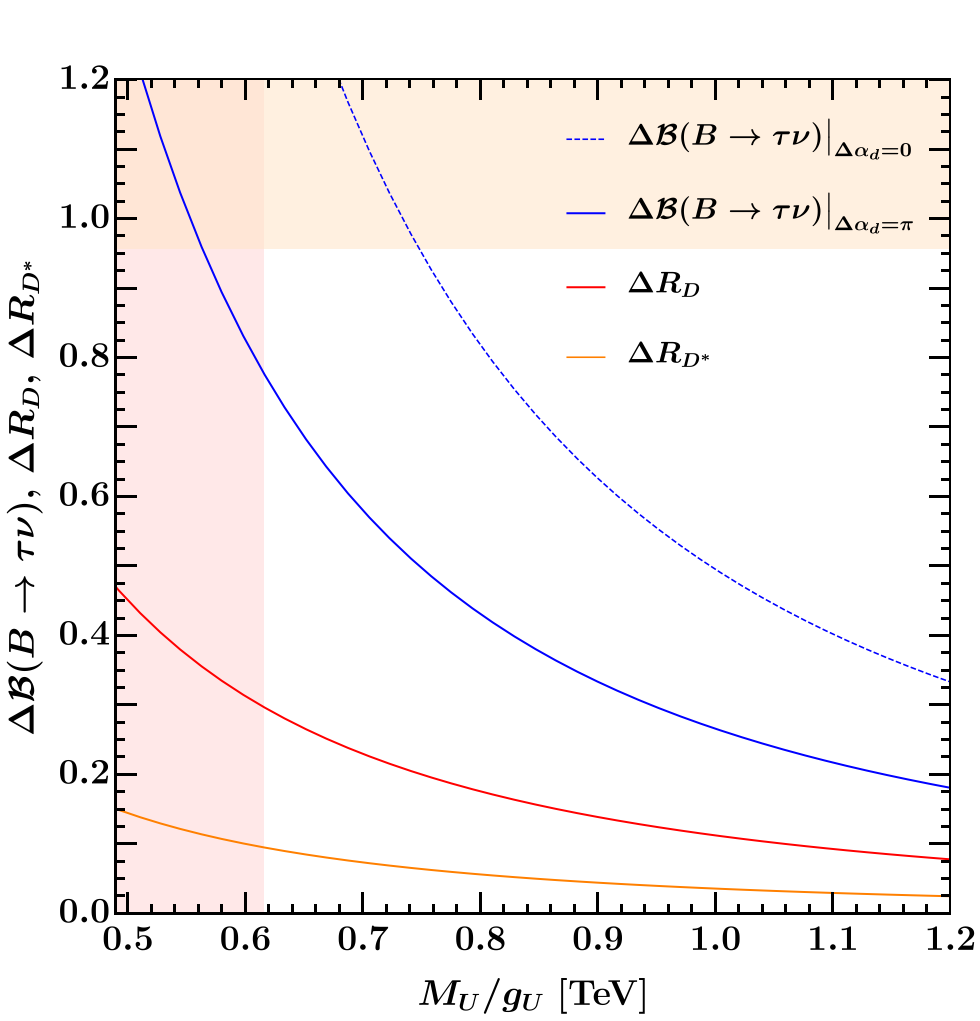}
\caption{NP enhancements in $\mathcal{B}(B\to\tau\nu),\, R_D$ and $R_{D^*}$ as function of $M_U/g_U$. We use the following inputs: $s_b=0.15\,|V_{ts}|$ (left), $s_b=0.10\,|V_{ts}|$ (right), $\phi_b= \pi/2$. The red and orange bands correspond, respectively, to the 95\% CL exclusion limits from LFU tests in $\tau$ decays and from $\mathcal{B}(B\to\tau\nu)$.}\label{fig:CC}
\end{figure}

The violation of LFU in $b\to c \ell \nu$  transitions, measured via the  ratios $R_D$ and $R_{D^*}$, 
sets the scale of NP (or the preferred value of $C_U$).
In the $\PSC$ model NP effects in $b\to c(u)\tau\nu$ transitions are described by the following effective operators
\begin{equation}\label{eq:bctanuLag}
\mathcal{L}(b\to u_i \tau \bar{\nu})=-\frac{4G_F}{\sqrt{2}}\left(\eftWCF{\nu edu}{V, LL}{333i}^*(\overline{\tau}_L\gamma^\mu\nu_{L3})(\overline{u}_L^{\,i}\gamma_{\mu}b_L)+\eftWCF{\nu edu}{S, RL}{333i}^* (\overline{\tau}_R\,\nu_{L3})(\overline{u}^{\,i}_Lb_R)\right)\, ,
\end{equation}
where $i=1(2)$ for up (charm) quarks. At $\Lambda=M_U$ we have to a good approximation
\begin{align}
\eftWCFS{\nu edu}{S, RL}{M_U}{333i}=2\,\eftWCFS{\nu edu}{V, LL}{M_U}{333i}\approx\,2\,C_U\,V_{ib}^*\,.
\end{align}
The RG running (due to QCD) introduces an important correction to the scalar operator contributions. To account for these effects we define the following RG factor 
\begin{align}
\eftWCFS{\nu edu}{S, RL}{m_b}{333i}=\,\etaF\,\eftWCFS{\nu edu}{S, RL}{M_U}{333i}\,.
\end{align}
Using DsixTools~\cite{Celis:2017hod} (see also~\cite{Aebischer:2017gaw,Gonzalez-Alonso:2017iyc}) we find $\etaF\approx1.8$ for $M_U=2$~TeV. On the other hand, the running of the vector operator
(which is a conserved current as far as QCD is concerned) is very small 
and will be neglected in the following discussion.

Due to the presence of a scalar operator, we predict departures from a pure $V-A$ structure, hence different NP contributions to $R_D$ and $R_{D^*}$. 
We define the relative NP contribution to these observables as
\begin{align}\label{eq:DRD}
\Delta R_{D^{(*)}}&=\frac{R_{D^{(*)}}}{R_{D^{(*)}}^{\rm SM}}-1\,.
\end{align}
Using the results in~\cite{Fajfer:2012vx} for the scalar form factors, we find the following simplified expressions
\begin{align}\label{eq:DRDCU}
\begin{aligned}
\Delta R_{D}&\approx 2\,C_U\times (1+1.5\,\etaF)\,,\\
\Delta R_{D^*}&\approx 2\,C_U\times (1+0.12\,\etaF)\,, 
\end{aligned}
\end{align}
which imply a $30\%$ ($10\%$) NP effect in $R_{D}$ ($R_{D^*}$) for $C_U \approx 0.04$, i.e.~a value around the upper bound of the LFU constraint in Eq.~(\ref{eq:LFUbound}).

The (non-standard) contributions to $\mathcal{B}\left(B_c\to\tau\nu\right)$ induced by the scalar operator is chirally enhanced, 
yielding an enhancement of $\mathcal{O}(100\%)$ compared to the SM prediction. 
However, given the low experimental accuracy in this observable, this does not 
pose any significant bound on the model.
Similarly, the modification of the $B_c$ lifetime, which has been shown to introduce important constraints on explanations  of the $b\to c\tau\nu$ anomalies based on pure scalar operators~\cite{Alonso:2016oyd}, is well below the experimental limit. 

Given the approximate $\mathrm{U(2)_q}$ symmetry, similar NP effects are also expected in $b\to u \ell \nu$.
So far, the most relevant measurement involving these transition is $\mathcal{B}\left(B\to\tau\nu\right)$. 
In analogy to the case $R(D^{(*)})$ case, we define
\begin{align}
\Delta \mathcal{B}\left(B\to\tau\nu\right)&=\frac{\mathcal{B}\left(B\to\tau\nu\right)}{\mathcal{B}\left(B\to\tau\nu\right)^{\rm SM}}-1\,.
\end{align}
Using the current experimental value~\cite{Patrignani:2016xqp} and the result from UTfit~\cite{Bona:2017cxr} for the SM prediction, we find
\begin{align}
\Delta \mathcal{B}\left(B\to\tau\nu\right)&=0.35 \pm 0.31\,.
\end{align}

In our model, we obtain 
\begin{align}
\Delta\mathcal{B}\left(B\to\tau\nu\right)&\approx \left|1+C_U\left[1+c_d\,s_b\,e^{i\phi_b}\,\frac{V_{tb}^*}{|V_{ts}|}\,\Lambda_u\right]\left(1+\etaF\,\frac{2\,m_B^2}{m_\tau(m_b+m_u)}\right)\right|^2-1\,.
\end{align}
Also in  this case scalar contributions are chirally enhanced and we typically expect large NP effects. However, similarly to $D$--$\bar D$ mixing, 
in the limit where $\Delta\alpha_d\to\pi$ (and analogously for $\alpha_d\to\pi$) the large phase in $ \Lambda_u$, together with the values of $s_b$ and $\phi_b$ required to explain the deficit in $\Delta B=2$ transitions, yields a significant attenuation of the NP enhancement. The possible range of deviations from the SM is illustrated in Figure~\ref{fig:CC}.

Contrary to $B$ decays, LFU breaking effects in charged-current $K$ and $D$ decays are strongly CKM suppressed (relative to the 
corresponding SM amplitudes) and do not lead to significant constraints.

\subsection{$b\to s\ell\ell$ and $b\to s\nu\nu$}\label{subsec:b2sll}
The violation of LFU in $b\to s \ell \ell$  transitions, measured via the  ratios $R_K$ and $R_{K^*}$, 
sets the amount of $\Ufive$ breaking in the model which is not directly related to the Yukawa couplings. 
After imposing the constraints from $\Delta F=2$ observables, the $Z^\prime$-mediated contributions to $b\to s\ell\ell$ amplitudes 
turn out to be well below those mediated by the vector leptoquark. This is because the $\Delta F=2$ constraints require the effective 
$bsZ^\prime$ coupling to be either very small in size or almost purely imaginary (hence with a tiny interference with the SM contribution). 
As a result, the following approximate relations hold (assuming $\phi_\tau=0$ and $\epsilon_U$ real):
\begin{align}\label{eq:bsllWET}
\begin{aligned}
\mathrm{Re}\,(\Delta \cC_9^{\mu\mu }) &\approx  -\, \mathrm{Re}\,(\Delta \cC_{10}^{\mu\mu}) \approx  -\frac{2\,\pi}{\alpha_{\rm em} }
\, \frac{\stau\,\epsilon_U}{|V_{ts}|}\,C_U \,,\\
\mathrm{Re}\,(\Delta \cC_9^{\tau\tau }) &\approx  -\, \mathrm{Re}\,(\Delta \cC_{10}^{\tau\tau}) \approx  \frac{2\,\pi}{\alpha_{\rm em} }
\, \frac{\stau\,\epsilon_U}{|V_{ts}|}\,C_U \,,
\end{aligned}
\end{align}
where 
$\Delta \cC_i^{\alpha\alpha }  = \cC_i^{\alpha\alpha } -\cC_i^{\rm SM}$, 
and $\Delta \cC_9^{ee} \approx  \Delta \cC_{10}^{ee} \approx  0$. 
Hence, the deviations from unity in the LFU ratios $R_K$ and $R_{K^*}$
can be expressed as~\cite{Capdevila:2017bsm,Celis:2017doq}
 \begin{align}
 \begin{aligned}
\Delta R_K &=& 1 - \left.R_K\right|_{[1,\,6]~{\rm GeV}^2}  ~\approx~   & 0.23\,\Delta \cC_9^{\mu\mu} - 0.23\, \Delta \cC_{10}^{\mu\mu}  \approx  0.46\, \Delta \cC_9^{\mu\mu}~,\\
\Delta R_{K^*} &=& 1 -  \left.R_{K^*} \right|_{[1.1,\,6]~{\rm GeV}^2} ~\approx~ &    0.20\, \Delta \cC_9^{\mu\mu} -0.27\,\Delta \cC_{10}^{\mu\mu}  \approx  0.47\, \Delta \cC_9^{\mu\mu} ~.
\end{aligned}
\end{align}
Contrary to other models aiming at a combined explanation of the anomalies, we predict 
$\mathrm{Re}\,(\Delta \cC_{9,10}^{\mu\mu})$ and $\mathrm{Re}\,(\Delta \cC_{9,10}^{\tau\tau})$ 
 to be of similar size.
 This is a consequence of the different $\mathrm{U(2)}^5$ breaking structure discussed in Section~\ref{sect:MainYuk}.

Another key difference with respect to the existing 
 literature  is the presence of right-handed leptoquark currents. 
 These generate the following scalar and pseudo-scalar contributions:\footnote{Given that the leading RG effects for the scalar operators are dominated by QCD, the RG running factor for $\cC_{S,P}$ and $\eftWC{\nu edu}{S, RL}$ remains the same to a very good approximation.} 
\begin{align}\label{eq:scalarOp}
\begin{aligned}
\mathcal{C}_S^{\mu\mu}&=-\,\mathcal{C}_P^{\mu\mu}\approx \frac{4\,\pi}{\alpha_{\rm em}V_{tb}V_{ts}^*}\,C_U\,\etaF\,\epsilon_U\,\ttaumuR\,,\\[2pt]
\mathcal{C}_S^{\tau\tau}&=-\,\mathcal{C}_P^{\tau\tau}\approx -\frac{4\,\pi}{\alpha_{\rm em}V_{tb}V_{ts}^*}\,C_U \,\etaF\,\left[\epsilon_U\, \stau\,e^{i\phi_\tau}+s_b\,e^{i\phi_b}\right]\,.
\end{aligned}
\end{align}
While the effect of these operators is negligible in chirally-allowed transitions, this is not the case for  $P\to\ell\ell$ decays (see appendix~\ref{app:obs}). In particular, the enhancement of scalar amplitudes 
is enough to overcome the mass suppression of the right-handed rotation angle $\ttaumuR$ in $\mathcal{C}_{S,P}^{\mu\mu}$. 
Setting $\Delta\mathcal{C}_{9}^{\mu\mu}=-0.6$, as required by the central value of the $R_K$ and $R_{K^*}$ anomalies, 
and using the latest LHCb measurement of $\mathcal{B}(B_s\to\mu\mu)=3.02(65)\times10^{-9}$~\cite{Aaij:2017vad}, 
 we find the following bounds at $95\%$~CL on the right-handed mixing in the lepton sector:
\begin{align}
\left|  \ttaumuR / \ttaumu  \right|   \leq 0.013~, \qquad\qquad~  0.04 \leq\ttaumuR / \ttaumu   \leq 0.07~.
\end{align}
The second solution corresponds to a destructive interference between a large NP amplitude and the SM, yielding 
$\mathcal{B}(B_s\to\mu\mu)$ close to the SM expectation.  As we discuss in the following section, this accidental cancellation 
is disfavored by LFV constraints. Therefore, we focus on the first solution, which requires the $\mu$--$\tau$ right-handed mixing angle to be 
slightly smaller than what we expect in  absence of dimension-7 operators
($| \ttaumuR / \ttaumu | =m_\mu/m_\tau=0.06$), but it is still natural.

We also expect relatively large NP enhancement in $\mathcal{B}(B_s\to\tau\tau)$, dominated by the chirally-enhanced scalar contributions in~\eqref{eq:scalarOp}. Setting $\Delta\mathcal{C}_{9}^{\mu\mu}=-0.6$ and $C_U=0.04$, and assuming $\phi_b\approx\pi/2$ and $\phi_\tau \approx0$
we find
\begin{align}
\frac{\mathcal{B}(B_s\to\tau\tau)}{\mathcal{B}(B_s\to\tau\tau)^{\rm SM}}&\approx 5+45  \left(\frac{s_b}{0.1\,|V_{ts}|}\right)^2\,,
\end{align}
where $\mathcal{B}(B_s\to\tau\tau)^{\rm SM}=\left(7.73\pm0.49\right)\times10^{-7}$~\cite{Bobeth:2013uxa}. We stress the strong correlation between the 
possible NP contribution to $\Delta B=2$ amplitudes discussed  in Section~\ref{subsec:DF2} (controlled by $|s_b|$)
and a large enhancement in $\mathcal{B}(B_s\to\tau\tau)$. 

Finally, we mention that $b\to s\nu\nu$ transitions do not get significantly modified in this framework.  On the one hand, due to its coupling structure, the vector leptoquark does not contribute at tree-level to such transitions. On the other hand,  the $Z^\prime$ contribution is negligible because of the 
constraints on the  $bsZ^\prime$ coupling, as already discussed in the $b\to s\ell\ell$ case.

\subsection{LFV processes}
\label{subsec:LFV}

\begin{figure}[t]
\centering
\includegraphics[width=0.45\textwidth]{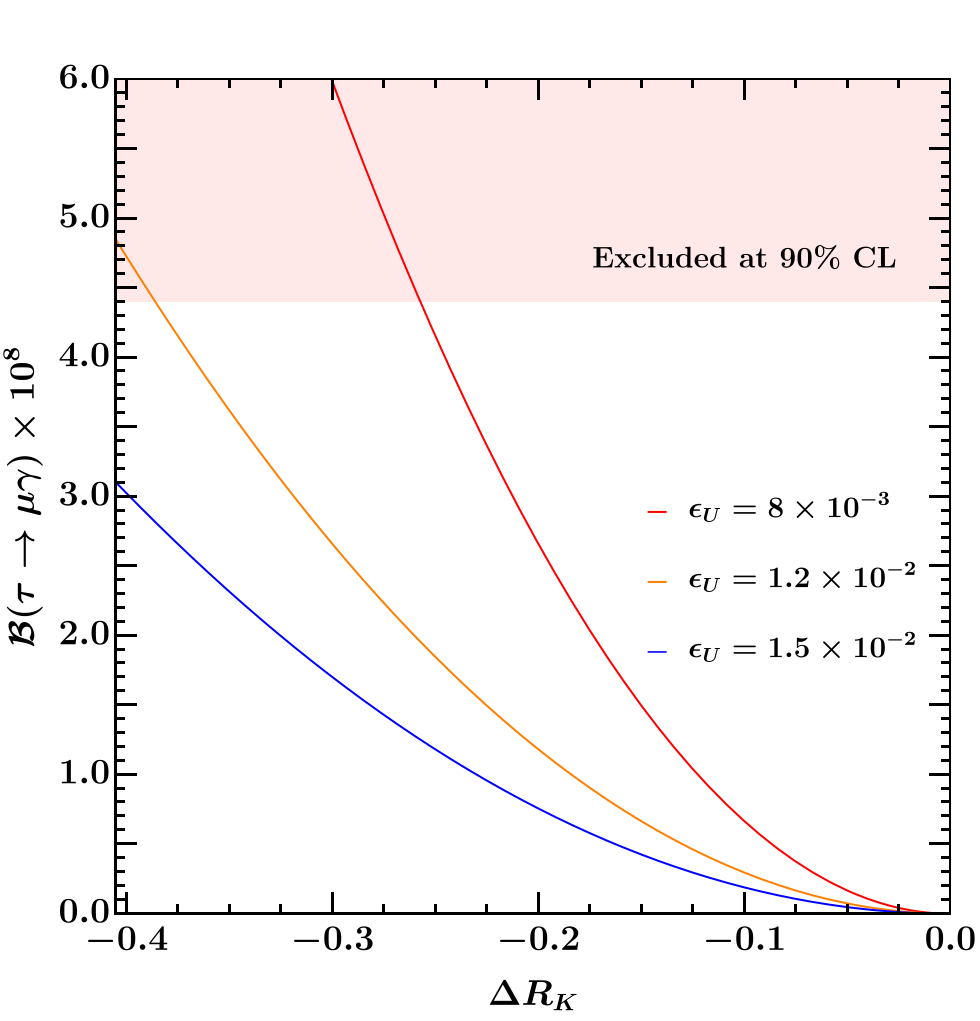}\qquad\quad\includegraphics[width=0.45\textwidth]{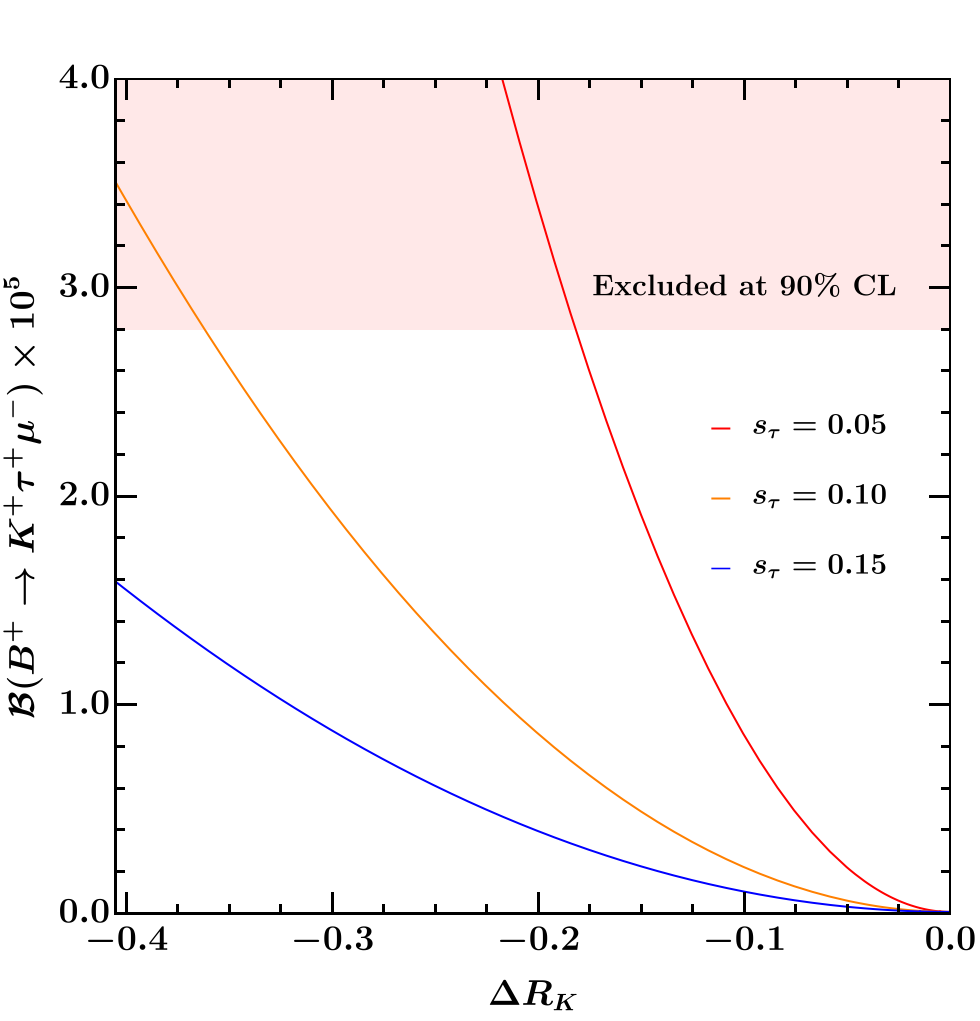}
\caption{Left: $\mathcal{B}(\tau\to\mu\gamma)$ as function of the NP shift in $R_K$ for different values of $\epsilon_U$.
Right: $\mathcal{B}(B^+\to K^+\tau^+\mu^-)$ as function of the NP shift in $R_K$ for different values of $\stau$.}
\label{fig:LFVpheno}
\end{figure}

We finally turn to LFV processes.
Given the unambiguous prediction of a large $\tau \to \mu$ effective coupling, they represent a striking signature of the model.

In $b\to s\ell\ell^\prime$ transitions the dominant contribution is 
mediated by the leptoquark, leading to 
\begin{align}
\begin{aligned}
\Re(\cC_9^{\tau\mu}) &\approx  -\Re(\cC_{10}^{\tau\mu}) \approx  -\frac{\mathrm{Re}\,(\Delta \cC_9^{\mu\mu})}{\ttaumu}\,,\quad\qquad
\Re(\mathcal{C}_S^{\tau\mu}) =-\Re(\mathcal{C}_P^{\tau\mu}) \approx - \frac{2\,\etaF\,\mathrm{Re}\,(\Delta \cC_9^{\mu\mu})}{\ttaumu}\,.
\end{aligned}
\end{align}
Due to the $\stau^{-1}$ enhancement, large NP contributions in $\mathcal{B}(B_s\to\tau\mu)$ and in $\mathcal{B}(B\to K\tau\mu)$ are expected. 
In the former case the effect is further reinforced by the chiral-enhancement of scalar amplitudes, leading to
\begin{align}
\begin{aligned}
\mathcal{B}(B_s\to\tau^+\mu^-) &\approx  2  \times 10^{-4}\, \left(\frac{\Delta R_K}{0.3}\right)^2\left( \frac{0.1}{\ttaumu}\right)^2\,,\\[5pt]
\mathcal{B}(B\to K^*\tau^+\mu^-)&\approx  1.5 \times 10^{-6}\, \left(\frac{\Delta R_K}{0.3}\right)^2\left( \frac{0.1}{\ttaumu}\right)^2\,,\\[5pt]
\mathcal{B}(B^+\to K^+\tau^+\mu^-)&\approx  2 \times 10^{-5}\, \left(\frac{\Delta R_K}{0.3}\right)^2\left( \frac{0.1}{\ttaumu}\right)^2\,,
\end{aligned}
\end{align}
with $\mathcal{B}(B^-\to K^-\tau^-\mu^+)=\mathcal{B}(B^+\to K^+\tau^+\mu^-)$ and 
$\mathcal{B}(B^+\to K^+\tau^-\mu^+)\approx\mathcal{B}(B_s\to\tau^-\mu^+)\approx 0$, and similarly for the $K^*$ channel. NP effects in the latter are predicted to be smaller because, contrary to the $K$ channel, 
the scalar contributions are suppressed in this case.
While there are no experimental constraints in $B_s\to\tau\mu$ so far, the model prediction for $B^+\to K^+\tau^+\mu^-$ lies close to the
current experimental limit by BaBar: $\mathcal{B}(B^+\to K^+\tau^+\mu^-)<2.8\times10^{-5}$~(90\% CL)~\cite{Lees:2012zz}. In figure~\ref{fig:LFVpheno} (right) we
show the predicted values of $\mathcal{B}(B^+\to K^+\tau^+\mu^-)$ as a function of the NP shift in $R_K$ and for different benchmark values of $\stau$. 
We also note that, contrary to other proposed solutions to the anomalies, in our model the $s\tau U$ coupling is very small, resulting in a negligible contribution to the 
$\tau\to\phi\mu$ decay rate.

In purely leptonic decays the most interesting observable is $\tau\to\mu\gamma$. Radiative LFV decays are generated at the one loop level, both by $Z^\prime$ and $U$ loops.
The leptoquark yields the largest contribution due to its larger couplings  and the $m_b$-enhancement of the loop function. 
From the explicit one-loop calculation (see appendix~\ref{app:LFV}), we find 

which is just below the current experimental limit set by Babar: $\mathcal{B}(\tau \to \mu \gamma)<4.4 \times 10^{-8}\;(90\%\,\mathrm{CL})$~\cite{Aubert:2009ag}. In figure~\ref{fig:LFVpheno} (left) we show the prediction for $\mathcal{B}\left(\tau\to\mu\gamma\right)$ as a function of the NP contribution to $R_K$ for different values of $\epsilon_U$. The model also predicts a sizable NP contribution to $\tau\to3\mu$, mediated by a tree-level $Z^\prime$ exchange. 
We obtain the following approximate expression
\begin{align}
\mathcal{B}\left(\tau\to3\mu\right)&\approx C_{Z^\prime}^2\, \stau^2\left[ 28\,(\stau^2+\epsilon_\ell)^2-38\left(\frac{g_1}{g_4}\right)^2\left(\stau^2+\epsilon_\ell-2\left(\frac{g_1}{g_4}\right)^2\right)\right]\,.
\end{align}
For typical values of the model parameters, this contribution lies about one order of magnitude below the current experimental limit by Belle: $\mathcal{B}\left(\tau\to3\mu\right)<1.1\times 10^{-8}\;(90\%\,\mathrm{CL})$~\cite{Hayasaka:2010np}. However, this conclusion is strongly dependent 
on the precise value of $\ttaumu$. 

Purely leptonic LFV transitions of the type $\mu\to e$  are controlled by the mixing angle $s_e$ in Eq.~\eqref{eq:Le}. We find that the most stringent constraint on this angle is obtained, at present, by the experimental bound on  $\mu\to3e$ set by the Sindrum Collaboration: 
$\mathcal{B}\left(\mu\to3e\right)<1.0\times10^{-13}~(90\%\,\mathrm{CL})$~\cite{Bellgardt:1987du}. 
Similarly to $\tau \to 3\mu$, also $\mu\to3e$ is dominated by the tree-level exchange 
of the $Z^\prime$, which yields
\begin{align}
\begin{aligned}
\mathcal{B}\left(\mu\to3e\right)&\approx 420\,C_{Z^\prime}^2\left(\frac{g_1}{g_4}\right)^4\,s_e^2\left(\epsilon_l+\stau^2\right)^2 \\ 
&\approx (1-10) \times 10^{-14} \left(\frac{s_e}{0.01}\right)^2\left(\frac{\epsilon_l+\stau^2}{0.02}\right)^2\,.
\end{aligned}
\end{align}
where the range in the second numerical expression reflects the uncertainty on the $Z'$ mass and couplings.
Assuming $\epsilon_l\sim\epsilon_U\sim\mathcal{O}(10^{-2})$, and taking natural values for the other parameters, we find
\begin{align}
s_e\lesssim 10^{-2}\,,
\label{eq:boundL12}
\end{align}
consistently with the EFT estimate derived in~\cite{Bordone:2017anc}.\footnote{Despite stringent, the bound 
on $s_e$ in \eqref{eq:boundL12} is not unnatural. 
The benchmark for subleading $\mathrm{U(2)}_\ell$ breaking terms not aligned to the second generation is 
provided by $(m_e/m_\mu)^{1/2} \approx 7 \times 10^{-2}$.} Another important constraint on $s_e$,
which however depends also on $\ttaumuR$, is provided by $\mu\to e\gamma$. As in $\tau\to\mu\gamma$, contributions 
to this observable appear in our model at one loop, with the dominant effect being mediated by the leptoquark. We find
\begin{align}
\mathcal{B}(\mu\to e\gamma)&\approx6 \times 10^{-13}\left(\frac{\Delta R_K}{0.3}\right)^2\left(\frac{0.01}{\epsilon_U}\right)^2\left(\frac{s_e}{0.01}\right)^2\left(\frac{\left|\ttaumuR\right|}{0.01}\right)^2\,,
\end{align}
to be compared with the bound by the MEG Collaboration: $\mathcal{B}(\mu\to e\gamma)<4.2\times10^{-13}$~(90\% CL)~\cite{TheMEG:2016wtm}.
Other limits on this angle are significantly weaker. In particular, from the $Z'$ contribution to 
$\bar \mu e \bar d d$ effective operators, which are constrained by $\mu\to e$ conversion~\cite{Carpentier:2010ue,Giudice:2014tma},
we get $s_e\lesssim 10^{-1}$. 
 
On the other hand, the leading contribution to $\bar \mu e \bar d d^{(\prime)}$  effective operators is due to the 
leptoquark exchange, and the dominant constraint is set by  $K_L\to\mu e$~\cite{Giudice:2014tma}.
In this case the amplitude is (formally) independent from $s_e$, but it depends on the subleading $\mathrm{U(2)}_\ell$ breaking parameter
$\Delta\epsilon_U$, defined in Eq.~(\ref{eq:DeltaU}):
\begin{align}
\mathcal{B}( K_L  \to \mu^\pm e^\mp )  & \approx0.8\times 10^{-5} \, (\Delta \epsilon_U)^2\,\left(\frac{\Delta R_K}{0.3}\right)^2\left( \frac{0.1}{\ttaumu}\right)^2\,.
\end{align}
Using the current experimental bound by the BNL Collaboration, $\mathcal{B}( K_L  \to \mu^\pm e^\mp )=0.47\times10^{-11}~(90\%\,\mathrm{CL})$~\cite{Ambrose:1998us}, we find
\begin{align}
\Delta \epsilon_U\lesssim 6\times 10^{-4}\,.
\end{align}
This  bound is consistent with the naive estimate of this parameter, 
$\Delta\epsilon_U =  \cO(\epsilon_U s_e s_d)$,  provided $s_e$ satisfies the bound in Eq.~(\ref{eq:boundL12}).

%%%%%%%%%%%%%%%%%%%%%%%%%%%%%%%%%%%%%%%%%%%%%%%%%%%%%%%%%%%%%%%%%%
\section{Low-energy fit and discussion}\label{sect:Fit}

In order to precisely quantify the quality of the proposed model in the description of the anomalies, we perform a fit to low-energy data. We work in the minimal breaking scenario presented in Section~\ref{sect:MainYuk} and set $\Delta\alpha_d=\pi$ to minimize undesired NP contributions in $\mathcal{B}(B\to\tau\nu)$ and $\Delta F=2$ transitions, as discussed in Section~\ref{sect:pheno}.  We also restrict ourselves to the case $s_e=0$, hence to vanishing LFV in $\mu\to e$ transitions, given that this parameter has no impact on the description of the anomalies. Under these assumptions, the following model parameters have a relevant impact at low energies: $\omega_1,\omega_3,\,\stau,\epsilon_R^e,s_b,\phi_b,\epsilon_U$.\footnote{In order to remove marginally relevant parameters we fix $\epsilon_q=\epsilon_\ell=\epsilon_U$. We have checked explicitly that departing from this restriction, while keeping $\epsilon_q$ and $\epsilon_\ell$ within their expected range, has no effect on fit results. We also set $\phi_\tau$ to zero 
and treat $\epsilon_U$ and $\epsilon_R^e$ as a real parameters, since these extra phases do not introduce any interesting features. Finally, we conservatively assume $\epsilon_R^d=0$; a non-zero value for this parameter would slightly improve the agreement with $\Delta F=2$ data.} The first two are related to the NP scale while, the other five control the breaking of the $\Ufive$ symmetry. We perform a Bayesian estimation for these parameters using the log-likelihood
\begin{align}
\log\mathcal{L}=-\frac{1}{2}\sum_{i\in \rm obs}\left(\frac{x^{\PSC}_i-x^{\rm exp}_i}{\sigma_i}\right)^2\,,
\end{align}
constructed from the observables listed in Tables~\ref{tab:LFV_exp},~\ref{tab:LFU_exp},~\ref{tab:semileptonic} and~\ref{tab:hadronic} and using the expressions in appendix~\ref{app:obs} for the model predictions. For the CKM matrix elements we take the values reported in the NP fit from UTFit and for the remaining  input parameters we use PDG values~\cite{Patrignani:2016xqp}. For the Bayesian analysis we use the nested sampling algorithm implemented in the public package 
MultiNest~\cite{Feroz:2007kg,Feroz:2008xx,Feroz:2013hea}. The resulting posterior probabilities are analysed using the Markov Chain sample analysis tool GetDist~\cite{GetDist}. In the analysis we consider flat priors in all the parameters  for the following ranges\footnote{Since the observables considered in the fit are not sensitive to the individual signs of  $\epsilon_U$ and $\stau$ but only to their product, there is a double degeneracy in the fit. We remove this degeneracy by considering both $\epsilon_U$ and $\stau$ to be positve.}
\begin{align}
\begin{aligned}
\omega_1&\in\left[0.3,1.5\right]~\mathrm{TeV},&\qquad\omega_3&\in\left[0.3,1.5\right]~\mathrm{TeV},&\qquad \stau &\in\left[0,0.15\right]\,,\\[5pt]
s_b&\in\left[-0.1,0.1\right],&\phi_b&\in\left[0,\pi\right],&\epsilon_R^e &\in\left[-0.01,0.01\right]\,,\\[5pt]
\epsilon_U&\in\left[0,0.02\right]\,.
\end{aligned}
\end{align}
\begin{figure}[t]
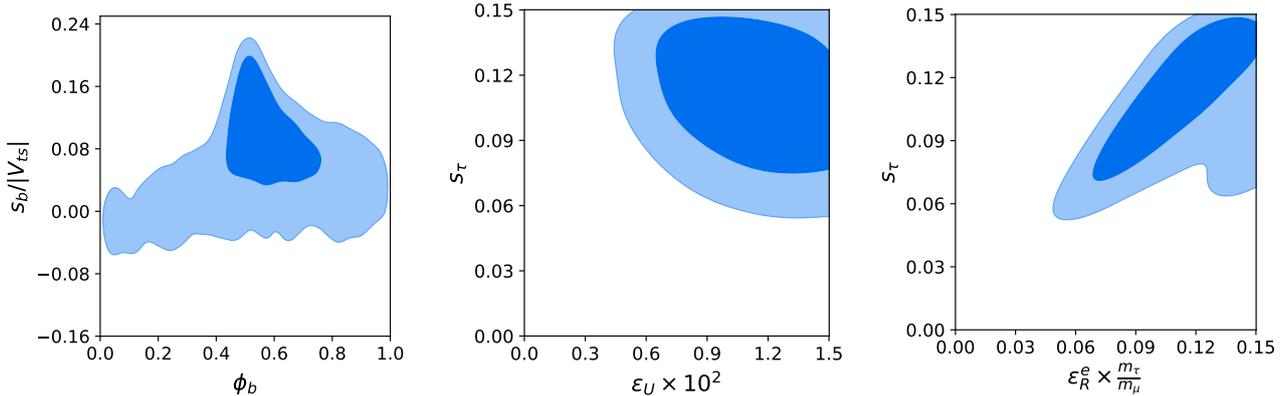

\centering
\includegraphics[width=0.325\textwidth]{./sbVSphib.png}~~~\includegraphics[width=0.325\textwidth]{./epsUVSstau.png}~~\includegraphics[width=0.325\textwidth]{./epsRSstau.png}
\caption{$68\%$ (dark blue) and $95\%$ (light blue) posterior probabilities of $\phi_b$ and $s_b$ (left), $\epsilon_U$ and $\stau$ (mid), and of $\epsilon_R^e$ and $\stau$ (right).}
\label{fig:fitparam}
\end{figure}
We obtain the following $68\%$ probability ranges for the model parameters extracted from the marginalized posterior probabilities 
\begin{align}
\omega_1&=1.0\pm  0.3~\mathrm{TeV},&\qquad\omega_3&=1.2\pm 0.2~\mathrm{TeV},&\qquad \stau &=0.11\pm 0.03,\nonumber\\[2pt]
s_b&=(0.09\pm 0.06)\,|V_{ts}|,&\phi_b&=\left(0.55\pm 0.15\right)\pi,& \epsilon_R^e &=(0.11\pm 0.03)\,\frac{m_\mu}{m_\tau}\,,\nonumber\\[2pt]
\epsilon_U&=(1.2\pm 0.3)\times 10^{-2}\,.
\end{align}
In figure~\ref{fig:fitparam}, we show the $68\%$ and $95\%$ two-dimensional posterior probabilities for $s_b$ and $\phi_b$, $\epsilon_U$ and $\stau$, and for $\epsilon_R^e$ and $\stau$. As can be seen, there is a clear correlation between the phase $\phi_b$ and the maximum allowed value for $s_b$. We also find that positive values of $s_b$ are preferred. This behaviour is expected from the discussion in the previous section: while the size of $s_b$ and preferred value for $\phi_b$ are connected to the (negative) NP contribution to $\Delta F=2$, the preference for a positive $s_b$ is related to the partial cancellations in $D-\bar D$ mixing and $\mathcal{B}(B\to\tau\nu)$. On the other hand, the anti-correlation between $\epsilon_U$ and $\stau$ can be easily understood from the fact that the NP contribution in $b\to s\ell\ell$ transitions is proportional to the product of these two parameters, i.e. $\mathrm{Re}\,(\Delta \cC_9^{\mu\mu }) \approx  -\, \mathrm{Re}\,(\Delta \cC_{10}^{\mu\mu}) \propto C_U\,\stau\,\epsilon_U$. Finally, we find a significant
 correlation between $\epsilon_R^e$ and $\stau$. 
As shown in the previous section, a mild cancellation (at the level of $20\%$) among these two parameters is required to ensure a sufficiently small $\ttaumuR$, as indicated by  
$\mathcal{B}(B_s\to\mu\mu)$ and $\mathcal{B}(\mu\to e\gamma)$.
Note that, beside the smallness of $s_b$ compared to $|V_{ts}|$,
the other three mixing parameters ($\epsilon_U$, $\stau$, and $\epsilon_R^e$) 
turn out to have magnitudes in good agreement with their natural parametric size.

\begin{figure}[p]
\centering
\includegraphics[width=0.45\textwidth]{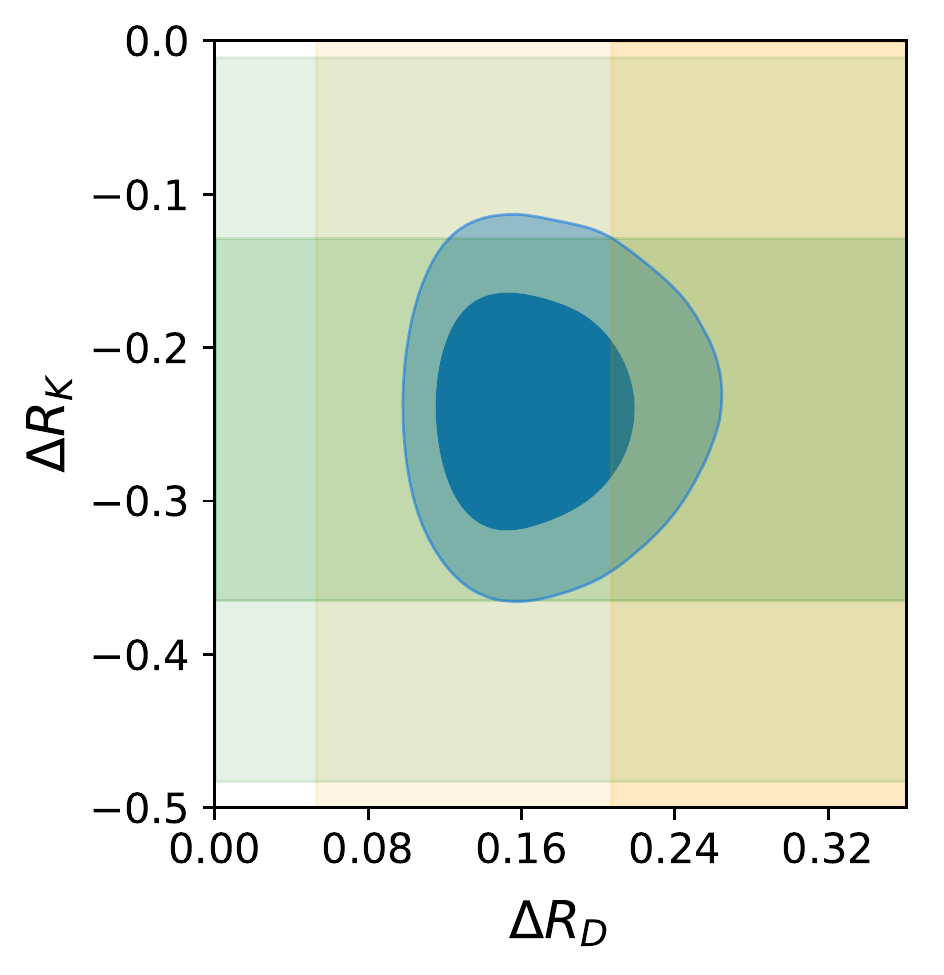}~~~\includegraphics[width=0.45\textwidth]{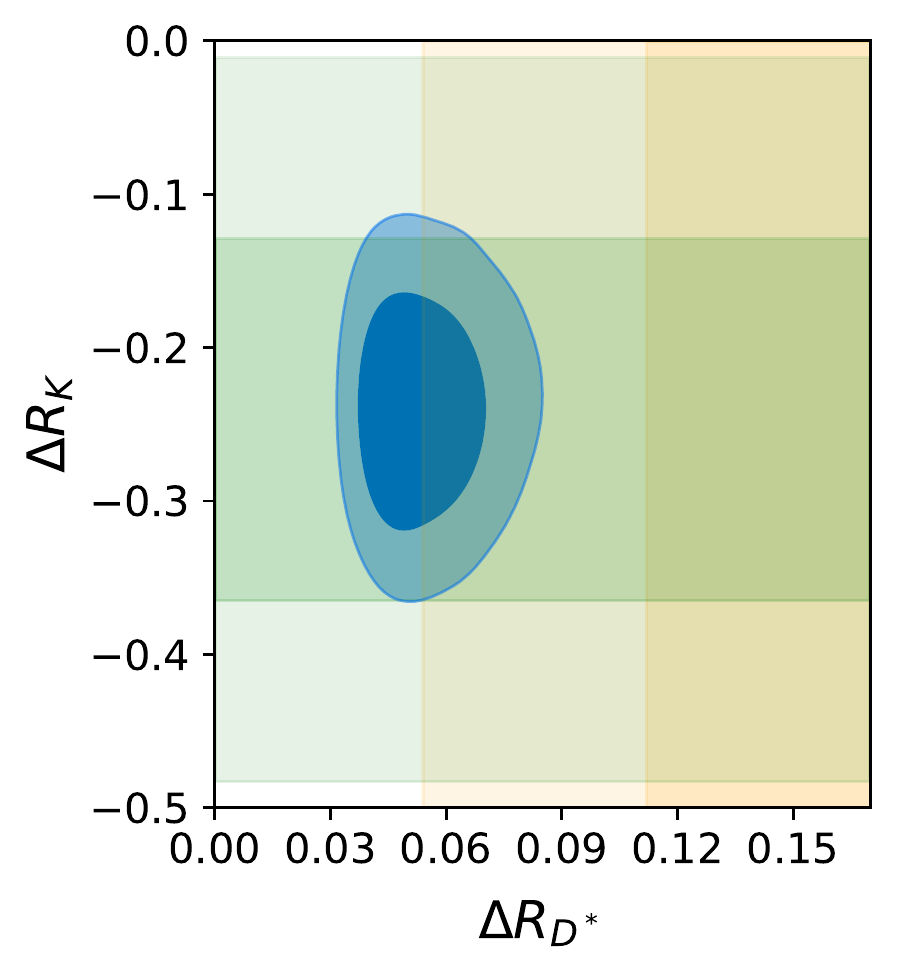}
\caption{$68\%$ (dark blue) and $95\%$ (light blue) posterior probabilities of the NP shifts in $R_{D^{*}}$ vs. 
$\Delta R_K$. The experimental values at $1\sigma$ ($2\sigma$) are indicated by the dark (light) coloured bands.}
\label{fig:anomalies}
\end{figure}

Concerning low energy observables, we reach similar conclusions to those already discussed in Section~\ref{sect:pheno} in terms of simplified   analytical expressions. In Figure~\ref{fig:anomalies} we show the $68\%$ and $95\%$ posterior probabilities for $\Delta R_{D^{(*)}}$ and $\Delta R_K$. 
As can be seen, the model can fully accommodate the anomalies in $b\to s\ell\ell$. However, as anticipated in Section~\ref{subsec:b2ctaunu}, the complete explanation of the $R_{D^{(*)}}$ anomalies within this framework is limited by  LFU tests in $\tau$ decays. From the fit we obtain a NP enhancement of around $7\%$--$8\%$ for $R_{D^*}$ and $18\%$--$22\%$ for $R_D$.

\begin{figure}
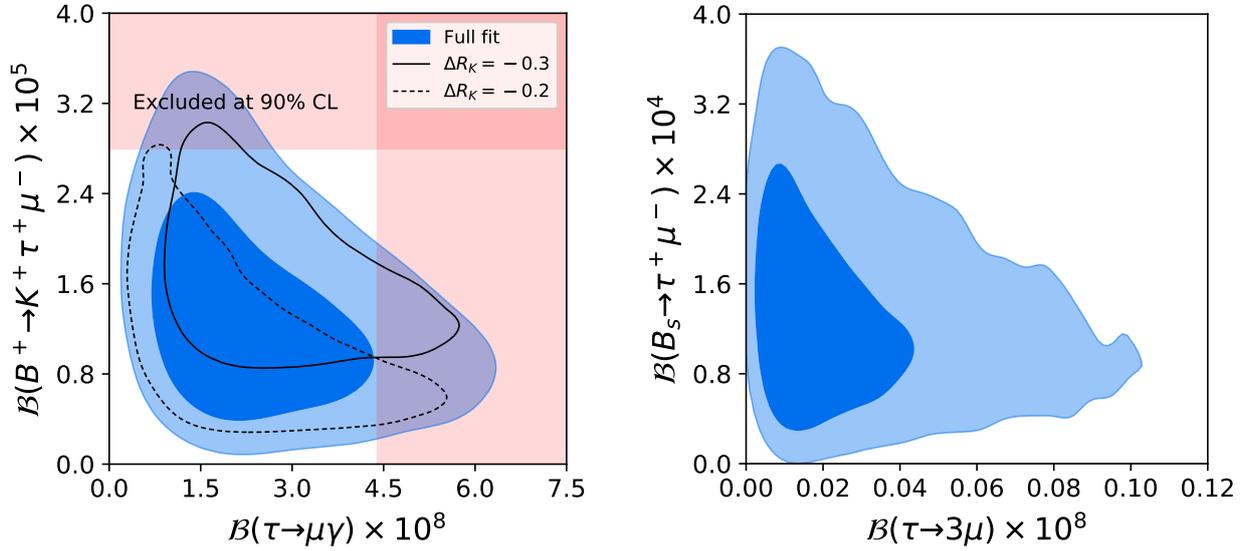

\centering
\includegraphics[width=0.47\textwidth]{./LFVfit2.png}~~~\includegraphics[width=0.48\textwidth]{./LFVfit1.png}
\caption{Left: $68\%$ (dark blue) and $95\%$ (light blue) posterior probabilities of $\mathcal{B}(\tau\to\mu\gamma)$ and $\mathcal{B}(B^+\to K^+ \tau^+\mu^-)$ from the global fit. The black lines denote the $95\%$ posterior probabilities  fixing $\Delta R_K=-0.3$ (solid) and $\Delta R_K=-0.2$ (dashed).
The red bands show the 90\% CL exclusion limits for these observables. 
Right:  $68\%$ (dark blue) and $95\%$ (light blue) posterior probabilities of $\mathcal{B}(\tau\to3 \mu)$ and $\mathcal{B}(B_s \to \tau^+\mu^-)$ 
from the global fit.
}
\label{fig:LFVfit}
\end{figure}

As already emphasized in Section~\ref{subsec:LFV},
in our setup the explanation of the anomalies implies large LFV effects in $\tau\to\mu$ transitions, in particular in $\tau\to\mu\gamma$,
$\tau\to3\mu$, $B\to K\tau\mu$, and $B_s\to \tau\mu$. Interestingly, we find that the NP effects in $\tau\to\mu\gamma$ are anti-correlated to those in $B_s\to \tau\mu$ (and $B\to K\tau\mu$), allowing us to directly connect the product of these LFV rates to the NP enhancement in $R_{D^{(*)}}$ and $b\to s\ell\ell$. More precisely, we find the following relations among NP observables
\begin{align}\label{eq:Anom2LFV}
\begin{aligned}
\left(\frac{\Delta R_D}{0.2}\right)^2\left(\frac{\Delta R_K}{0.3}\right)^2&\approx3\left[\frac{\mathcal{B}(B\to K\tau^+\mu^-)}{3\times10^{-5}}\right]\left[\frac{\mathcal{B}(\tau\to\mu\gamma)}{5\times10^{-8}}\right]  \\[5pt]
 &\approx\left[\frac{\mathcal{B}(B_s\to \tau^+\mu^-)}{1\times10^{-4}}\right]\left[\frac{\mathcal{B}(\tau\to\mu\gamma)}{5\times10^{-8}}\right]\,,
\end{aligned}
\end{align}
which hold almost independently from any model parameter. This is illustrated in Figure~\ref{fig:LFVfit} (left) where we show the $68\%$ and $95\%$ posterior probabilities for $\mathcal{B}(\tau\to\mu\gamma)$ and $\mathcal{B}(B\to K\tau\mu)$. We see that the model predictions for these two observables are close to their experimental bounds shown in the red bands, as implied by the expressions in \eqref{eq:Anom2LFV}.
A partial anti-correlation is present also between  $\tau\to3 \mu$ and LFV in $B$ decays, as illustrated in Figure~\ref{fig:LFVfit} (right).
However, in this case the effect is diluted 
by the uncertainty on $Z^\prime$ mass and couplings, which are not strongly constrained by other observables.

As a final comment, it is worth stressing that this low-energy fit does not pose stringent constraints on the masses of the heavy vector bosons.
The low-energy observables constrain only the effective Fermi couplings in Eq.~(\ref{eq:effFermi}), or $\omega_{1,3}$. Still, we can derive 
a well-defined range for vector boson masses taking into account that   $g_U \gg g_c$: 
setting $2.5 \leq g_U \leq 3.0$, the masses of $Z'$, $U$, and $G'$ range between 2 and 3 TeV.

%%%%%%%%%%%%%%%%%%%%%%%%%%%%%%%%%%%%%%%%%%%%%%%%%%%%%%%%%%%%%%%%%%
\section{Conclusions}
%%%%%%%%%%%%%%%%%%%%%%%%%%%%%%%%%%%%%%%%%%%%%%%%%%%%%%%%%%%%%%%%%%

The main idea behind the $\PSC$ model is that the 
flavor universality of strong, weak, and electromagnetic 
interactions observed at low energies is only a low-energy property: 
the ultraviolet completion of the SM is a theory where gauge interactions  are 
completely flavor non-universal, with each fermion family being charged under its own gauge group.
The motivation for this hypothesis, and the explicit construction of the $\PSC$ model presented in Ref.~\cite{Bordone:2017bld} 
is twofold: it explains the pattern of anomalies recently observed in $B$ meson decays 
and, at the same time, the well-known hierarchical structure of quark and lepton mass matrices. These two phenomena 
turn out to be  closely connected: they both follow from the dynamical breaking of the flavor non-universal gauge
structure holding at high energies down to the SM.

On general grounds, low-energy observables put very stringent constraints on flavor non-universal interactions
mediated by TeV-scale bosons, as expected in the $\PSC$ model.  
In this paper we have presented a comprehensive analysis of such constrains, 
and the corresponding implications for future low-energy measurements. 
As far as the constraints are concerned, we confirm 
the main conclusions of Ref.~\cite{Bordone:2017bld}: i)~the model is in very good agreement with all existing bounds, 
without significant tuning of its free parameters; ii)~the model could account for the $B$ anomalies, 
reaching the $1\sigma$ range of all the present measurements with the exception of $R_{D^*}$, 
where the maximal allowed deviation from the SM does not exceed the 10\% level.  
In addition, we have shown that the model 
can slightly improve the description of $\Delta F=2$ observables with respect to the SM.

The most interesting aspect of this analysis is related to the possible implications of  the $\PSC$ model 
in view of future low-energy measurements. We have shown that a remarkable feature is the prediction  of 
sizeable rates for  LFV processes of the type $\tau\to\mu$, both in $B$ decays (such as 
$B\to K \tau \mu$ and  $B_s\to\tau\mu$) as well as in $\tau$ decays (most notably 
$\tau \to \mu\gamma$ and $\tau \to 3\mu$).  The fact that the $B$ anomalies could naturally 
imply large LFV effects in $B$ decays was first pointed out  
in Ref.~\cite{Glashow:2014iga}, on the basis of general considerations. 
The $\PSC$ model provides an explicit realization of this mechanism, predicting in addition 
a strict anti-correlation between $\tau \to \mu\gamma$  and $b\to s \tau\mu$ transitions, illustrated in Figure~\ref{fig:LFVfit},
that can be viewed as a distinctive signature.  As we have shown in Section~\ref{subsec:LFV}, 
also $\mu \to 3 e$,  $\mu \to e\gamma$,  and $K_L \to \mu e$ decays could be close to their present exclusion limits; 
however, this conclusion is less strict given the uncertainty on the $\mu\to e$ mixing, which
is not constrained by the anomalies. 

Besides LFV processes, we have shown that the model predicts interesting non-standard effects 
in $\Delta F=1$ and  $\Delta F=2$ observables, with non-trivial correlations. Particularly relevant 
and distinctive are the predictions for the violations of LFU in charged currents illustrated in Figure~\ref{fig:CC}:
the presence of right-handed currents implies $\Delta R_{D} \approx 2.6\, \Delta R_{D^*}$ and 
a possible large enhancement of $\mathcal{B}(B\to \tau\nu)$ ranging from $30\%$ up to $100\%$ 
of the SM prediction.

Most of the predictions for low-energy observables presented in this work differ with respect to what is expected in 
other models proposed for a combined explanation of the $B$ anomalies.
The corresponding measurements would therefore be of great value in shedding light on the dynamics behind the anomalies,  
if unambiguously confirmed as due to physics beyond the SM, and clarify their possible link to the 
origin of quark and lepton masses.

\subsubsection*{Acknowledgments}

We thank L. Di Luzio, M. Nardecchia, A. Greljo and M. K\"onig for useful comments and discussions. 
This research was supported in part by the Swiss National Science Foundation (SNF) under contract 200021-159720.

\bigskip 
\appendix

\section{Structure of the SM Yukawa couplings in $\PSC$}
\label{app:Yukawa}

Within our model the complete set of Yukawa couplings, i.e.~the couplings of the chiral fermions to the scalar field responsible for the breaking 
of the electroweak symmetry, is generated only  after the 
 $\mathrm{SM}_{1+2} \times \mathrm{PS}_3 \to \mathrm{SM}$ symmetry breaking. 
Below such scale, adopting the SM notation, we define the couplings as
\be
\cL  =   {\bar q}^i_L  (Y_d)_{ij} d^i_R\, \phi
+ {\bar q}^i_L  (Y_u)_{ij} u^i_R\,  \phi^c
+ {\bar e}^i_L  (Y_e)_{ij} e^j_R\, \phi  {\rm ~+~h.c.}~,
\label{eq:LY}
\ee
where $i,j=1,2,3$ and  $\phi$ is the effective SM Higgs field (normalized such that $\langle \phi^\dagger \phi \rangle = v^2/2$, with $v=246\GeV$).

As discussed in Section~\ref{sect:MainYuk}, we can decompose each Yukawa coupling as follows
\be
Y_f = y_3^f 
\begin{pmatrix}
 \epsilon^f_{LR}\, X_{LR} & \epsilon^{f}_L\, V_L  \\[2pt]  
\epsilon^f_{R}\, V^\intercal_R  & 1
\end{pmatrix}~,
\ee
where $V_L$ and $V_R$ are unit vectors in the $\mathrm{U(2)}_{q+\ell}$ and $\mathrm{U(2)}_{u+d+e}$ space, 
and $X_{LR}$ is a $2\times 2$ non-hermitian matrix satisfying ${\rm Tr}(X_{LR} X_{LR}^{\dagger})=1$.
Since non-vanishing $\epsilon^f_L$,  $\epsilon^f_{LR}$, $\epsilon^f_R$,
are induced  by  operators with $d=5$,~$6$,~$7$, respectively, on general grounds we expect 
$|\epsilon^f_R| \ll |\epsilon^f_{LR}| \ll |\epsilon^f_L| \ll 1~$.

Without loss of generality, we can work in the flavor basis where 
\be
V_L \to \hat n_2 = \left(\begin{array}{c} 0 \\ 1 \end{array}\right)~, 
\label{eq:VLbasis}
\ee
i.e.~in the basis where the  left-handed second generation is defined by the orientation in flavor space of the link fields  $\Omega_{3}$ and $\Omega_1$. This is what we conventionally define as the {\em interaction basis} for the left-handed doublets.
We can use the freedom on the 
right-handed sector to set $X_{LR}$ in the form $ U \times \textrm{diag}\left(0, 1 \right)$, where $U$ is a unitary matrix.
 The null eigenvalue of $X_{LR}$,
corresponding to the limit of  massless first generation, can be lifted by introducing additional link fields, with subleading VEVs. 
The inclusion of such terms effectively amount to change $X_{LR}$ into a Yukawa-dependent term 
$\Delta_{f}$ of the form 
\be
X_{LR}  \to   U_f^\dagger\,  \Delta_f~,  \qquad  \Delta_f=\textrm{diag}\left(\frac{m_f^1}{m_f^2}, 1 \right)~,
\label{eq:VRbasis}
\ee
where $U_f$ is a (complex) unitary matrix. In the limit where the $\Ufive$ breaking in the right-handed sector 
is induced by a single field ($\Phi_R$ in the minimal set-up), then $d=6$ and $d=7$ terms are aligned 
in the right-handed sector.  This implies $V_R \to \hat n_2$
in the basis defined by Eq.~(\ref{eq:VRbasis}). 

In such basis, the quark Yukawa matrices assume the explicit form
\be
\label{Yudef}
Y_u= y_t \left(\begin{array}{cc}
  \epsilon_{LR}^u\, U_u^\dagger\,  \Delta_u & \epsilon^u_{L}\, \hat n_2 \\[2pt]  
  \epsilon^u_R\,  \hat n_2^\intercal & 1
\end{array}\right),  \qquad 
Y_d= y_b \left(\begin{array}{cc}
 \epsilon^d_{LR}\, U_d^\dagger\, \Delta_d & \epsilon^d_L\, \hat n_2 \\[2pt]  
 \epsilon^d_R\,  \hat n^\intercal_2  & 1
\end{array}\right)~.
\ee
Following the discussion of CP phases in Ref.~\cite{Barbieri:2011ci}, without loss of generality 
we can set $\epsilon^f_{LR}$ to be real 
(contrary to  $\epsilon^f_L$ and $\epsilon^f_R$) and decompose the $2\times 2$ matrix $U_{f}$ as
\begin{equation}
 U_{f}=
\left(\begin{array}{cc}
c_f & s_f\,e^{i\alpha_f} \\
-s_f\,e^{-i\alpha_f} & c_f
\end{array}\right).
\end{equation}
In the following we assume that $s_f \ll 1$,
as naturally implied by the absence of fine-tuning in deriving the CKM matrix.

In the phenomenological analysis we employ the down-type quark and the charged-lepton mass-eigenstate basis for the SM fermions,
where the $\mathrm{SU(2)_L}$ structure of the left-handed doublets is given by
\begin{align}
\label{eq:down_basis}
q_L^i=
\begin{pmatrix}
V_{ki}^*\, u_k\\
d_i
\end{pmatrix}
\,,\qquad \ell_L^i=
\begin{pmatrix}
\nu_i\\
e_i
\end{pmatrix}\, ,
\end{align}
with $V_{ki}$ being the elements of CKM matrix. 
We move  from the interaction basis to this basis  by performing the rotation
\be
\left. q_L \right|_{\rm int} = L_d \times \left. q_L \right|_{\rm d-basis} ~, \qquad 
\left. \ell_L \right|_{\rm int} = L_e \times \left. \ell_L \right|_{\rm e-basis} ~.
\ee
More generally, we denote by $X_a$ (with $X=L,R$ and $a=u,d,e$) the unitary matrices
that bring the Yukawa couplings in diagonal form (starting from the interaction basis),
\be
 L_u^\dagger Y_u R_u  = {\rm diag}(y_u,y_c,y_t)~ ,\quad 
 L_d^\dagger Y_d R_d  = {\rm diag}(y_d,y_s,y_b)~,  \quad 
 L_e^\dagger Y_e R_e  = {\rm diag}(y_e,y_\mu,y_\tau)~,
\ee
where the $y_i$ are  real and positive and $V_{\rm CKM}=L_u^\dagger L_d$.

The $X_a$ have non-trival flavor-blind phases
$[{\rm det}(X_a)= e^{i\phi^X_a}$]. The electroweak symmetry 
implies $\phi^L_d=\phi^L_u$, and three relative phases  corresponding 
to unbroken global symmetries (hypercharge, lepton number, baryon number)
are unobservable.
Of the two remaining phases one combination affects the relative 
phase between the leptoquark couplings $\beta_q$ and $\beta_d$, 
and is potentially observable. 
Following Ref.~\cite{Bordone:2017bld}, we fix this phase by the condition $(\beta_q)_{33}=-(\beta_d)_{33}$
which allows us to maximize the contribution to $\Delta R_{D}$. 
Having fixed this phase, in the following we set ${\rm det}(X_a)=1$.

\paragraph*{Left-handed rotations in the quark sector.}
To a very good approximation, the left-handed diagonalization matrices have the form
\be
 L_{d}^\dagger  =   R_{12} (s_d; \alpha_d) \times  R_{23} (s_b; \phi_b )~,
 \qquad 
L_{u}^\dagger  =   R_{12} (s_u; \alpha_u) \times  R_{23} ( s_t; \phi_t)~,
 \ee
 where 
\be
R_{12} (s_d; \alpha_d) = \left(\begin{array}{cc}
 U_{d} & 0 \\  0 & 1 
 \end{array}\right)~, \quad
R_{23} (s_b; \phi_b) =  \left(\begin{array}{ccc}
 1   & 0  & 0  \\
 0   &  c_b &  s_b\,  e^{i\phi_{b}}   \\
 0   & -s_b\,  e^{-i\phi_{b}} & c_b
\end{array}\right)~, 
\ee
with $s_b/c_b = |\epsilon^d_L|$ and $\phi_{b}={\rm arg}(\epsilon^d_L)$, 
and similarly for the up sector. As we discuss next, three  out of the four real mixing parameters 
($s_b, s_d, s_t, s_u$) appearing in these matrices can be expressed in 
terms of CKM elements. Concerning the four phases ($\phi_b, \alpha_d, \phi_t, \alpha_u$),
one is unphysical and one can be expressed in terms of the CKM phase $\gamma$.

The CKM matrix is $V_{\rm CKM}= L_u^\dagger  L_d$, implying
\bea
V_{\rm CKM} &=&
 \left(\begin{array}{cc}
 U_{u} & 0 \\ 0 & 1 
 \end{array}\right) \times  R_{23} (s;\xi) 
\times 
 \left(\begin{array}{cc}
 U^\dagger_{d} & 0 \\ 0 & 1 
 \end{array}\right) 
\eea
where $(s/c) e^{i\xi} = s_b\, e^{-i\phi_{b}} -s_t\, e^{-i\phi_{t}} $. 
To match this structure 
with the standard CKM parametrization, we rephase it by imposing real 
 $V_{ud}$, $V_{us}$, $V_{cb}$, $V_{tb}$, and  $V_{cs}$ (which is real at the 
level of approximation we are working, namely up to corrections of $\ord{\lambda^2}$ relative to the leading term for each of CKM entry), 
obtaining 
\be
 V_{\rm CKM}=\left(\begin{array}{ccc}
 1- \lambda^2/2 &  \lambda & s_u\, s\, e^{-i \delta}  \\
-\lambda & 1- \lambda^2/2   & c_u\, s  \\
-s_d\, s \,e^{i (\delta+\alpha_u - \alpha_d)} & -s\, c_d & 1 \\
\end{array}\right),
\label{eq:CKMstand}
\ee
where the phase  $\delta$ and the  real and positive parameter $\lambda$, 
are defined by 
\be
\lambda\, e^{i \delta} = s_u\, c_d - c_u\, s_d\, e^{-i (\alpha_u - \alpha_d)} ~.
\label{eq:lambda}
\ee
Hence it follows that the three mixing angles $s_u$, $s_d$, and $s$ can be determined 
completely in terms of three independent CKM elements: 
\be
 s  = | s_t  - s_b e^{i(\phi_{t}-\phi_b)} | =   |V_{cb}|~, \qquad 
 \frac{s_u}{c_u} = \frac{|V_{ub}|}{|V_{cb}|}~, \qquad 
 \frac{s_d}{c_d} = - \frac{|V_{td}|}{|V_{ts}|}~.
\label{eq:CKMconst}
\ee
As far as the phases are concerned,  we find 
\be
\delta = - {\rm arg} (V_{ub}) \equiv  \gamma ~, \qquad  
\alpha_u - \alpha_d  = {\rm arg}(V_{td}) +  {\rm arg}(V_{ub})  \approx -\pi/2~,
\label{eq:phigamma}
\ee
where the last relation follows, to a very good accuracy, from the numerical values of the CKM inputs.

Flavor mixing in the left-handed sector is therefore controlled by the matrix $L_d$
that contains only three free parameters (the real mixing angle $s_b$ and the  
unconstrained phases $\phi_b$ and $\alpha_d$):
\be
 L_{d}  =   R_{23} (-s_b ; \phi_b ) R_{12} (-s_d; \alpha_d)  =
  \left(\begin{array}{ccc}
 c_d   &  -s_d\,e^{i\alpha_d}  & 0  \\
 s_d\,e^{-i\alpha_d}   &  c_d &  -s_b\,  e^{i\phi_{b}}   \\
s_d\,s_b\,e^{-i(\alpha_d+\phi_b)}   & s_b\, c_d\,e^{-i\phi_{b}} & 1
\end{array}\right)~, 
\ee
where $s_d$ is fixed by Eq.(\ref{eq:CKMconst}) and, 
consistently with the approximations so far performed, we have set $c_b=1$.

\paragraph*{Right-handed rotations in the quark sector.}
The structure of the right-handed rotation matrices is simpler, being confined 
to the 2-3 sector in the limit where we neglect tiny terms of $\ord{m_f^1/m_f^3,(\epsilon^f_{LR})^2}$. We find
\bea
R_d  &=&   \left(\begin{array}{ccc}
 1 & 0 & 0  \\
 0 & 1 &  \epsilon_R^d + \frac{m_s}{m_b} s_b\,  e^{i\phi_{b}} \\
 0 &  - (\epsilon^d_R)^* - \frac{m_s}{m_b} s_b\, e^{-i\phi_{b}} & 1
\end{array}\right) \equiv   \left(\begin{array}{ccc}
 1 & 0 & 0  \\
 0 & 1 &  \tbsR \\
 0 &  - (\tbsR)^* & 1
\end{array}\right)~,  
\label{eq:RHrot1}
\\
R_u  &=&   \left(\begin{array}{ccc}
 1 & 0 & 0  \\
 0 & 1 &  \epsilon^u_R + \frac{m_c}{m_t} s_t\,  e^{i\phi_{t}} \\
 0 &  - (\epsilon^u_R)^* - \frac{m_c}{m_t} s_t\, e^{-i\phi_{t}} & 1
\end{array}\right) \equiv   \left(\begin{array}{ccc}
 1 & 0 & 0  \\
 0 & 1 &  \theta^R_{tc} \\
 0 &  - (\theta^R_{tc})^* & 1
\end{array}\right)~.
\eea
Note that if we neglect the effect of d=7 effective operators 
(i.e.~for $\epsilon^{u,d}_R \to 0$), these matrices do not contain additional free parameters 
(i.e.~they are completely determined in terms of angles and phases appearing already in the left-handed sector).

\paragraph*{Rotations in the lepton sector.}
Given the model-dependence on the neutrino mass matrix,
 in the left-handed sector we cannot eliminate parameters in terms of known mixing angles; moreover, the strong constraints
on the $\mu \to e$ transitions imply that the 1-2 mixing terms are very small.  Proceeding as above, and neglecting higher-order terms 
in the 1-2 mixing, we thus decompose the left-handed rotation mixing matrix as 
\be\label{eq:Le}
 L_{e}  =    
  \left(\begin{array}{ccc}
 1   &   s_e\, e^{i \alpha_e} & 0  \\
 - s_e\, e^{-i \alpha_e } &  1& \stau\, e^{i\phi_\tau}  \\
   s_e \stau\, e^{-i(\alpha_e+\phi_\tau) }   &  - \stau\, e^{-i\phi_\tau} & 1
\end{array}\right)~.
\ee
In the right-handed sector, proceeding in full analogy with the quark case we get 
\be\label{eq:RHrot}
R_e  =   \left(\begin{array}{ccc}
 1 & 0 & 0  \\
 0 & 1 &  \epsilon_R^e - \frac{m_\mu}{m_\tau}\,\stau\, e^{i\phi_\tau} \\
 0 &  - (\epsilon^e_R)^* + \frac{m_\mu}{m_\tau}\, \stau\, e^{-i\phi_\tau}  & 1
\end{array}\right) \equiv   \left(\begin{array}{ccc}
 1 & 0 & 0  \\
 0 & 1 &  \ttaumuR \\
 0 &  - (\ttaumuR)^* & 1
\end{array}\right)~.
\ee

\section{Generation of the $\Ufive$-breaking effective operators}
\label{sect:Omegaeffops}  

An example of dynamical generation of the  $\Ufive$-breaking effective operators appearing in  
$\mathcal{L}_{\rm \Omega}^{d=5}$ and $\mathcal{L}_{\rm \Omega}^{d=6}$, defined in Eqs.~\eqref{eq:d5spurions} and~\eqref{eq:d6eps},
is obtained by introducing a pair of vector-like fermions, $\chi^i_{L/R}\sim\left(\mathbf{4},\mathbf{2},\mathbf{1}\right)_3$, $i=1,2$,
coupled to the SM leptons and quarks via
\bea
-\mathcal{L}_\chi \supset M_\chi\, \bar\chi^i_L\chi^i_R + \lambda_1\,\bar\ell^{\,2}_L\Omega_1 \chi^2_R+\lambda_3\, \bar q^{\,i}_L\Omega_3 \chi^i_R  + \lambda_H \,\bar\chi^2_L H_1 \Psi^3_R + \lambda^\prime_H \, \bar\chi^2_L H_1^c \Psi^3_R+ {\rm h.c.}~,
\label{eq:chiLag}
\eea
where $\Psi^3_R$ denotes the complete right-handed multiplet charged under $\mathrm{PS_3}$.
Assuming the vector-like fermions to be heavy, we can integrate them out obtaining the following tree-level expressions for the 
coefficients of the $\mathcal{L}_{\rm \Omega}^{d=5}$ operators:
\be
\frac{y_{q3}}{\Lambda_{23}}  =\frac{\lambda_3\lambda_H}{M_\chi}~,\qquad 
\frac{y_{\ell3}}{\Lambda_{23}}=\frac{\lambda_1\lambda_H}{M_\chi}~,\qquad 
\frac{y_{q3}^\prime}{\Lambda_{23}}  =\frac{\lambda_3 \lambda^\prime_H}{M_\chi}~,\qquad 
\frac{y_{\ell3}^\prime}{\Lambda_{23}}=\frac{\lambda_1\lambda^\prime_H}{M_\chi}~.
\ee
Similarly, in the case of the  $\mathcal{L}_{\rm \Omega}^{d=6}$ operators we get
\be
\epsilon_U = c_{q\ell}  \frac{  \omega_1 \omega_3}{\Lambda^2_{23}} = \frac{\lambda_1^*\lambda_3\, \omega_1 \omega_3}{2M_\chi^2}~, \quad
\epsilon_{\ell} =c_{\ell\ell} \frac{ \omega_1^2   }{\Lambda^2_{23}} =  \frac{|\lambda_1|^2\,\omega_1^2}{2M_\chi^2}~, \quad
\epsilon_{q} = c_{qq} \frac{ \omega_3^2   }{\Lambda^2_{23}} = \frac{|\lambda_3|^2\,\omega_3^2}{2M_\chi^2}~. \quad
\ee
If the vector-like mass is of $\mathcal{O}(\Lambda_{23})$, namely $M_\chi = {\rm~few} \times 10~\mathrm{TeV}$,  
then the $\lambda_i$ should assume $\cO(1)$ values  
to recover numerically correct entries for the Yukawa couplings. In this case the 
$\epsilon_i$ turn out to be of $\mathcal{O}(10^{-3})$. 
Alternatively, lowering the vector-like mass to  $M_\chi = \mathcal{O}(1~\mathrm{TeV})$, which is still compatible with 
high-energy phenomenology,\footnote{As suggested in~\cite{Greljo:2018tuh}, this option has the advantage of increasing 
the width of the TeV-scale vectors, hence alleviating the bounds from direct searches on these particles.}
the $\lambda_i$ turn out to be of $\cO(10^{-1})$ and the 
$\epsilon_i$ can rise up to $\mathcal{O}(10^{-2})$. We thus conclude that the natural range for the  parameters 
controlling the $\Ufive$ breaking  of the TeV-scale vectors  is $10^{-3} \lsim  | \epsilon_{\ell,q,U} |  \lsim 10^{-2}$.
 
In the limit $\lambda_i\to0$, the inclusion of the vector-like fermions enlarges the flavor symmetry of the model to $\Ufive \times \mathrm{U(2)_\chi}$.
The minimal breaking structure for the spurions discussed in section~\ref{sect:MainYuk} is achieved by choosing 
the coupling $\lambda_3$ to leave the subgroup $\mathrm{U(2)_{q+\chi}}$ unbroken.\footnote{While $X_{q\ell}\not=0$ 
necessarily implies a breaking of $\Ufive$, more precisely a breaking of 
$\mathrm{U(2)}_{q}\times \mathrm{U(2)}_\ell$, this is not the case for $X_{\ell\ell}$ and $X_{qq}$: the latter break $\Ufive$ only if they 
are not proportional to the identity matrix.}  This subgroup is however broken in other sectors, in particular by the couplings of the vector-like fermions to the Higgs.  
As a result, the minimal breaking structure receives subleading corrections when  considering products of more spurions, see Sections~\ref{sec:noYukBreak} 
and~\ref{sec:modelTeV} for a more detailed discussion.

\section{Wilson coefficients of the SMEFT}

\begin{table}[t]
\centering
\renewcommand{\arraystretch}{1.5}
\begin{tabular}{||l|l||}
\hline\hline
$ \left[Q_{\nu \nu}\right]_{ \alpha \beta \gamma \delta}= (\overline{\nu}_R^\alpha \gamma_\mu \nu_R^\beta) (\overline{\nu}_R^\gamma \gamma^\mu \nu_R^\delta)$ &
$ \left[Q_{\nu e}\right]_{ \alpha \beta \gamma \delta} = (\overline{\nu}_R^\alpha \gamma_\mu \nu_R^\beta) (\overline{e}_R^\gamma \gamma^\mu e_R^\delta)$  \\
$ \left[Q_{\nu u}\right]_{ \alpha \beta ij} = (\overline{\nu}_R^\alpha \gamma_\mu \nu_R^\beta) (\overline{u}_R^i \gamma^\mu u_R^j)$  &
$ \left[Q_{\nu d}\right]_{ \alpha \beta ij} = (\overline{\nu}_R^\alpha \gamma_\mu \nu_R^\beta) (\overline{d}_R^i \gamma^\mu d_R^j)$  \\
$\left[Q_{\ell \nu u q}\right]_{ \alpha \beta i j} = (\overline{\ell}_L^\alpha \nu_R^\beta) (\overline{u}_R^i q_L^j) $  &
$\left[Q_{e \nu u d}\right]_{\alpha\beta i j} = (\overline{e}_R^\alpha \gamma_\mu \nu_R^\beta) (\overline{u}_R^i \gamma_\mu d_R^j) $ \\
$\left[Q_{\ell \nu}\right]_{ \alpha \beta \gamma \delta} = (\overline{\ell}_L^\alpha \gamma_\mu \ell_L^\beta) (\overline{\nu}_R^\gamma \gamma^\mu \nu_R^\delta) $  &
$\left[Q_{q \nu}\right]_{ij \alpha \beta} = (\overline{q}_L^i \gamma_\mu q_L^j) (\overline{\nu}_R^\alpha \gamma^\mu \nu_R^\beta) $ \\
$\left[Q_{\phi \nu}\right]_{\alpha\beta}=(\phi^\dagger \,i\overleftrightarrow{D}_{\!\!\!\mu}\,\phi)(\overline{\nu}_R^\alpha \gamma^\mu \nu_R^\beta)$ &  \\
      \hline\hline
\end{tabular}
\caption{\small Dimension-six operators containing right-handed}  neutrinos. \label{eq:operators_nuR}
\end{table}

In Tables~\ref{tab:no4ferm},~\ref{tab:4ferm} and~\ref{tab:4ferm_nuR} we provide the matching conditions of the $Z^\prime$, $G^\prime$ and $U$ to the SMEFT, following the prescriptions described in Section~\ref{sec:Model2SMEFT}. We list the operators including right-handed neutrinos in Table~\ref{eq:operators_nuR}, while for the other operators we use the same basis as in~\cite{Grzadkowski:2010es}.

\begin{table}[p] 
\centering
\renewcommand{\arraystretch}{1.5}
\begin{tabular}{||l|l||l|l||l|l||} 
\hline \hline
\multicolumn{2}{||c||}{$X^3$} & 
\multicolumn{2}{c||}{$\phi^6$~ and~ $\phi^4 D^2$} &
\multicolumn{2}{c||}{$\psi^2\phi^3$}\\
\hline
$Q_G$                & - &  $Q_\phi$       & - & $\left[Q_{e\phi}\right]_{\alpha\beta}$           & - \\
$Q_{\widetilde G}$          & - &   $Q_{\phi\Box}$ & $\eftZp{\phi\Box}=4\,\Big(\frac{g_1}{g_4}\Big)^4$ & $\left[Q_{u\phi}\right]_{ij}$           & - \\
$Q_W$                & - &    $Q_{\phi D}$   & $\eftZp{\phi D}=16\,\Big(\frac{g_1}{g_4}\Big)^4$ & $\left[Q_{d\phi}\right]_{ij}$           & - \\
$Q_{\widetilde W}$          & - &&&&\\    
\hline \hline
\multicolumn{2}{||c||}{$X^2\phi^2$} &
\multicolumn{2}{c||}{$\psi^2 X\phi$} &
\multicolumn{2}{c||}{$\psi^2\phi^2 D$}\\ 
\hline
$Q_{\phi G}$     & - & $Q_{eW}$               & - &   $\left[Q_{\phi\ell}^{(1)}\right]_{\alpha\beta}$      & $\eftZpF[(1)]{\phi\ell}{ \alpha \beta}=12\,\Big(\frac{g_1}{g_4}\Big)^2\cZl[\alpha\beta]$\\
$Q_{\phi\widetilde G}$         & - &  $Q_{eB}$        & - &$\left[Q_{\phi\ell}^{(3)}\right]_{\alpha\beta}$ & - \\
$Q_{\phi W}$     & - & $Q_{uG}$        & - &$\left[Q_{\phi e}\right]_{\alpha\beta}$            & $\eftZpF{\phi e}{ \alpha \beta}=12\,\Big(\frac{g_1}{g_4}\Big)^2\,\cZe[\alpha\beta]$\\
$Q_{\phi\widetilde W}$         & - & $Q_{uW}$               & -  &$\left[Q_{\phi q}^{(1)}\right]_{ij}$      & $\eftZpF[(1)]{\phi q}{ij}=-4\,\Big(\frac{g_1}{g_4}\Big)^2\,\cZq[ij]$ \\
$Q_{\phi B}$     & - & $Q_{uB}$        & - & $\left[Q_{\phi q}^{(3)}\right]_{ij}$ & -\\
$Q_{\phi\widetilde B}$         & - & $Q_{dG}$        & -  &$\left[Q_{\phi u}\right]_{ij}$            & $\eftZpF{\phi u}{ij}=-4\,\Big(\frac{g_1}{g_4}\Big)^2\,\cZu[ij]$\\
$Q_{\phi WB}$     & - &  $Q_{dW}$               & -  & $\left[Q_{\phi d}\right]_{ij}$            & $\eftZpF{\phi d}{ij}=-4\,\Big(\frac{g_1}{g_4}\Big)^2\,\cZd[ij]$\\
$Q_{\phi\widetilde WB}$ & - & $Q_{dB}$  & - &  $\left[Q_{\phi u d}\right]_{ij}$   & -\\
\hline \hline
\end{tabular}
\caption{\small Wilson coefficients of operators other than four-fermion ones. }\label{tab:no4ferm}
\end{table}
\begin{table}[p]
\centering
\renewcommand{\arraystretch}{1.5}
\begin{tabular}{||l|l||l|l||}
\hline\hline
\multicolumn{2}{||c||}{$(\bar RR)(\bar RR)$} & \multicolumn{2}{c||}{$(\bar LR)(\bar RL)$ and $(\bar LR)(\bar LR)$}\\
\hline
$\left[Q_{\nu \nu}\right]_{ \alpha \beta \gamma \delta}$ & $\eftZpF{\nu \nu}{ \alpha \beta \gamma \delta} = 9\, \cZnu[\alpha \beta] \cZnu[ \gamma \delta]$ & $\left[Q_{\ell \nu u q}\right]_{ \alpha \beta i j}$~[*] & $\eftUF{\ell \nu u q}{\alpha \beta i j}= -2\, \cUu[i\beta]  \cUqs[j \alpha] $\\[5pt]
\hhline{~~==}
$\left[Q_{\nu e}\right]_{ \alpha \beta \gamma \delta}$ & $\eftZpF{\nu e}{ \alpha \beta \gamma \delta} = 18\, \cZnu[\alpha \beta] \cZe[ \gamma \delta]$ & \multicolumn{2}{c||}{$(\bar LL)(\bar RR)$} \\
\hhline{~~--}
$\left[Q_{\nu u }\right]_{\alpha\beta i j}$ & $\eftUF{\nu u}{\alpha\beta i j} =\cUu[i \beta]\cUus[j \alpha]$ & $ \left[Q_{\ell \nu}\right]_{ \alpha \beta \gamma \delta}$   & $\eftZpF{\ell \nu}{ \alpha \beta \gamma \delta} =  18\,\cZl[\alpha \beta] \cZnu[ \gamma \delta]$ \\
 & $\eftZpF{\nu u}{ \alpha \beta  ij} = -6\, \cZnu[\alpha \beta] \cZu[ij]$ & $\left[Q_{q \nu}\right]_{ij \alpha \beta}$  & $\eftZpF{q\nu}{ ij  \alpha \beta} =  -6\,\cZq[ij] \cZnu[\alpha \beta] $\\[5pt]
 \hhline{~~==}
$\left[Q_{\nu d}\right]_{ \alpha \beta \gamma \delta}$ & $\eftZpF{\nu d}{ \alpha \beta ij} = -6\, \cZnu[\alpha \beta] \cZd[ ij]$ & \multicolumn{2}{c||}{$\phi^2\psi^2$} \\
\hhline{~~--}
$\left[Q_{e \nu u d}\right]_{\alpha\beta i j}$~[*] & $\eftUF{e \nu u d}{\alpha\beta i j} =\cUu[i \beta]\cUds[j \alpha]$ & $\left[Q_{q \nu}\right]_{\alpha\beta}$ & $ \eftZpF{\phi \nu}{ \alpha \beta}=12\, \left(g_Y/g_4\right)^2\cZnu[\alpha\beta] $\\[5pt]
\hhline{====}
\end{tabular}
\caption{\small Wilson coefficients of four-fermion operators involving right-handed neutrinos.
For the operators denoted with a [*], 
the hermitian conjugate has to be considered as well. 
 \label{tab:4ferm_nuR}}
\end{table}

\begin{table}[p]
\centering
\renewcommand{\arraystretch}{1.5}
\begin{tabular}{||l|l||l|l||}
\hline\hline
\multicolumn{2}{||c||}{$(\bar LL)(\bar LL)$} & \multicolumn{2}{c||}{$(\bar LR)(\bar RL)$ and $(\bar LR)(\bar LR)$} \\
\hline
$\left[Q_{\ell\ell}\right]_{\alpha \beta \gamma \delta}$  & $\eftZpF{\ell\ell}{\alpha \beta \gamma \delta}= 9\,\cZl[\alpha \beta] \cZl[\gamma \delta]$  & $\left[Q_{\ell edq}\right]_{\alpha \beta  ij}$ ~[*]  & $\eftUF{\ell edq}{\alpha \beta i j}= -2\,\cUd[i\beta]  \cUqs[j \alpha] $ \\
$\left[Q_{qq}^{(1)}\right]_{ijkl}$  & $\eftGF[(1)]{qq}{ijkl} = \frac{1}{4}\, \cGq[il]\cGq[kj] -\frac{1}{6}\, \cGq[ij]\cGq[kl]$ & $\left[Q_{quqd}^{(1)}\right]_{ijkl}$ &  -  \\
 & $\eftZpF[(1)]{qq}{ijkl} = \cZq[ij] \cZq[kl]$ & $\left[Q_{quqd}^{(8)}\right]_{ijkl}$ & - \\
$\left[Q_{qq}^{(3)}\right]_{ijkl}$  &  $\eftGF[(3)]{qq}{ijkl} = \frac{1}{4}\, \cGq[il]\cGq[kj] $ &$\left[Q_{\ell equ}^{(1)}\right]_{\alpha\beta ij}$ & - \\
$\left[Q_{\ell q}^{(1)}\right]_{\alpha \beta i j}$  & $\eftUF[(1)]{\ell q}{\alpha \beta i j} = \frac{1}{2}\, \cUq[i \beta]\cUqs[j \alpha]$ & $\left[Q_{\ell equ}^{(3)}\right]_{\alpha\beta ij}$ & -  \\ 
  & $\eftZpF[(1)]{\ell q}{\alpha \beta i j} = -6 \, \cZl[\alpha \beta]\cZq[i j]$ &  &   \\ 
$\left[Q_{\ell q}^{(3)}\right]_{\alpha \beta i j}$  & $\eftUF[(3)]{\ell q}{\alpha \beta i j} = \eftUF[(1)]{\ell q}{\alpha \beta i j}$ &  &  \\[5pt]
\hline\hline
\multicolumn{2}{||c||}{$(\bar RR)(\bar RR)$} & \multicolumn{2}{c||}{$(\bar LL)(\bar RR)$}\\
\hline
$\left[Q_{ee}\right]_{ \alpha \beta \gamma \delta}$  &  $\eftZpF{e e}{ \alpha \beta \gamma \delta} = 9\, \cZe[\alpha \beta] \cZe[ \gamma \delta]$ & $\left[Q_{\ell e}\right]_{ \alpha \beta \gamma \delta}$   & $\eftZpF{\ell e}{ \alpha \beta \gamma \delta} =  18\,\cZl[\alpha \beta] \cZe[ \gamma \delta]$  \\
$\left[Q_{uu}\right]_{ijkl}$  & $\eftGF{uu}{ijkl} =  \frac12 \cGu[il]\cGu[kj] -\frac16 \cGu[ij]\cGu[kl]$ & $\left[Q_{\ell u}\right]_{ \alpha \beta i j}$   & $\eftZpF{\ell u}{ \alpha \beta ij} =  -6\,\cZl[\alpha \beta] \cZu[ij]$ \\
  & $\eftZpF{uu}{ijkl} =  \cZu[ij] \cZu[kl]$ & $\left[Q_{\ell d}\right]_{ \alpha \beta i j}$   & $\eftZpF{\ell d}{ \alpha \beta ij} =  -6\,\cZl[\alpha \beta] \cZd[ij]$\\
$\left[Q_{dd}\right]_{ijkl}$  & $\eftGF{dd}{ijkl} =  \frac12 \cGd[il]\cGd[kj] -\frac16 \cGd[ij]\cGd[kl]$ &  $\left[Q_{qe}\right]_{ij \alpha \beta}$  & $\eftZpF{qe}{ ij  \alpha \beta} =  -6\,\cZq[ij] \cZe[\alpha \beta] $ \\
 & $\eftZpF{dd}{ijkl} =  \cZd[ij] \cZd[kl]$ &  $\left[Q_{qu}^{(1)}\right]_{ijkl}$      & $\eftZpF{qu}{ ijkl} =  2\,\cZq[ij] \cZu[kl] $\\
$\left[Q_{eu}\right]_{\alpha\beta ij}$  &  $\eftZpF{eu}{ \alpha \beta ij} = -6\, \cZe[\alpha \beta] \cZu[ ij]$  &$\left[Q_{qu}^{(8)}\right]_{ijkl}$ &  $\eftGF[(8)]{qu}{ijkl} =  2\, \cGq[ij]\cGu[kl]$\\
$\left[Q_{ed}\right]_{\alpha\beta i j}$ &  $\eftUF{ed}{\alpha\beta i j} = \cUd[i \beta]\cUds[j \alpha]$ & $\left[Q_{qd}^{(1)}\right]_{ijkl}$ & $\eftZpF{qd}{ ijkl} =  2\,\cZq[ij] \cZd[kl] $\\
 & $\eftZpF{ed}{ \alpha \beta ij} = -6\, \cZe[\alpha \beta] \cZd[ ij]$ &$\left[Q_{qd}^{(8)}\right]_{ijkl}$ & $\eftGF[(8)]{qu}{ijkl} =  2\,\cGq[ij]\cGd[kl]$ \\
$\left[Q_{ud}^{(1)}\right]_{ijkl}$ & $\eftZpF[(1)]{ud}{ijkl} = 2\, \cZu[ij] \cZd[ kl]$ & & \\
$\left[Q_{ud}^{(8)}\right]_{ijkl}$ & $\eftGF[(8)]{ud}{ijkl} =  2\, \cGu[ij]\cGd[kl] $ & & \\[5pt]
\hline\hline
\end{tabular}
\caption{\small Wilson coefficients of four-fermion operators. For the operators denoted with a [*], 
the hermitian conjugate has to be considered as well. \label{tab:4ferm}}
\end{table}

\section{Low energy observables and NP contributions}
\label{app:obs}

In this section we list all the low-energy observables considered in the phenomenological analysis together with their theory expressions and experimental values. The expressions for the low-energy observables are parametrised in terms of the WCs of the LEFT, for which we use the operator basis introduced in Ref.~\cite{Jenkins:2017jig}. The matching conditions between the SMEFT WCs and those of the LEFT can be found in App.~C of Ref.~\cite{Jenkins:2017jig}.

\subsection{LFV observables}
\label{app:LFV}

The full list of experimental values for the LFV observables included in the fit is provided in Table~\ref{tab:LFV_exp}. In what follows we describe the corresponding theory expressions.

\paragraph*{$\ell_\alpha \to \ell_\beta\ell_\gamma\bar\ell_\gamma$.} LFV decays of the type $\ell_\alpha \to \ell_\beta\ell_\gamma\bar\ell_\gamma$ are described in our model by the effective Lagrangian
\begin{align}
\begin{aligned}
\mathcal{L}(\ell_\alpha \to \ell_\beta\ell_\gamma\bar\ell_\gamma)=& -\frac{4 G_{F}}{\sqrt{2}} \left( \eftWCF{ee}{V, LL}{\beta\alpha\gamma\gamma} (\bar\ell^\beta_{L} \gamma_\mu{\ell}_L^\alpha)(\bar\ell^\gamma_{L} \gamma^\mu{\ell}_L^\gamma) + \eftWCF{ee}{V, RR}{\beta\alpha\gamma\gamma} (\bar\ell^\beta_R \gamma_\mu{\ell}_R^\alpha)(\bar\ell^\gamma_R \gamma^\mu{\ell}_R^\gamma)\right.\\
&\left. \qquad \;\;\; +\eftWCF{ee}{V, LR}{\gamma\gamma\beta\alpha} (\bar\ell^\beta_R \gamma_\mu{\ell}_R^\alpha)(\bar\ell^\gamma_L \gamma^\mu{\ell}_L^\gamma)+ \eftWCF{ee}{V, LR}{\beta\alpha\gamma\gamma}  (\bar\ell^\beta_L \gamma_\mu{\ell}_L^\alpha)(\bar\ell^\gamma_R \gamma^\mu{\ell}_R^\gamma)\right)\,.
\end{aligned}
\end{align}
Using the expressions in \cite{Kuno:1999jp,Brignole:2004ah}, we find the following result for the branching ratio for $\ell_\alpha \to \ell_\beta\ell_\beta\bar\ell_\beta$:
\begin{equation}\label{eq:tauTo3mu}
\frac{\mathcal{B}(\ell_\alpha\to\ell_\beta\ell_\beta\bar\ell_\beta)}{\mathcal{B}(\ell_\beta \to \ell_\beta \overline{\nu}_\beta \nu_\alpha)_{\rm SM}} =  \left( 2 \left| \eftWCF{ee}{V, LL}{\beta\alpha\beta\beta}  \right|^2 + 2 \left| \eftWCF{ee}{V, RR}{\beta\alpha\beta\beta}  \right|^2 +  \left| \eftWCF{ee}{V, LR}{\beta\beta\beta\alpha}  \right|^2 +  \left| \eftWCF{ee}{V, LR}{\beta\alpha\beta\beta}   \right|^2 \right) \,.
\end{equation}

\paragraph*{$\ell_\alpha \to \ell_\beta \gamma$.} In our model these processes receive the dominant contributions from one-loop amplitudes mediated by the leptoquark and the $b$ quark. In spite of the loop suppression, the presence of both left- and right-handed leptoquark couplings gives rise to contributions that are $m_b$-enhanced. Considering only the enhanced contributions we find
\begin{align}\label{eq:tauTomuGam}
\begin{aligned}
\mathcal{B}(\tau \to \mu \gamma) &\approx\frac{1}{\Gamma_\tau} \frac{\alpha}{256\pi^4}\frac{m_{\tau}^{3}\,m_b^2}{v^4}\,C_U^2\,\stau^2\,,\\[5pt]
\mathcal{B}(\mu \to e \gamma) &\approx\frac{1}{\Gamma_\mu} \frac{\alpha}{256\pi^4}\frac{m_{\mu}^{3}\,m_b^2}{v^4}\,C_U^2\,\stau^2\,s_e^2\,|\ttaumuR|^2\,.
\end{aligned}
\end{align}
On the other hand, we have that $\mathcal{B}(\tau \to e \gamma)$ is parametrically suppressed with respect to $\mathcal{B}(\tau \to \mu \gamma)$ and thus does not give any relevant constraint.

\begin{table}[tp]
\renewcommand{\arraystretch}{1}
\setlength{\tabcolsep}{9pt}
\centering
\begin{tabular}{@{}clcc@{}}
\toprule[1.6pt] 
\multicolumn{3}{l}{\bf Purely leptonic LFV transitions} \\ 
 \midrule 
 Observable & Experiment & EFT \\ 
  \midrule 
$\mathcal{B}(\tau\to3\mu)$ & $0(7)\cdot 10^{-9}$~\cite{Amhis:2016xyh} & \eqref{eq:tauTo3mu}  \\[3pt]  
$\mathcal{B}(\mu\to3 e)$ & $0(5)\cdot 10^{-13}$~\cite{Bellgardt:1987du}  & \eqref{eq:tauTo3mu}  \\[3pt]  
$\mathcal{B}(\tau\to\mu\gamma)$ & $0(3)\cdot 10^{-8}$~\cite{Amhis:2016xyh}  & \eqref{eq:tauTomuGam}  \\[3pt]  
$\mathcal{B}(\mu\to e\gamma)$ & $0(6)\cdot 10^{-14}$~\cite{TheMEG:2016wtm}  & \eqref{eq:tauTomuGam}  \\[3pt]  
\midrule[1.6pt] 
\multicolumn{3}{l}{\bf  Semileptonic LFV transitions} \\ 
 \midrule
Observable & Experiment & EFT \\ 
\midrule 
$\mathcal{B}(B\to\tau^\pm e^\mp)$ & $0.0(1.7)\cdot 10^{-5}$~\cite{Aubert:2008cu}  & \eqref{eq:P2ellellp}  \\[3pt]  
$\mathcal{B}(B\to\mu^\pm e^\mp)$ & $0.0(1.5)\cdot 10^{-9}$~\cite{Aaij:2013cby}  & \eqref{eq:P2ellellp}  \\[3pt]
$\mathcal{B}(K_L\to\mu^\pm e^\mp)$ & $0.0(2.9)\cdot 10^{-12}$~\cite{Ambrose:1998us}  &  \eqref{eq:P2ellellp}\\[3pt]
$\mathcal{B}(B^+\to K^+\tau^+\mu^-)$ & $0.0(1.7)\cdot 10^{-5}$~\cite{Lees:2012zz} &  \eqref{eq:B2Kll}\\
\bottomrule[1.6pt] 
\end{tabular}
\caption{\small   List of observables involving LFV transitions.}
\label{tab:LFV_exp}
\end{table} 

\paragraph*{$P\to\ell\ell^\prime$ and $B\to K^{(*)}\tau\mu$.}  The leptoquark generally yields large contributions to leptonic and semileptonic LFV meson decays. To describe these processes it is useful to match the Wilson coefficients of the LEFT into the commonly used weak effective Hamiltonian
\begin{equation}
\mathcal{H}_{\rm WET} \supset - \frac{4G_F}{\sqrt2}\,\frac{e^2}{16\pi^2}\,V_{ti}V_{tj}^* \sum_i \Big[ \mathcal{C}_i\,\mathcal{O}_i + h.c. \Big]\,,
\label{HWET}
\end{equation}
where the operators are defined as
\begin{align}\label{eq:WET_op}
\begin{aligned}
\mathcal{O}_9^{ij,\alpha\beta}&=\left(\bar d_j\gamma_\mu P_L\,d_i\right)\left(\overline{e}_\alpha\gamma^\mu e_\beta\right)\,,    &\mathcal{O}_{9^\prime}^{ij,\alpha\beta} &=  \left(\bar d_j\gamma_\mu P_R\,d_i\right)\left(\overline{e}_\alpha\gamma^\mu e_\beta\right)\,,\\[5pt] 
\mathcal{O}_{10}^{ij,\alpha\beta}&= \left(\bar d_j\gamma_\mu P_L\,d_i\right)\left(\overline{e}_\alpha\gamma^\mu\gamma_5 e_\beta\right)\,,   &\mathcal{O}_{10^\prime}^{ij,\alpha\beta}&= \left(\bar d_j\gamma_\mu P_R\,d_i\right)\left(\overline{e}_\alpha\gamma^\mu\gamma_5 e_\beta\right)\,,    \\[5pt]
\mathcal{O}_{S}^{ij,\alpha\beta}&= (\bar d_j P_R\,d_i )( \bar e_\alpha e_\beta) \,,  &\mathcal{O}_{S^\prime}^{ij,\alpha\beta}&= (\bar d_j P_L d_i)( \bar e_\alpha e_\beta) \,,  \\[5pt] 
\mathcal{O}_{P}^{ij,\alpha\beta}&= (\bar d_j P_R\,d_i)( \bar e _\alpha \gamma_5 e_\beta) \,,  &\mathcal{O}_{P^\prime}^{ij,\alpha\beta}&= (\bar d_j P_L d_i )( \bar e_\alpha  \gamma_5 e_\beta) \,,\\[5pt]
\mathcal{O}_{\nu}^{ij,\alpha\beta}&=\left(\bar d_j\gamma_\mu P_L d_i\right)\left(\overline{\nu}_\alpha\gamma^\mu (1-\gamma_5) \nu_\beta\right)\,,    &\mathcal{O}_{\nu^\prime}^{ij,\alpha\beta} &=  \left(\bar d_j\gamma_\mu P_R\,d_i\right)\left(\overline{\nu}_\alpha\gamma^\mu (1-\gamma_5) \nu_\beta\right)\,,\\[5pt] 
\end{aligned}
\end{align}
with $P_{L,R}=1/2(1\mp\gamma_5)$. We have 
\begin{align}\label{eq:WETmatching}
\begin{aligned}
\mathcal{C}_9^{ij,\alpha\beta}&=-\frac{2\pi}{\alpha\,V_{ti}V_{tj}^*}\left(\eftWCF{ed}{V,LL}{\alpha\beta ji}+\eftWCF{de}{V, LR}{ji\alpha\beta}\right)  
	& &  \hskip -1.2 cm 
	 +~ \mathcal{C}_9^{\rm SM}\, \delta_{\alpha\beta} \,,   \\
\mathcal{C}_{10}^{ij,\alpha\beta}&=\frac{2\pi}{\alpha\,V_{ti}V_{tj}^*}\left(\eftWCF{ed}{V, LL}{\alpha\beta ji}-\eftWCF{de}{V, LR}{ji\alpha\beta}\right) 
	& &  \hskip -1.2 cm
	+~ \mathcal{C}_{10}^{\rm SM}\, \delta_{\alpha\beta}\,,   \\
\mathcal{C}_{9^\prime}^{ij,\alpha\beta}&=-\frac{2\pi}{\alpha\,V_{ti}V_{tj}^*}\left(\eftWCF{ed}{V, LR}{\alpha\beta ji}+\eftWCF{ed}{V, RR}{\alpha\beta ji}\right)\, 
& \mathcal{C}_{10^\prime}^{ij,\alpha\beta}&=\frac{2\pi}{\alpha\,V_{ti}V_{tj}^*}\left(\eftWCF{ed}{V, LR}{\alpha\beta ji}-\eftWCF{ed}{V, RR}{\alpha\beta ji}\right)\,,\\
\mathcal{C}_{S}^{ij,\alpha\beta}&=-\frac{2\pi}{\alpha\,V_{ti}V_{tj}^*}\eftWCF{ed}{S, RL}{\beta\alpha ij}^*\,,& \mathcal{C}_{S^\prime}^{ij,\alpha\beta}&=-\frac{2\pi}{\alpha\,V_{ti}V_{tj}^*}\eftWCF{ed}{S, RL}{\alpha\beta ji}\,,\\
\mathcal{C}_{P}^{ij,\alpha\beta}&=\frac{2\pi}{\alpha\,V_{ti}V_{tj}^*}\eftWCF{ed}{S, RL}{\beta\alpha ij}^*\,,& \mathcal{C}_{P^\prime}^{ij,\alpha\beta}&=-\frac{2\pi}{\alpha\,V_{ti}V_{tj}^*}\eftWCF{ed}{S, RL}{\alpha\beta ji}\,,\\
\mathcal{C}_{\nu}^{ij,\alpha\beta}&=-\frac{2\pi}{\alpha\,V_{ti}V_{tj}^*}\eftWCF{\nu d}{V, LL}{\alpha\beta ji} 
~+~ \mathcal{C}_\nu^{\rm SM}\, \delta_{\alpha\beta} \,,
 & \mathcal{C}_{\nu^\prime}^{ij,\alpha\beta}&=-\frac{2\pi}{\alpha\,V_{ti}V_{tj}^*}\eftWCF{\nu d}{V, LR}{\alpha\beta ji}\,. 
\end{aligned}
\end{align}
with the SMEFT Wilson coefficients evaluated at the low-energy scale. Throughout the paper we will omit the quark indices whenever they refer to $b\to s$ transitions, i.e. when $ij=bs$. Using this effective Hamiltonian, we can write the branching fraction for the LFV leptonic decay of a neutral pseudo-scalar meson with valence quarks $i$ and $j$, $P_{ij}$, as
\begin{align}\label{eq:P2ellellp}
\mathcal{B}(P_{ij}\to \ell_\alpha^- \ell_\beta^+) &= \frac{\tau_P}{64\pi^3}\frac{\alpha^2 G_F^2}{m_P^3}\,f_P^2\, |V_{ti}V_{tj}^\ast|^2\, \lambda^{1/2}(m_P^2,m_{\ell_\alpha}^2,m_{\ell_\beta}^2)\nonumber\\
&\times\Bigg{\lbrace} [m_P^2-(m_{\ell_\alpha}-m_{\ell_\beta})^2]\, \Bigg|(m_{\ell_\alpha}+m_{\ell_\beta})(\mathcal{C}_{10}^{ij,\alpha\beta}-\mathcal{C}_{10^\prime}^{ij,\alpha\beta})+\frac{m_P^2}{m_i+m_j}(\mathcal{C}_{P}^{ij,\alpha\beta}-\mathcal{C}_{P^\prime}^{ij,\alpha\beta}) \Bigg{|}^2\nonumber \\
&+ [m_P^2-(m_{\ell_\alpha}+m_{\ell_\beta})^2]\Bigg|(m_{\ell_\alpha}-m_{\ell_\beta})(\mathcal{C}_{9}^{ij,\alpha\beta}-\mathcal{C}_{9^\prime}^{ij,\alpha\beta})+\frac{m_P^2}{m_i+m_j}(\mathcal{C}_{S}^{ij,\alpha\beta}-\mathcal{C}_{S^\prime}^{ij,\alpha\beta}) \Bigg{|}^2\Bigg{\rbrace},
\end{align}
where the $P$ decay constant is defined as $\langle 0| \,\bar q_i\,\gamma_\mu\gamma_5\,q_j|P(p)\rangle=ip_\mu\,f_P$.
\begin{table}[tp]
\hspace{-20pt}
\small
\renewcommand{\arraystretch}{1}
\begin{tabular}{|c|c|c|c|c|c|c|c|c|}
\hline
 & $c^{9+}_{K^{(*)}}$ & $c^{10+}_{K^{(*)}}$ & $c^{9-}_{K^{(*)}}$ & $c^{10-}_{K^{(*)}}$ & $c^S_{K^{(*)}}$ & $c^P_{K^{(*)}}$ & $c^{S9}_{K^{(*)}}$ & $c^{P10}_{K^{(*)}}$ \\[5pt]
\hline
$K$ & $\;9.6 \pm 1.0\;$ & $\;10.0 \pm 1.3\;$ & $0$ & $0$ & $13.6\pm0.9$& $14.6\pm1.0$&$12.4\pm0.9$ &$15.2\pm1.2$ \\
$K^*$ & $\;3.0 \pm 0.8\;$ & $\;2.7 \pm 0.7\;$ & $\;16.4 \pm 2.1\;$ & $\;15.4 \pm 1.9\;$ &- &- &- &-\\
\hline
\end{tabular} 
\caption{\small  Hadronic coefficients for the $B\to K^{(*)}\tau\mu$ decay.}\label{tab:B2Ktaumu}
\end{table}
For the branching fraction of the LFV semileptonic decay $B\to K^{(*)}\tau\mu$ we have:
\begin{align}\label{eq:B2Kll}
\begin{aligned}
\mathcal{B}(B\to K^{(*)}\tau^+\mu^-) &= 10^{-9} \left(c^{9+}_{K^{(*)}}\left|\mathcal{C}_9^{\tau\mu} + \mathcal{C}_{9^\prime}^{\tau\mu} \right|^2 + c^{10+}_{K^{(*)}}\left|\mathcal{C}_{10}^{\tau\mu} + \mathcal{C}_{10^\prime}^{\tau\mu} \right|^2 + c^{9-}_{K^{(*)}}\left|\mathcal{C}_9^{\tau\mu}-\mathcal{C}_{9^\prime}^{\tau\mu} \right|^2\right. \\[2pt]
 &\quad  + c^{10-}_{K^{(*)}}\left|\mathcal{C}_{10}^{\tau\mu} -\mathcal{C}_{10^\prime}^{\tau\mu} \right|^2 +  c^S_{K^{(*)}}\left|\mathcal{C}_S^{\tau\mu} + \mathcal{C}_{S^\prime}^{\tau\mu} \right|^2 +c^P_{K^{(*)}}\left|\mathcal{C}_P^{\tau\mu} + \mathcal{C}_{P^\prime}^{\tau\mu}  \right|^2 \\[4pt]
 &\quad +\left.  a^{S9}_{K^{(*)}}\,\Re[(\mathcal{C}_S^{\tau\mu} + \mathcal{C}_{S^\prime}^{\tau\mu})^* (\mathcal{C}_9^{\tau\mu}-\mathcal{C}_{9^\prime}^{\tau\mu})]+c^{P10}_{K^{(*)}}\,\Re[(\mathcal{C}_P^{\tau\mu} + \mathcal{C}_{P^\prime}^{\tau\mu})^*(\mathcal{C}_{10}^{\tau\mu} -\mathcal{C}_{10^\prime}^{\tau\mu})]
 \right)\,.
 \end{aligned}
\end{align}
The $a^i_{K^{(*)}}$ coefficients are given Table~\ref{tab:B2Ktaumu}. They have been computed using the lattice inputs in~\cite{Bouchard:2013pna}, and have been cross checked against~\cite{Crivellin:2015era} for those involving the $\cC_{9(10)}$ operators only. Note that for the $K^*$ channel the scalar contributions are expected to be negligible and hence we do not provide them.

\subsection{LFU tests in charged lepton decays and at LEP}\label{app:LFU}

Strong tests of LFU can be derived from the precise measurements of purely leptonic and semi-hadronic $\tau$ decays. 
Here we use the results from the HFLAV~\cite{Amhis:2016xyh}

\paragraph*{Leptonic decays.} Stringent tests of LFU can be obtained from ratios of leptonic lepton decays such as
\begin{align}
\left(\frac{g_\tau}{g_\mu}\right)_{\ell}=\left[\frac{\mathcal{B}(\tau\to e\, \nu\bar\nu)_{\rm exp}/\mathcal{B}(\tau\to e\, \nu\bar\nu)_{\rm SM}}{\mathcal{B}(\mu\to e\, \nu\bar\nu)_{\rm exp}/\mathcal{B}(\mu\to e\, \nu\bar\nu)_{\rm SM}} \right]^{\frac{1}{2}}\,,
\end{align}
and analogously for the other leptons. These ratios can be written in terms of the effective Lagrangian:

\begin{align}
\begin{aligned}
\mathcal{L}(\ell\to \ell^\prime\nu\bar\nu)=& -\frac{4 G_{F}}{\sqrt{2}} \left( \eftWCF{\nu e}{V, LL}{\rho \sigma \alpha\beta} (\overline{\nu}^{\rho}_{L} \gamma^{\mu} \nu_{L}^{\sigma})(\overline{\ell}_{L}^{\alpha} \gamma^{\mu} \ell^{\beta}_{L}) +  \eftWCF{\nu e}{V, LR}{\rho \sigma \alpha\beta} (\overline{\nu}_{L}^{\rho} \gamma^{\mu} \nu_{L}^{\sigma})(\overline{\ell}_{R}^{\alpha} \gamma^{\mu} \ell^{\beta}_{R}) \right)\,,
\end{aligned}
\end{align}
yielding the following expressions:
\begin{align}\label{eq:LFU_leptonic}
\begin{aligned}
\left(\frac{g_\tau}{g_\mu}\right)_{\ell}& = \left[\frac{\sum_{\rho \sigma}\left(|\delta_{\rho 3}\delta_{\sigma 1}+\eftWCF{\nu e}{V, LL}{\rho \sigma 13} |^2 +  |\eftWCF{\nu e}{V, LR}{\rho \sigma 13} |^2\right)}{\sum_{\rho \sigma}\left(|\delta_{\rho 2}\delta_{\sigma 1}+\eftWCF{\nu e}{V, LL}{\rho \sigma 12 } |^2 +  |\eftWCF{\nu e}{V, LR}{\rho \sigma 12 } |^2\right)}\right]^\frac{1}{2}\,,\\[5pt]
\left(\frac{g_\tau}{g_e}\right)_{\ell}& = \left[\frac{\sum_{\rho \sigma}\left( |\delta_{\rho 3}\delta_{\sigma 2}+\eftWCF{\nu e}{V, LL}{\rho \sigma 23} |^2 +  |\eftWCF{\nu e}{V, LR}{\rho \sigma 23} |^2\right)}{\sum_{\rho \sigma}\left( |\delta_{\rho 2}\delta_{\sigma 1}+\eftWCF{\nu e}{V, LL}{\rho \sigma 12 } |^2 +  |\eftWCF{\nu e}{V, LR}{\rho \sigma 12 } |^2\right)}\right]^\frac{1}{2}\,,\\[5pt]
\left(\frac{g_\mu}{g_e}\right)_{\ell}& = \left[\frac{\sum_{\rho \sigma}\left(|\delta_{\rho 3}\delta_{\sigma 2}+\eftWCF{\nu e}{V, LL}{\rho \sigma 23} |^2 +  |\eftWCF{\nu e}{V, LR}{\rho \sigma 23} |^2\right)}{\sum_{\rho \sigma} \left(|\delta_{\rho 3}\delta_{\sigma 1}+\eftWCF{\nu e}{V, LL}{\rho \sigma 13 } |^2 +  |\eftWCF{\nu e}{V, LR}{\rho \sigma 13 } |^2\right)}\right]^\frac{1}{2}\,.
\end{aligned}
\end{align}

\paragraph*{Hadronic decays.} LFU violation in hadronic $\tau$ decays can be tested by ratios such as 
\begin{align}
\left( \frac{g_\tau}{g_\mu} \right)_h = \left[\frac{\mathcal{B}(\tau \to h \nu)}{\mathcal{B}(h \to \mu \overline{\nu})}  \frac{2 m_h m_\mu^2 \tau_h}{(1 + \delta R_{\tau/h}) m_\tau^3 \tau_\tau} \left(\frac{1- m_\mu^2/m_h^2}{1- m_h^2/m_\tau^2}\right)^2\right]^{\frac{1}{2}}\,.
\end{align}
The decay $\tau^{-} \to h^{-} \nu$, with $h^{-}=d_{i} \overline{u}_{j}$, is described by the Lagrangian
\begin{align}
\mathcal{L}(\tau \to h \nu )= - \frac{ 4 G_{F}}{\sqrt{2}}  \sum_\rho\left(\delta_{\rho 3} V^{*}_{ji} + \eftWCF{\nu edu}{V, LL}{\rho 3 ij}\right)(\overline{\nu}_L^{\,\rho} \gamma^{\mu} \tau_L)(\overline{d}_L^{\,i} \gamma_{\mu} u_L^j) + \eftWCF{\nu edu}{S,RL}{\rho 3 ij}(\overline{\nu}_L^{\,\rho}  \tau_R)(\overline{d}_R^{\,i} u_L^j)\,,
\end{align} 
where we included also the SM contribution. The branching ratio for the process reads
\begin{align}
\mathcal{B}(\tau \to h \nu ) =  \frac{1}{16 \pi^{2}} G_{F}^{2} \tau_{h} f_{h}^{2} m_{\tau}^{3}  \left( 1 - \frac{m_h^{2}}{m_{\tau}^{2}}  \right)^{2}  \left| \delta_{\rho3} V^{*}_{ji} +\eftWCF{\nu e d u}{V, LL}{\rho 3ij }   +  \frac{m_h^{2}}{m_\tau (m_{d_{i}} + m_{u_{j}})} \eftWCF{\nu e d u}{S,RL}{\rho 3ij} \right|^2\,,
\end{align}
and analogously for $\mathcal{B}(h \to \mu \nu )$. Thus we find the following theoretical predictions for $(g_\tau/g_\mu)_\pi $ and $(g_\tau/g_\mu)_K$
\begin{align} 
\label{eq:LFU_semileptonic}
\begin{aligned}
\left( \frac{g_\tau}{g_\mu} \right)_\pi &=\left(\frac{\sum_\rho \left| \delta_{\rho3} V^*_{ud} +\eftWCF{\nu e d u}{V, LL}{\rho311}  +  \frac{m_\pi^{2}}{m_\tau (m_{d} + m_{u})}  \eftWCF{\nu e d u}{S,RL}{\rho311}\right|^2}{\sum_\rho\left|   \delta_{\rho2} V_{ud} +\eftWCF{\nu e d u}{V, LL}{\rho211}^\ast  +  \frac{m_\pi^{2}}{m_\mu(m_{d} + m_{u})}  \eftWCF{\nu e d u}{S,RL}{\rho211}^\ast \right|^2}\right)^\frac{1}{2}\,,\\[5pt]
\left( \frac{g_\tau}{g_\mu} \right)_K &=\left(\frac{\sum_\rho\left|   \delta_{\rho3} V^{*}_{us} +\eftWCF{\nu edu}{V, LL}{\rho321}  +  \frac{m_K^{2}}{m_\tau( m_{s} + m_{u})} \eftWCF{\nu e d u}{S,RL}{\rho321}\right|^2}{\sum_\rho\left| \delta_{\rho2} V_{us} +\eftWCF{\nu e d u}{V, LL}{\rho221}^\ast +  \frac{m_K^{2}}{m_\mu( m_{s} + m_{u})} \eftWCF{\nu e d u}{S,RL}{\rho221}^\ast \right|^2}\right)^\frac{1}{2}\,.
\end{aligned}
\end{align}
Due to the flavor structure of the model, tree-level leptoquark contributions in the hadronic $\tau$ vs $\mu$ ratios are found to be much smaller than those induced by the $m_t$-enhanced leptoquark loop. As a consequence, we find $(g_\tau/g_\mu)_\ell\approx(g_\tau/g_\mu)_\pi\approx(g_\tau/g_\mu)_K$ to a good extent. Similar tests with hadronic $\tau$ vs $e$ ratios can also we performed. These are less precise and do not yield relevant constraints. 

\bigskip
We also use the results of the fit in~\cite{Efrati:2015eaa} to account for the bounds on precision $Z$- and $W$-pole measurements at LEP. The experimental measurements we use in the fit for the LFU tests described in this section are summarized in Table~\ref{tab:LFU_exp}.

\begin{table}[tp]
\renewcommand{\arraystretch}{1}
\setlength{\tabcolsep}{9pt}
\centering
\begin{tabular}{@{}ccccc@{}}
\toprule[1.6pt] 
\multicolumn{4}{l}{\bf LFU tests in lepton decays} \\ 
 \midrule
 Observable & Experiment~\cite{Amhis:2016xyh} & Correlation & SM & EFT \\
 \midrule
$\left(\frac{g_\tau}{g_\mu}\right)_\ell$ & $1.0010(15)$ &  \multirow{5}{*}
 { $\left[
\begin{array}{ccccc}
\cdot & \cdot & \cdot & \cdot & \cdot \\[8.5pt]
0.53 & \cdot & \cdot & \cdot & \cdot\\[8.5pt]
-0.49 & 0.48 & \cdot & \cdot & \cdot\\[8.5pt]
0.24 & 0.26 & 0.02 & \cdot & \cdot\\[8.5pt]
0.11 & 0.10 & -0.01 & 0.06 & \cdot\\
\end{array}
\right]$} & 1. & \\[5pt]
$\left(\frac{g_\tau}{g_e}\right)_\ell$ & $1.0029(15)$ & & 1. & \eqref{eq:LFU_leptonic}\\[5pt]
$\left(\frac{g_\mu}{g_e}\right)_\ell$ & $1.0019(14)$ & & 1. &\\[5pt]
$\left(\frac{g_\tau}{g_\mu}\right)_\pi$ & $0.9961(27)$ & & 1. &  \multirow{2}{*}{\eqref{eq:LFU_semileptonic}}\\[2pt]
$\left(\frac{g_\tau}{g_\mu}\right)_K$ & $0.9860(70)$ & & 1. &  \\
\midrule[1.6pt] 
\multicolumn{4}{l}{\bf Z/W coupling modifications} \\ 
 \midrule
\multicolumn{4}{l}{We use the results of the fit in~\cite{Efrati:2015eaa}}\\
\bottomrule[1.6pt] 
\end{tabular}
\caption{\small   List of observables involving LFV transitions and LFU tests.}
\label{tab:LFU_exp}
\end{table} 

\subsection{$\Delta F=1$ semi-leptonic processes}

\paragraph*{$b\to s$ transitions.} 
We describe the NP contributions to $b\to s\ell\bar\ell$ and $b\to s\nu\bar\nu$ transitions in terms of the effective operators in \eqref{eq:WET_op}. The model predicts $\mathrm{Re}\,(\cC_9^{\alpha\alpha }) \approx  -\, \mathrm{Re}\,(\cC_{10}^{\alpha\alpha})$ to a very good approximation so we use fit results in~\cite{Capdevila:2017bsm} (see also~\cite{Altmannshofer:2017yso,Ciuchini:2017mik,Geng:2017svp,DAmico:2017mtc,Alok:2017sui,Hurth:2017sqw}) for this NP hypothesis. In order to analyse possible departures given by the scalar operators we also consider the $B_q\to\ell\ell$~($q=s,d$) channels separately. We have
\begin{align}\label{eq:Bs2ll}
\mathcal{B}(B_q\to\ell^-\ell^+) &= \left.\mathcal{B}(B_q\to \ell^-\ell^+)\right|_{\rm SM} \Bigg{\lbrace} \Bigg|\, \frac{\mathcal{C}_{10}^{bq,\ell\ell}-\mathcal{C}_{10^\prime}^{bq,\ell\ell}}{\mathcal{C}_{10}^{\rm SM}}+\frac{m_{B_q}^2}{2m_\ell(m_b+m_q)}\frac{\mathcal{C}_{P}^{bq,\ell\ell}-\mathcal{C}_{P^\prime}^{bq,\ell\ell}}{\mathcal{C}_{10}^{\rm SM}}\Bigg|^2 \nonumber \\
&\quad+\frac{m_{B_q}^2-4m_\ell^2}{m_{B_q}^2}\, \Bigg|\frac{m_{B_q}^2}{2m_\ell(m_b+m_q)}\frac{\mathcal{C}_{S}^{bq,\ell\ell}-\mathcal{C}_{S^\prime}^{bq,\ell\ell}}{\mathcal{C}_{10}^{\rm SM}} \Bigg|^2\Bigg{\rbrace}.
\end{align}
with the experimental and SM values listed in Table~\ref{tab:semileptonic}. The branching fraction of the $B\to K^{(*)}\nu\bar\nu$ decays are given by
\begin{align}\label{eq:BKnunu}
\begin{aligned}
\left.\mathcal{B}(B\to K^{(*)}\nu\bar\nu)\right|_{\rm \frac{exp}{SM}} &= \frac{\sum_{\alpha \beta} \left|  \mathcal{C}_{\nu}^{\alpha\beta} + \mathcal{C}_{\nu^{\prime}}^{\alpha\beta} \right|^{2} }{ 3 \left| \mathcal{C}_{\nu}^{SM} \right|^{2}}\,,
\end{aligned}
\end{align}
with the SM Wilson coefficient $\mathcal{C}_{\nu}^{SM}\approx-6.35$~\cite{Brod:2010hi,Buras:2014fpa}.

\paragraph*{$s\to d$ transitions.}  Here we focus only in $s\to d\nu\bar\nu$ decays. Since right-handed rotations involving the light families are negligible, the NP Lagrangian relevant for the $s \to d \nu\bar\nu$ transition reads 
\begin{align}
\mathcal{L}(s \to d \nu\bar\nu)=-\frac{4G_F}{\sqrt2}\,\eftWCF{\nu d}{V, LL}{\alpha\beta21}(\overline{\nu}_L^{\,\beta}\gamma^\mu\nu_L^{\alpha})(\overline{s}_L\gamma_{\mu}d_L)\,.
\end{align} 
Constraints on the Wilson coefficients above can be obtained from the measurements of $\mathcal{B}(K^+\to\pi^+\nu\bar\nu)$ and $\mathcal{B}(K_L\to\pi^0\nu\bar\nu)$, whose experimental values (with symmetrized errors) and SM predictions are collected in Table~\ref{tab:semileptonic}. The NP predictions in terms of the EFT (assuming NP only in $\nu_\tau$) can be extracted from~\cite{Bordone:2017lsy} and read
\begin{align}\label{eq:K2pinunuLEFT}
\begin{aligned}
\mathcal{B}(K^+\to\pi^+\nu\bar\nu)&=\left.\mathcal{B}(K^+\to\pi^+\nu\bar\nu)\right|_{SM}\left(\frac{2}{3}+\frac{1}{3}\left|1-\frac{2\,\eftWCF{\nu d}{V, LL}{3321}}{(\alpha/\pi)\,V_{ts}^*V_{td}\,C_{sd,\tau}^{\rm SM, eff}}\right|^2\right)\,,\\[5pt]
\mathcal{B}(K_L\to\pi^0\nu\bar\nu)&=\left.\mathcal{B}(K_L\to\pi^0\nu\bar\nu)\right|_{SM}\left(\frac{2}{3}+\frac{1}{3}\left|1+\frac{2\,\eftWCF{\nu d}{V, LL}{3321}}{(\alpha/\pi)\,V_{ts}^*V_{td}\,(X_t/s_W^2)}\right|^2\right)\,,
\end{aligned}
\end{align}
where $C_{sd,\tau}^{\rm SM, eff}\approx-8.5\, e^{0.11i}$ (including the long-distance contributions), and $X_t/s_W^2\approx6.4$. Given that the bounds from $K_L$ decays are way less stringent than those from the $K^+$, we implement only the latter in the fit.

\paragraph*{$b\to c\,(u)$ transitions.} In our setup, these transitions are described by the following effective operators:
\begin{equation}
\mathcal{L}(b\to u_i \ell \bar{\nu})=-\frac{4G_F}{\sqrt{2}}\left(\eftWCF{\nu edu}{V, LL}{\alpha\beta3i}^*(\overline{\ell}_L^{\,\beta}\gamma^\mu\nu_L^{\alpha})(\overline{u}_L^{\,i}\gamma_{\mu}b_L)+\eftWCF{\nu edu}{S, RL}{\alpha\beta3i}^* (\overline{\ell}_R^{\,\beta}\,\nu_L^{\alpha})(\overline{u}^{\,i}_Lb_R)\right)\, ,
\end{equation}
where $i=1,2$ for a $u$ or a $c$ quark respectively. We define the LFU ratios $R_{D^{(*)}}^{\ell\ell^\prime}$ as
\begin{align}
R_{D^{(*)}}^{\ell\ell^\prime}=\frac{\mathcal{B}(B\to D^{(*)}\ell\nu)}{\mathcal{B}(B\to D^{(*)}\ell^\prime\nu)}\,,
\end{align}
for which we find the following expression in terms of the EFT Wilson coefficients
\begin{equation}\label{eq:RDs}
\begin{aligned}
R_{D^*}^{\ell_\alpha\ell_\beta}&=\left.R_{D^*}^{\ell_\alpha\ell_\beta}\right|_{\rm SM} \left[1+2\,\mathrm{Re}\left\{\frac{\eftWCF{\nu edu}{V, LL}{\alpha\alpha 32}^*}{V_{cb}}\right\}+f^{S}_{D^{*}}(\ell_\alpha)\,\mathrm{Re}\left\{\frac{\eftWCF{\nu edu}{S, RL}{\alpha\alpha 32}^*}{V_{cb}}\right\}-(\alpha\to\beta)\right] \,,\\
R_{D}^{\ell_\alpha\ell_\beta}&=\left.R_{D}^{\ell_\alpha\ell_\beta}\right|_{\rm SM} \left[1+2\,\mathrm{Re}\left\{\frac{\eftWCF{\nu edu}{V, LL}{\alpha\alpha 32}^*}{V_{cb}}\right\}+f^{S}_{D}(\ell_\alpha)\,\mathrm{Re}\left\{\frac{\eftWCF{\nu edu}{S, RL}{\alpha\alpha 32}^*}{V_{cb}}\right\}-(\alpha\to\beta)\right] \,.
\end{aligned}
\end{equation}
The hadronic information on the scalar contributions is encoded in $f_{S}^{D^{(*)}}(\ell_\alpha)$. In our model, scalar contributions with taus are sizeable while those involving light leptons are negligible. For the tau channel we have~\cite{Fajfer:2012vx}
\begin{align}
f^{S}_{D^*}(\tau)= 0.12\,,
\qquad\qquad
f^{S}_D(\tau)= 1.5\,.
\end{align}
In order to constrain $e-\mu$ universality in $B\to D\ell\nu$ and $B\to D^*\ell\nu$ we use the $V_{cb}$ determinations in~\cite{Jung:2018lfu} instead of $R_{D^{(*)}}^{\mu e}$. The former also include the information on the differential distributions and therefore lead to stronger constraints than the ones on the branching ratios alone. We construct the following universality ratios, analogous to $R_{D^{(*)}}^{\mu e}$,
\begin{align}
V_{D^{(*)}}^{\mu e}=\frac{V_{cb}^{B\to D^{(*)}\mu\nu}}{V_{cb}^{B\to D^{(*)}e\nu}}\,.
\end{align}
Since we expect scalar contributions involving light leptons to be suppressed, we find
\begin{align}\label{eq:Vcb_mue}
V_D^{\mu e}=V_{D^*}^{\mu e}\approx 1 + 2\,\mathrm{Re}\left\{\frac{\eftWCF{\nu edu}{V, LL}{2232}^*}{V_{cb}}\right\}-2\,\mathrm{Re}\left\{\frac{\eftWCF{\nu edu}{V, LL}{1132}^*}{V_{cb}}\right\}\,.
\end{align}
Finally, defining the ratio of inclusive $B$ decays into charm states as
\begin{align}
R_{X_c}^{\tau\ell}=\frac{\mathcal{B}(B\to X_c\tau\nu)}{\mathcal{B}(B\to X_c\ell\nu)}\,,
\end{align}
and neglecting the light-lepton scalar contribution, we have
\begin{equation}
\label{eq:btoc_inclusive}
R_{X_c}^{\tau\ell}= \left.R_{X_c}^{\tau\ell}\right|_{\rm SM}\left[1+2\,\mathrm{Re}\left\{\frac{\eftWCF{\nu edu}{V, LL}{3332}^*}{V_{cb}}\right\}+0.427  \ \mathrm{Re}\left\{\frac{\eftWCF{\nu edu}{S, RL}{33 32}^*}{V_{cb}}\right\}-2\,\mathrm{Re}\left\{\frac{\eftWCF{\nu edu}{V, LL}{\ell\ell32}^*}{V_{cb}}\right\}\right]\,,
\end{equation}
with $\left.R_{X_c}^{\tau\ell}\right|_{\rm SM}=0.212\pm0.003$ and where we used the results in \cite{Celis:2016azn} for the scalar contributions.

\bigskip
The only important constraint in $b\to u\ell\nu$ transitions is given by the $B\to\tau\bar\nu_\tau$ branching fraction. For $B_q\to\tau\bar\nu_\tau~(q=u,\,c)$, we have
\begin{align}\label{eq:Btaunu}
\begin{aligned}
\mathcal{B}(B_q\to\tau\bar\nu)&=\left.\mathcal{B}(B_q\to\tau\bar\nu_\tau)\right|_{\rm SM}\sum_\rho\left\vert \delta_{\rho3}+ \frac{\eftWCF{\nu edu}{V, LL}{\rho33q}^*}{V_{qb}}+\frac{m_{B_q}^2}{(m_b+m_q) m_\tau}\frac{\eftWCF{\nu edu}{S, RL}{\rho33q}^*}{V_{qb}} \right\vert^2 \,.
\end{aligned}
\end{align}
In the fit we use $\left.\mathcal{B}(B\to\tau\bar\nu_\tau)\right|_{\rm SM}=0.807(61)$~\cite{Bona:2017cxr} for the SM value.

\begin{table}[t]
\hspace{-15pt}
\renewcommand{\arraystretch}{1}
\setlength{\tabcolsep}{11pt}
\begin{tabular}{@{}ccccr@{}}
\toprule[1.6pt]  
 \multicolumn{1}{l}{\bf $\boldsymbol{b \to s}$ transitions} &&& & \\ 
  \midrule 
$\mathcal{C}_9^{\mu\mu}=-\mathcal{C}_{10}^{\mu\mu}$& $-0.62(13)$~\cite{Capdevila:2017bsm} & & \eqref{eq:WETmatching} \\ 
 \midrule
 Observable & Experiment &   & SM  & EFT \\ 
\midrule
$\mathcal{B}(B_s\to \mu^-\mu^+)$ & $3.02(65)\times10^{-9}$~\cite{Aaij:2017vad} & & $3.65(23)\times10^{-9}$~\cite{Bobeth:2013uxa} & \multirow{2}{*}{\eqref{eq:Bs2ll}}\\
$\mathcal{B}(B\to \mu^-\mu^+)$ & $1.6(1.1)\times10^{-10}$~\cite{Aaij:2017vad} & & $1.06(9)\times10^{-10}$~\cite{Bobeth:2013uxa} & \\[2pt]
$\mathcal{B}(B_s\to \tau^-\tau^+)$ & $0.0(3.4)\times10^{-3}$~\cite{Aaij:2017xqt} & & $7.73(49)\times10^{-7}$~\cite{Bobeth:2013uxa} & \multirow{2}{*}{\eqref{eq:Bs2ll}}\\
$\mathcal{B}(B\to \tau^-\tau^+)$ & $0.0(1.1)\times10^{-3}$~\cite{Aaij:2017xqt} & & $2.22(19)\times10^{-8}$~\cite{Bobeth:2013uxa} & \\
$\left.\mathcal{B}(B\to K^{(*)}\nu\bar\nu)\right|_{\rm \frac{exp}{SM}}$ & $0.0(2.2)$~\cite{Lutz:2013ftz,Buras:2014fpa} &   & $1.$ & \eqref{eq:BKnunu} \\
\midrule
Coefficient & Fit &  & SM & EFT \\ 
  \midrule[1.6pt]
\multicolumn{1}{l}{\bf $\boldsymbol{s \to d}$ transitions} &&& & \\ 
 \midrule 
 Observable & Experiment &   & SM & EFT \\ 
   \midrule  
$\mathcal{B}(K^+\to\pi^+\nu\bar\nu)\times10^{11}$ & $17.8(11.0)$~\cite{Artamonov:2008qb}&&$8.4(1.0)$~\cite{Buras:2015qea}&\eqref{eq:K2pinunuLEFT}\\[2pt]
$\mathcal{B}(K_L\to\pi^0\nu\bar\nu)\times10^{11}$ & \multicolumn{2}{l}{\quad\;$<2.6\times10^3$~(90\% CL)~\cite{Ahn:2009gb}} &$3.4(0.6)$~\cite{Buras:2015qea}&\eqref{eq:K2pinunuLEFT}\\
\midrule[1.6pt] 
\multicolumn{1}{l}{\bf $\boldsymbol{b \to c}$ transitions} &&& & \\ 
\midrule
Observable & Experiment & Correlation & SM & EFT \\ 
\midrule
$\left.V_{cb}^{\mu e}\right|_D$ & $1.004(42)$~\cite{Jung:2018lfu} &  & $1.$ & \multirow{2}{*}{\eqref{eq:Vcb_mue}} \\
$\left.V_{cb}^{\mu e}\right|_{D^*}$ & $0.97(4)$~\cite{Jung:2018lfu} &   & $1.$ &  \\ 
\midrule
$R_D^{\tau\ell}$ & $0.407(46)$~\cite{Amhis:2016xyh} & \multirow{2}{*}{$-0.20$} & $0.299(3)$~\cite{Bigi:2016mdz} & \multirow{2}{*}{\eqref{eq:RDs}} \\ 
$R_{D^*}^{\tau\ell}$ & $0.304(15)$~\cite{Amhis:2016xyh} &   & $0.260(8)$~\cite{Bigi:2017jbd} & \\[5pt]
$R_{X_c}^{\tau\ell}$ &  $0.228(30)$~\cite{Barate:2000rc,Kamali:2018fhr} &   & $0.212(3)$~\cite{Mannel:2017jfk} & \eqref{eq:btoc_inclusive}\\
\midrule[1.6pt]
\multicolumn{1}{l}{\bf $\boldsymbol{b \to u}$ transitions} &&& & \\ 
 \midrule 
 Observable & Experiment &   & SM & EFT \\ 
  \midrule  
  $\mathcal{B}(B\to \tau\bar\nu)$ & $1.09(24)\times 10^{-4}$~\cite{Patrignani:2016xqp} &   & $0.807(61)\times 10^{-4}$~\cite{Bona:2017cxr} & \eqref{eq:Btaunu} \\[1mm] 
\bottomrule[1.6pt] 
\end{tabular}
\caption{\small   List of observables involving semileptonic transitions.}
\label{tab:semileptonic}
\end{table}

\subsection{$\Delta F=1$ non-leptonic processes}
A relevant constraint is obtained by time-dependent CP-violating 
asymmetries probing the weak phases of non-leptonic $b\to s$ 
amplitudes. The relevant effective Lagrangian reads 
\begin{equation}
\label{eq:Delta_F=1}
\begin{aligned}
\mathcal{L}^{(b_L \to s_L)}_{\Delta F=2} = &  -\frac{4G_F}{\sqrt{2}} \sum_a \cC_a O_a = -\frac{4G_F}{\sqrt{2}} \bigg[    
 	\eftWCF{dd}{V, LL}{bsii}(\overline{b}_L\gamma^{\mu}s_L) (\overline{d}_L^{\,i}\gamma_{\mu}d_L^i)    \\
& 	+\eftWCF{du}{V1, LL}{bsii}(\overline{b}_L\gamma^{\mu}s_L) (\overline{u}_L^{\,i}\gamma_{\mu}u_L^i) 
	+\eftWCF{du}{V8, LL}{bsii}(\overline{b}_L\gamma^{\mu}\,T^a\,s_L)(\overline{u}_L^{\,i}\gamma_{\mu}T^a\,u_L^i)   \\
&  	+\eftWCF{dd}{V1, LR}{bsii}(\overline{b}_L\gamma^{\mu}s_L)(\overline{d}_R^{\,i}\gamma_{\mu}d_R^i)    
  	+\eftWCF{dd}{V8, LR}{bsii}(\overline{b}_L\gamma^{\mu}\,T^a\,s_L)(\overline{d}_R^{\,i}\gamma_{\mu}T^a\,d_R^i)   \\
& 	+\eftWCF{du}{V1, LR}{bsii}(\overline{b}_L\gamma^{\mu}s_L)(\overline{u}_R^{\,i}\gamma_{\mu}u_R^i)  
 	+\eftWCF{du}{V8, LR}{bsii}(\overline{b}_L\gamma^{\mu}\,T^a\,s_L)(\overline{u}_R^{\,i}\gamma_{\mu}T^a\,u_R^i)~\bigg]~.  \\
\end{aligned}
\end{equation}
For a given exclusive transition of the type $B_{s,d} \to F$ we can write 
\be
\cA(B_q \to F)  \approx   \cA(B_q \to F)_{\rm SM}  e^{i \Delta \phi_q^{[F]} } ~, \qquad 
\Delta \phi_q^{[F]} = \sum_a (b_q^{[F]})_{\cC_a}  \times \Im\left[  \frac{ \cC_a}{ V_{ts} V_{tb^*}} \right]~,
\ee
where the $(b_q^{[F]})_{\cC_a}$ are real parameters encoding the RG evolution from the weak scale down to $m_b$ and the hadronix 
matrix elements of various four-quark operators. 

The phase shift $\Delta \phi_q^{[F]}$ is directly constrained  by the CP-violating 
asymmetries. In particular, in the clean case of $B_d \to \psi K$ one finds 
\be
\left| \Delta\phi_d^{[\phi K]} \right|_{\rm exp}  =  \left| \frac{ \sin(2\beta)_{\phi K} -   \sin(2\beta)_{\psi K}  }{  \sin(2\beta)_{\psi K}  } \right| =  0.07 \pm 0.15~.
\ee
Following the analysis of Ref.~\cite{Buchalla:2005us}, in this case the dominant non-vanishing coefficients are 
\be
\left(b_q^{[\phi K]}\right)_{\eftWCF{dd}{V, LL}{bsss}}   \approx  \left(b_q^{[\phi K]}\right)_{\eftWCF{dd}{V1, LR}{bsss}} \approx -45~, \qquad 
\left(b_q^{[\phi K]}\right)_{\eftWCF{dd}{V8, LL}{bsss}}   \approx  -4~.
\ee

\subsection{$\Delta F=2$ transitions}
The Lagrangian that contributes to $\Delta F=2$ in the down sector is given by
\begin{equation}
\label{eq:Delta_F=2}
\begin{aligned}
\mathcal{L}_{\Delta F=2}= -\frac{4G_F}{\sqrt{2}} \bigg[&\eftWCF{dd}{V, LL}{ijij}(\overline{d}_L^{\,i}\gamma^{\mu}d_L^j)(\overline{d}_L^{\,i}\gamma_{\mu}d_L^j)+\eftWCF{dd}{V, RR}{ijij}(\overline{d}_R^{\,i}\gamma^{\mu}d_R^j)(\overline{d}_R^{\,i}\gamma_{\mu}d_R^j) \\
+&\eftWCF{dd}{V1, LR}{ijij}(\overline{d}_L^{\,i}\gamma^{\mu}d_L^j)(\overline{d}_R^{\,i}\gamma_{\mu}d_R^j)+\eftWCF{dd}{V8, LR}{ijij}(\overline{d}_L^{\,i}\gamma^{\mu}\,T^a\,d_L^j)(\overline{d}_R^{\,i}\gamma_{\mu}T^a\,d_R^j)\bigg] \,,
\end{aligned}
\end{equation}
where $T^a$ are the generators of $SU(3)_c$. In order to study neutral meson mixing it is convenient to reexpress this operators in terms of the basis used in~\cite{Buras:2001ra}. After fierzing the operator $\eftOp{dd}{V8, LR}$ we find
\begin{equation}
\label{eq:Delta_F=2_Buras}
\begin{aligned}
\mathcal{L}_{\Delta F=2}= -\frac{4G_F}{\sqrt{2}} \bigg[&\eftWCF{dd}{V, LL}{ijij}\left[Q_1^{\rm VLL}\right]_{ijij}+\eftWCF{dd}{V, RR}{ijij}\left[Q_1^{\rm VRR}\right]_{ijij} \\
+&\Big(\eftWCF{dd}{V1, LR}{ijij}-\frac{1}{6}\eftWCF{dd}{V8, LR}{ijij}\Big)\left[Q_1^{\rm LR}\right]_{ijij}-\eftWCF{dd}{V8, LR}{ijij}\left[Q_2^{\rm LR}\right]_{ijij}\bigg] \,.
\end{aligned}
\end{equation}

\paragraph*{$B_{s,d}$--$\bar B_{s,d}$ mixing.}
The hadronic matrix elements for the operators relevant to $B_q$--$\bar B_q$ mixing ($q=d,s$) are 
conventionally decomposed as follows
\begin{align}
\begin{aligned}
\langle \bar{B}_q^0|Q_1^{\rm VLL}(\mu)|B_q^0\rangle =& \  \frac{1}{3} m_{B_q} f_{B_q}^2\, B_q^{\rm VLL}(\mu)\,, \\
\langle \bar{B}_q^0|Q_1^{\rm LR}(\mu)|B_q^0\rangle =&  -\frac{1}{6} R^1_q(\mu) m_{B_q} f_{B_q}^2\, B_q^{\rm LR1}(\mu)\,, \\
\langle \bar{B}_q^0|Q_2^{\rm LR}(\mu)|B_q^0\rangle =& \  \frac{1}{4} R^2_q(\mu) m_{B_q} f_{B_q}^2\, B_q^{\rm LR2}(\mu)\,.
\end{aligned}
\end{align} 
Here the so-called bag parameters $B_i^{a}(\mu)$, which are expected to be one in the vacuum saturation approximation, can be calculated in lattice QCD. The latest lattice determinations can be found in~\cite{Bazavov:2016nty}  and are shown in Table~\ref{tab:bag_Bmix_last}.\footnote{We stress that even though~\cite{Bazavov:2016nty} and~\cite{Buras:2001ra} adopt different conventions for the definition of the hadronic matrix elements, the matching between the different definitions of bag factors is consistent and unambiguous. In particular, the bag factors in Table~\ref{tab:bag_Bmix_last} have a one to one matching with the ones used in Eqs.~(7.28)-(7.30) of~\cite{Buras:2001ra}.} The chirality factors $R_i(\mu)$ are defined as~\cite{Gabbiani:1996hi}
\begin{equation}
R^1_q(\mu)=\left[\frac{m_{B_q}}{m_b(\mu)+m_q(\mu)}\right]^2+\frac{3}{2}\,,\qquad R^2_q(\mu)=\left[\frac{m_{B_q}}{m_b(\mu)+m_q(\mu)}\right]^2+\frac{1}{6}\,,
\end{equation}
with $\mu$ denoting the low-energy scale.

\begin{table}[t]
\begin{center}
\renewcommand{\arraystretch}{1.2}
\setlength{\tabcolsep}{11pt}
\begin{tabular}{|c|c|}
\hline
\multicolumn{2}{|c|}{($\overline {\rm{MS}}$--BMU, $m_b$)}\tabularnewline
\hline
\hline
$B_d^{\rm LR1}/B_d^{\rm VLL}$ & $B_d^{\rm LR2}/B_d^{\rm VLL}$\tabularnewline
\hline
$1.06(11)$ & $1.14(10)$\tabularnewline
\hline
\hline
$B_s^{\rm LR1}/B_s^{\rm VLL}$ & $B_s^{\rm LR2}/B_s^{\rm VLL}$\tabularnewline
\hline
$0.990(75)$ & $1.073(68)$\tabularnewline
\hline
\end{tabular}
\caption{Bag parameters taken from~\cite{Bazavov:2016nty} [Fermilab/MILC Collaboration, 2016] and adjusted to Buras et al. operator basis.} 
\label{tab:bag_Bmix_last}
\end{center}
\end{table}    

In the SM only the operator $Q_1^{\text{VLL}}(\mu)$ contributes to the 
$\cM(B_q\to \bar B_q) \equiv \cM_{12}(B_q)$ amplitude. 
We normalize it such that the meson-antimeson mass splitting and 
the CP-violating phase of mixing amplitude are defined by 
\be
\Delta M_q = 2  | \cM_{12}(B_q) |~, \qquad 
\phi_{B_q} = {\rm arg} \left[  \cM_{12} (B_q) \right]~.
\ee
The explicit expression in the SM reads
\be
\cM_{12} (B_q)^{\rm SM} =\frac{G_F^2 M_W^2 M_{B_q}}{12\pi^2}\, S_0(x_t) (V_{tb}V_{tq}^*)^2 f_{B_q}^2 \hat\eta_B\, B^{\rm VLL}_q\,,
\ee
with $S_0(x_t) \approx 2.36853$ being the Inami-Lim function~\cite{Inami:1980fz}, and 
$\hat\eta_B\approx 0.842$~\cite{Buras:2001ra} accounting for the QCD running of the effective operator from the $m_t$ to the $m_b$ scale.  In the presence of NP, the expression of $\cM_{12} (B_q)$ is modified; factorizing the SM contribution,
we can generally decompose it as
\begin{equation}
\label{eq:deltaMsd}
 \cM_{12} (B_q) = \cM_{12} (B_q)^{\rm SM} \, \left[ 1+\frac{  \cM_{12}(B_q)^{\rm NP} }{\cM_{12}(B_q)^{\rm SM}}\right ]~.
\end{equation}
The NP modifications can be written in terms of the Wilson coefficients in~\eqref{eq:Delta_F=2} as follows
\begin{align}
\label{eq:deltaMsd_NPoSM}
&\frac{  \cM_{12}(B_q)^{\rm NP} }{
\cM_{12}(B_q)^{\rm SM}} =\frac{1}{(V_{tb}^* V_{tq})^2 R_\text{SM}^\text{loop}} \bigg[\left(\eftWCFS{dd}{V, LL}{\mu_t}{3q3q}+\eftWCFS{dd}{V, RR}{\mu_t}{3q3q}\right) \nonumber\\[2pt]
&\qquad + \frac{P_q^{\rm LR1}(\mu_b)}{P_q^{\rm VLL}(\mu_b)} \bigg(\eftWCFS{dd}{V1, LR}{\mu_t}{3q3q}-\frac{1}{6}\eftWCFS{dd}{V8, LR}{\mu_t}{3q3q}\bigg)-\frac{P_q^{\rm LR2}(\mu_b)}{P_q^{\rm VLL}(\mu_b)}\,\eftWCFS{dd}{V8, LR}{\mu_t}{3q3q}\bigg] \,,
\end{align}
where the SM factor reads
\begin{align}
R_\text{SM}^\text{loop}&=\frac{\sqrt{2}\,G_FM_W^2\, S_0(x_t)}{16\pi^2}=1.5987\times10^{-3}\,,
\end{align}
and where the $P_i^a$ coefficients contain the NNLO QCD corrections, computed in~\cite{Buras:2001ra}, and the bag factors. These are given by\footnote{Here we use the results from~\cite{Buras:2001ra}. In particular, Tables 1 and 2 [with $\alpha^{(5)}_s(M_Z)=0.118$], Eqs.~(7.28)-(7.30) and Eq.~(7.34) [with $m_b(\mu_b)+m_d(\mu_b)=\mu_b=4.4$~GeV and $m_B=5.28$~GeV].}
\begin{align}
\begin{aligned}
P_q^{\rm VLL}(\mu_b)&=0.842\,B_q^{\rm VLL}(\mu_b)\,,\\
P_q^{\rm LR1}(\mu_b)&=-0.663\,B_q^{\rm LR1}(\mu_b)-0.956\,B_q^{\rm LR2}(\mu_b)\,,\\
P_q^{\rm LR2}(\mu_b)&=0.030\,B_q^{\rm LR1}(\mu_b)+2.434\,B_q^{\rm LR2}(\mu_b)\,,
\end{aligned}
\end{align}
Using the results in Table~\ref{tab:bag_Bmix_last} for the bag factors we find
\begin{align}\label{eq:Pfactors}
\begin{aligned}
\frac{P_d^{\rm LR1}(\mu_b)}{P_q^{\rm VLL}(\mu_b)}&=-2.13(14)\,, &&& \frac{P_d^{\rm LR2}(\mu_b)}{P_q^{\rm VLL}(\mu_b)}&=3.33(29)\,,\\
\frac{P_s^{\rm LR1}(\mu_b)}{P_q^{\rm VLL}(\mu_b)}&=-2.00(10)\,, &&& \frac{P_s^{\rm LR2}(\mu_b)}{P_q^{\rm VLL}(\mu_b)}&=3.14(20)\,.
\end{aligned}
\end{align}
In Table~\ref{tab:hadronic} we provide the latest SM determinations and experimental values for mass differences and CP violating phases.

\paragraph*{CP violation in $K-\bar K$ and $D-\bar D$ mixing.}
The formalism for $K-\bar K$ mixing is  identical to that for $B_q$--$\bar B_q$ mixing but for trivial modfications.
The key difference is that in this case the magnitude of the amplitude is dominated by long-distance contributions.
Concerning the clean CP-violating observable $\epsilon_K$, we can write
\be
\Re (\epsilon_K ) = \frac{1}{ 2  \Delta M_K^{\rm exp} } \Im\left[ \cM_{12}(K) \right] 
= \Re (\epsilon_K)^{\rm SM}   +  \frac{1}{ 2 \Delta M_K^{\rm exp} }\,  \Im\left[ \cM_{12}(K)^{\rm NP} \right]  ~.
\ee
Since right-handed rotations involving the first family are negligible,
the NP correction assume the simple form
\bea
\label{eq:epsK}
 |\epsilon_K|^{\rm exp}  = |\epsilon_K|^{\rm SM}  +  \frac{2}{3}\, C_K  P_1^{\rm VLL}(\mu_K)~ \Im\left(\eftWCF{dd}{V, LL}{2121}(\mu_t) \right)~,
\eea
where we have used  $\Re (\epsilon_K ) = |\epsilon_K|/\sqrt{2}$, we have defined 
\be
  C_K  = \frac{ G_F M_K f_K^2 }{  \Delta M^{\rm exp}_K} = 4.23 \times 10^{7}~,
\ee
and we have introduced the factor $P_1^{\rm VLL}(\mu_K) =0.48$~\cite{Buras:2001ra} that encodes QCD corrections
and the bag parameter. 
As far as the magnitude of the amplitude is concerned, we can limit ourselves to impose the weaker constraint 
\be
\label{eq:DmK}
\left|  \frac{\Delta M_K^{\rm NP} }{ \Delta M^{\rm exp}_K } \right | = 
   \frac{8}{3 \sqrt{2}} C_K  P_1^{\rm VLL}(\mu_K)~\left|\eftWCF{dd}{V, LL}{2121}(\mu_t) \right| < 1~.
\ee

In the case of $D-\bar D$ mixing we can also neglect right-handed rotations and corresponding right-handed operators. 
Following the analysis of Ref.~\cite{Carrasco:2014uya}, the constraint following from the non-observation of CP-violation in this system  
can be expressed as
\be
\Im (C_1^D)  =  \frac{4 G_F}{ \sqrt{2} } \,\Im\left(\eftWCF{uu}{V, LL}{2121}(\mu_t) \right)    = (-0.03\pm0.46) \times 10^{-14}\ {\rm GeV}^{-2}~.
\label{eq:ImCDexp} 
\ee

\begin{table}[t]
\hspace{-8pt}
\renewcommand{\arraystretch}{1}
\setlength{\tabcolsep}{11pt}
\begin{tabular}{@{}ccccr@{}}
\toprule[1.6pt]  
 \multicolumn{1}{l}{\bf $\boldsymbol{\Delta F=2}$ transitions} &&& & \\ 
 \midrule
 Observable & Experiment &   & SM  & LEFT \\ 
\midrule
$\Delta M_d$ & $0.5065(19)\,\mathrm{ps}^{-1}$~\cite{Amhis:2016xyh} &   & $0.630(69)\,\mathrm{ps}^{-1}$~\cite{Bazavov:2016nty} & \eqref{eq:deltaMsd}-\eqref{eq:Pfactors} 
\\
$\Delta M_s$ & $17.757(21)\,\mathrm{ps}^{-1}$~\cite{Amhis:2016xyh} &   & $19.6(1.6)\,\mathrm{ps}^{-1}$~\cite{Bazavov:2016nty} & \eqref{eq:deltaMsd}-\eqref{eq:Pfactors} 
\\
$\sin(\phi_{B_s})$ & $ -0.021 \pm 0.031$~\cite{Amhis:2016xyh} &   & $- 0.036  \pm 0.001$~\cite{Bona:2017cxr}
&  \eqref{eq:deltaMsd}-\eqref{eq:Pfactors} 
\\
$\sin(\phi_{B_d})$ & $ -0.680 \pm 0.023$~\cite{Amhis:2016xyh} &   & $- 0.724 \pm 0.028$~\cite{Bona:2017cxr}
 & \eqref{eq:deltaMsd}-\eqref{eq:Pfactors} 
 \\
$10^3 \times | \epsilon_K |$ & $2.228 \pm 0.011$~\cite{Amhis:2016xyh} &   & $2.03 \pm 0.18$~\cite{Bona:2017cxr}  &   \eqref{eq:epsK} 
\\
$10^{14}  \times \Im ( C^D_1)$ &  $-0.03(46)~{\rm GeV}^{-2}$~\cite{Carrasco:2014uya} &   &  $0.$  &   \eqref{eq:ImCDexp}  
\\
\bottomrule[1.6pt] 
\end{tabular}
\caption{\small   List of observables involving hadronic transitions.}
\label{tab:hadronic}
\end{table}

\newpage
%%%%%%%%%%%%%%%%%%%%%%%%%%%%%%%%%%%%%%%%%%%%%%%%%%%%%%%%%%%%%%

%%%%%%%%%%%%%%%%%%%%%%%%%%%%%%%%%%%%%%%%%%%%%%%%%%%%%%%%%%%
\end{document}